\documentclass[11pt,a4paper]{article}
\usepackage{jheppub}

\usepackage{mathtools,slashed,array,float,multicol}
\usepackage[compat=1.1.0]{tikz-feynman}
\newcommand{\pslash}{p \!\!\!\!\; /}
\newcommand{\qslash}{{q \!\!\!/}} 
\newcommand{\Tr}{{\rm Tr}}
\newcommand{\mom}{\mathrm{RI'\mbox{-}MOM}}

\graphicspath{{Plots/}}
\title{\boldmath RI-(S)MOM to
	$\overline{\rm MS}$ conversion for $B_K$ at two-loop order}

\author[a]{Martin~Gorbahn,}
\author[b]{Sebastian~J{\"a}ger}
\author[c]{and Sandra~Kvedarait\.e}

\affiliation[a]{Department of Mathematical Sciences, University of Liverpool, Liverpool, L69 7ZL, UK}
\affiliation[b]{Department of Physics and Astronomy, University of Sussex, Falmer, Brighton BN1 9QH, United Kingdom}
\affiliation[c]{Departamento de Física Teórica y del Cosmos, Universidad de Granada, \\Campus de Fuentenueva, E–18071 Granada, Spain}

\emailAdd{Martin.Gorbahn@liverpool.ac.uk}
\emailAdd{S.Jaeger@sussex.ac.uk}
\emailAdd{skvedaraite@ugr.es}

\date{28 Nov 2024}

\abstract{

The Kaon bag parameter \( \hat{B}_K \) plays a critical role in constraining the parameters of the CKM matrix and in probing physics beyond the Standard Model.
In this work, we improve the precision of \( \hat{B}_K \) to next-to-next-to-leading order (NNLO) and provide
world averages for both $3$- and $4$-flavour theories.
In the course of this, as our main technical development, we carry out the two-loop matching between the RI-(S)MOM and $\overline{\mbox{MS}}$ schemes. 
Our world averages combine all available lattice data, including conversion between the 3- and 4-flavour theories
as appropriate. We obtain the result $\hat B_{K}^{(f=3)} = 0.7627(60)$, which comprises the complete set of $3$- and $4$-flavour
lattice results and can be used directly in phenomenological applications.
The error is dominated by lattice uncertainties and missing higher-order corrections
(residual scale dependence). Our averages include a PDG rescaling factor
of 1.28 reflecting a mild tension among the lattice inputs after inclusion of NNLO corrections in the scheme conversion and
matching across flavour thresholds.
Our averages imply an updated value
$|\epsilon_K|=2.171(65)_\text{pert.}(71)_\text{non-pert.}(153)_\text{param.} \times 10^{-3}$. We briefly discuss
applications of our results to $D$-meson mixing.

}

\begin{document}

\maketitle

\section{Introduction}

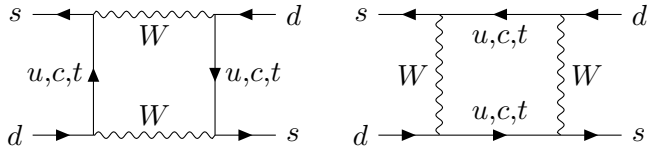
\begin{figure}[b]
	\centering
	\begin{tabular}{c c}
		\begin{tikzpicture} 
			\begin{feynman}[small] 
				\vertex (a) ;
				\vertex [left=0.8cm of a] (e) {$s$}; 
				\vertex [right=1.6 of a] (c) ; 
				\vertex [below=1.6cm of a] (b);
				\vertex [left=0.8cm of b] (f) {$d$};
				\vertex [right=1.6 of b] (d);
				\vertex [right=0.8cm of d] (h) {$s$};	
				\vertex [right=0.8cm of c] (g) {$d$};
				\vertex [right=1.6cm of a] (i);
				\vertex [right=1.6cm of b] (j);
				\node [right=0.8cm of b] (jj);
				
				\vertex [right=3cm of a] (t);
				\vertex [below=0.5cm of t] (r);
				\vertex [right=1cm of r] (s);
				
				\diagram* {
					(a) --[fermion] (e), (c) --[boson,edge label=$W$](a), (b) -- [fermion,edge label=$u\mbox{,}c\mbox{,}t$](a), (f) -- [fermion](b),
					(b) --[boson,edge label=$W$] (d), (c) --[fermion,edge label=$u\mbox{,}c\mbox{,}t$] (d), (g) -- [fermion](c),
					(d) -- [fermion](h)
				};
			\end{feynman}
		\end{tikzpicture}&
		\begin{tikzpicture} 
			\begin{feynman}[small]
				\vertex (a) ;
				\vertex [left=0.8cm of a] (e) {$s$}; 
				\vertex [right=1.6 of a] (c) ; 
				\vertex [below=1.6cm of a] (b);
				\vertex [left=0.8cm of b] (f) {$d$};
				\vertex [right=1.6 of b] (d);
				\vertex [right=0.8cm of d] (h) {$s$};	
				\vertex [right=0.8cm of c] (g) {$d$};
				\vertex [right=1.6cm of a] (i);
				\vertex [right=1.6cm of b] (j);
				\node [right=0.8cm of b] (jj);
				
				\vertex [right=3cm of a] (t);
				\vertex [below=0.5cm of t] (r);
				\vertex [right=1cm of r] (s);
				
				\diagram* {
					(a) -- [fermion](e), (c)--[fermion,edge label=$u\mbox{,}c\mbox{,}t$](a), (b) -- [boson,edge label=$W$](a), (f) -- [fermion](b),
					(b) --[fermion,edge label=$u\mbox{,}c\mbox{,}t$] (d), (c) --[boson,edge label=$W$] (d), (g) --  [fermion](c),
					(d) --[fermion] (h)
				};
			\end{feynman}
		\end{tikzpicture}\\
	\end{tabular}
	\caption{Leading contributions to $K^0$ - $\bar{K}^0$  mixing}
	\label{fig:mixd}
\end{figure}

The Kaon bag parameter $\hat B_K$ \cite{Buras:1990fn}
is one of the key quantities in helping us understand the neutral Kaon mixing as it enters a dominant contribution to indirect CP violation in the Kaon system $\epsilon_K$. Precise determination
of $\hat{B}_K$ allows us to constrain the CKM matrix and probe physics beyond the Standard Model,
with \cite{Brod:2019rzc,Buras:2008nn,Buras:2010pza,Buchalla:1995vs}
\begin{equation}
	|\epsilon_K| = \kappa_{\epsilon}C_\epsilon\hat{B}_K|V_{cb}|^2\lambda^2\bar{\eta}\times\left(|V_{cb}|^2(1-\bar{\rho})\eta_{tt}S_{tt}(x_t)-\eta_{ut}S_{ut}(x_c,x_t)\right)   \label{eq:epsilonK} \,.
\end{equation} 
 Here $S_{tt}(x_t)$ and $S_{ut}(x_c,x_t)$, with $x_{c,t} = m_{c,t}^2/M_W^2$,
are the leading-order weak-scale short distance contributions, due to the diagrams in Figure~\ref{fig:mixd}. $m_{c}$ and $m_{t}$ are the charm and top quark masses and $M_W$ is the $W$ boson mass. $\eta_{tt}$ and $\eta_{ut}$ comprise perturbative QCD and QED corrections to the aforementioned diagrams \cite{Buras:1989xd,Brod:2010mj,Brod:2011ty,Brod:2021qvc,Brod:2022har}, including the integration out of the bottom and charm quarks, as well as physics at intermediate scales. Together with the short distance contributions they give
a scale- and scheme-independent combination.
$\kappa_{\epsilon} = 0.94(2)(1+\delta_{m_c})$ denote subdominant corrections \cite{Buras:2010pza} not included in
$\hat{B}_K$, where $\delta_{m_c} = 0.010(3)$ incorporates an extended analysis of power corrections \cite{Ciuchini:2021zgf}. 
$V_{cb}$, $\lambda$, $\bar \rho$ and $\bar \eta$ represent the CKM inputs.
The remaining parametric input is collected in $ C_\epsilon=(G_F^2 F_K^2 M_K M_W^2)/(6\sqrt{2}\pi^2\Delta M_K)$, where $G_F$ is the Fermi constant, $F_K$ the Kaon decay constant, $M_K$ is the Kaon mass and $\Delta M_K$ is the Kaon mass difference.  Experimentally, $\epsilon_K$ has been precisely measured as $|\epsilon_K|_{\rm exp} = (2.228 \pm 0.011) \times 10^{-3}$~\cite{Workman:2022ynf}.
As for the theory prediction, the perturbative, non-perturbative and parametric uncertainties were found to be at around $3\%$,  $3.5\%$ and  $7\%$ respectively \cite{Brod:2019rzc}.

The remaining ingredient, the renormalization-scale and scheme-independent bag factor is a non-perturbative object
and can be defined as
\begin{equation}
   \hat B_K^{(f=3)} = \lim_{\mu \to \infty} B_K^{A}(\mu) \alpha_s(\mu)^{-2/9}, 
\end{equation}
where $\alpha_s$ is the strong coupling and $f$ is the number of flavours. $A$ denotes an arbitrary scheme, and $\mu$ represents an arbitrary renormalization scale.  Here
\begin{equation}
	\label{eq:1}
	B_K^{ A}(\mu) = \frac{\langle \bar{K}^0| Q^A(\mu) | K^0 \rangle}{\tfrac{8}{3}F_K^2 M_K^2} \,,
\end{equation}
where $\langle \bar{K}^0| Q^A(\mu) | K^0 \rangle$ is the matrix element of a unique dimension-six
$\Delta S=\Delta D=2$ operator, 
\begin{equation}\label{eq:Q}
Q = \big(\overline{s}_L^\alpha \gamma_{\mu} d_L^\alpha\big) 
	\big(\overline{s}_L^\beta \gamma^{\mu}d_L^\beta\big)\,,
\end{equation}
renormalized in the scheme $A$ and at the scale $\mu$. To any given order in perturbation theory, one has
\begin{equation}
   \hat B_K = C^{ A \to \rm RGI}_{B_K}(\mu) B_K^A(\mu)\,,
\end{equation}
with a conversion coefficient $C^{ A \to \rm RGI}_{B_K}(\mu) =  \alpha_s(\mu)^{-2/9}\left(1 + {\cal O}(\alpha_s(\mu))\right)$, where RGI stands for the renormalization-group-invariant bag parameter, which is here defined for three active flavours.
One can also write a formula for $\epsilon_K$ that avoids the perturbative treatment
of the charm scale, which then involves additional nonlocal non-perturbative contributions together with a 4-flavour
version of $\hat B_K$.

The current status of $\hat B_K$ is summarized by the average of $\hat{B}_K=0.7533(91)$~\cite{FlavourLatticeAveragingGroupFLAG:2024oxs}, based on a variety of Lattice calculations with 2+1
active flavours. These calculations use different intermediate schemes for the evaluation of the matrix elements,
which have to be transformed to the RGI bag parameter.

The objective of the present paper is to provide a result for $\hat B_K$ to next-to-next-to leading order (NNLO) accuracy, comprising all available
$f = 3$ and $f=4$ lattice calculations of $B_K$.
To this end, we compute the two-loop conversion
factor $C^{\rm (S)MOM \to \overline{\rm MS}}_{B_K}$ between the regularization invariant (symmetric) momentum-space subtraction (RI-(S)MOM) schemes~\cite{SchemesPhysRevD.80.014501, Martinelli:1994ty},
which can be implemented on the lattice,
and the modified minimal subtraction ($\overline{\rm MS}$) scheme. These conversion factors
can then be combined with existing computations for the matching across the charm threshold \cite{Brod:2010mj}
and for the conversion from the $\overline{\rm MS}$ scheme to RGI bag parameter, to provide the desired result. We will present
the average both in the form of a 3-flavour version $\hat B_K^{(f=3)}$, which can be directly used in
Eq.(\ref{eq:epsilonK}), and a 4-flavour version $\hat B_K^{(f=4)}$ which may be useful for future
phenomenological applications where the charm quark is treated non-perturbatively. We stress that each result
comprises all $3$- and $4$-flavour lattice inputs.

The remainder of this work is organized as follows. In Section~\ref{sec:opren} we review the momentum-space subtraction and $\overline{\mbox{MS}}$ schemes for operator renormalization. In Sections~\ref{sec:matchingcalc} and \ref{sec:loopcalc} we describe the calculation of the amplitude of an amputated four-point Green's function, first focusing on the overall set-up of the matching calculation, followed by a detailed discussion of the loop calculation and renormalization of the amplitude. Finally, in Section~\ref{sec:results} we present our results, including the NNLO conversion factors, their residual scale dependence, world averages for $\hat{B}_K$, and an updated prediction for
$\epsilon_K$ value. We also provide an RGI parameter for D-meson mixing.

\section{Operator renormalization and matching between schemes}\label{sec:opren} 

In this section we review the renormalization of four-quark operators in RI-(S)MOM and $\overline{\mbox{MS}}$ schemes. We also discuss the matching between the two schemes. For a comprehensive introduction to the treatment of four-quark operators we refer to \cite{Buchalla:1995vs, WeakHamBuras:1998raa} and references therein.

\subsection{Momentum-space subtraction schemes}\label{sec:MOM}

The regularization-invariant (RI) schemes are defined by imposing renormalization conditions on certain Lorentz- and colour-invariant linear combinations (``projections'') of two-point and four-point functions for certain
off-shell kinematics. Several variants of the RI schemes exist in the literature, which differ by the kinematics for the renormalization conditions and the choice of projectons used for the two-point and four-point functions. In this work, we are concerned with particular variants of the RI~schemes known as the RI-(S)MOM schemes.

To fix our notation, we write the connected fermion two-point function
$S^A$    in the following form
\begin{equation}
\label{fermion propagator}
(2\pi)^4 i S^A(p)_{\alpha \beta}\delta^4(p-q)\delta^{ij}=\int d^4 x d^4 y e^{i(p\cdot x - q\cdot y)}\langle 0|T\{\psi^{Ai}_{\alpha}(x) \bar{\psi}^{Aj}_\beta(y)\}|0\rangle \,,
\end{equation}
where $\psi$ are the quark fields, $i, j$ represent colour and $\alpha, \beta$ spinor indices.  $A$ labels the renormalization scheme.
From the two-point function we define two projected scalar two-point amplitudes 
\begin{equation}
\label{eq:RI-2pt-pslash}
\sigma^{(A,\slashed{q})} \equiv
\frac{1}{4\;p^2}{\rm Tr}\left[({S^A})^{-1}(p)\slashed{p}\right]
\end{equation}
and
\begin{equation}
\label{eq:RI-pt-gamma}
\sigma^{(A,\gamma_\mu)} \equiv
\frac{1}{16}{\rm Tr}\left[\gamma^{\mu} \frac{\partial ({S^A})^{-1}(p)}{\partial p^{\mu}}\right] \, .
\end{equation}

\begin{figure}[bt]
	\centering
	\includegraphics[trim=3.1cm 14cm 12cm 12cm, clip=true, width=6cm]{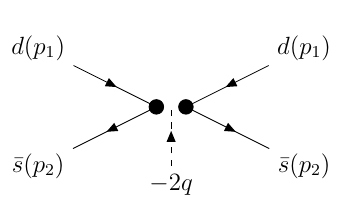}
	\caption{The RI-SMOM momentum subtraction point for a four-quark operator is defined by $p_1^2=p_2^2=q^2=-\mu^2$, while the RI-MOM momentum subtraction point fulfils $p_1^2=p_2^2=-\mu^2$ and $q^2=0$.
          The solid black lines with arrows indicate fermion and momentum flow, the dashed line with arrow indicates momentum flow into the vertex. }
	\label{Fig:SMOM}
\end{figure}

For any choice of field and operator renormalization, we
 define the amputated four-point function $\Lambda_{\rho \sigma \tau \nu}^{ijkl}$ with an operator insertion $Q$ as follows,
\begin{align}
\label{amputated green's function}
\int d^4x &d^4x_{1,2,3,4}  e^{-i(2 q\cdot x +p_1\cdot x_1 -p_2\cdot x_2+p_1\cdot x_3 -p_2\cdot x_4)} \langle
0|T\{\bar{d}^j_\beta(x_1)s^i_\alpha(x_2)\bar{d}^l_\delta(x_3)s^k_\gamma(x_4)Q(x)\}|0\rangle \nonumber\\
&=(2\pi)^4 S(p_1)_{\alpha \rho} S(p_2)_{\beta \sigma}S(p_3)_{\gamma \tau}S(p_4)_{\delta \nu}\Lambda_{\rho \sigma \tau \nu}^{ijkl}(Q)\delta^4 \big( 2q-2(p_1-p_2) \big) \,,
\end{align}
where we have specific quark fields $s$ and $d$ representing the strange and down quarks, and have
left the renormalization scheme implicit to avoid notational clutter.
The corresponding external momentum configuration is shown in Figure~\ref{Fig:SMOM}.

We define projected four-point functions  as
\begin{equation}
\label{eq:RIcond4}
\lambda^{(A, X)} \equiv \Lambda_{\alpha \beta \gamma \delta}^{A,ijkl} P_{(X),\alpha \beta \gamma \delta}^{ijkl} \equiv P_{(X)}(\Lambda^A)\,,
\end{equation}
where $X$ corresponds to the following two choices of projectors:
\begin{align}\label{eq:proj}
{P}^{ijkl}_{(\gamma_{\mu}),\alpha\beta\gamma\delta}&=\frac{(\gamma^\nu)_{\beta\alpha}(\gamma_\nu)_{\delta\gamma}+(\gamma^\nu\gamma^5)_{\beta\alpha}(\gamma_\nu\gamma^5)_{\delta\gamma}}{256N_c(N_c+1)}\delta_{ij}\delta_{kl} 
,\\
{P}^{ijkl}_{(\qslash),\alpha\beta\gamma\delta}&=\frac{(\slashed{q})_{\beta\alpha}(\slashed{q})_{\delta\gamma}+(\slashed{q}\gamma^5)_{\beta\alpha}(\slashed{q}\gamma^5)_{\delta\gamma}}{64q^2N_c(N_c+1)}\delta_{ij}\delta_{kl} 
,\label{eq:proj2}
\end{align}
where $N_c$ is the number of colours, $i,j,k,l$ are colour and $\alpha, \beta, \gamma, \delta$ are spinor indices~\cite{BK}.

The RI-MOM and RI-SMOM schemes are defined by imposing conditions on the projected amplitudes
at either of two momentum-space configurations (kinematics):
\begin{itemize}
	\item MOM kinematics, where $p_1^2=p_2^2=-\mu^2$ and $q^2 \equiv (p_1-p_2)^2 = 0$, 
	\item SMOM kinematics, where $p_1^2=p_2^2=q^2=-\mu^2$.
\end{itemize}
The individual schemes are defined by imposing
{\renewcommand{\jot}{1.5ex}
\begin{align}
&\mbox{RI-MOM:}&\,\,
  &\sigma^{(\rm RI-MOM, \gamma_\mu)} = 1,&
  &\lambda^{(\rm RI-MOM, \gamma_\mu)} = 1&
  &\mbox{at MOM kinematics}, \\
&\mbox{RI${}^\prime$-MOM:}&\,\,
  &\sigma^{(\rm RI'-MOM, \qslash)} = 1,&
  & \lambda^{(\rm RI'-MOM, \gamma_\mu)} = 1&
  &\mbox{at MOM kinematics}, \\
&\mbox{RI-SMOM:}&\,\,
  &\sigma^{(\rm RI-SMOM^{XY}, X)} = 1,&
  &\lambda^{(\rm RI-SMOM^{XY}, Y)} = 1&
  &\mbox{at SMOM kinematics},
\end{align}}
where $X,Y = \gamma_{\mu}, \qslash$ distinguishes four variants of the RI-SMOM scheme.

\subsection{$\overline{\mbox{MS}}$ scheme}\label{sec:Msbar}

The $\overline{\rm MS}$ schemes, within the context of dimensional regularisation, are constructed explicitly out of the bare operators. In general, the renormalized operators $Q_i^{\overline{\rm MS}}(\nu)$ involve several bare operators $Q_j$, which would lead to a renormalization
\begin{equation}\label{eq:Qrengen}
	Q_i^{\overline{\rm MS}}(\nu)=\sum_j Z_{ij}(\nu) Q_j,
\end{equation}
where $Z_{ij}(\nu)$ is a matrix of renormalization constants and $\nu$ is the $\overline{\rm MS}$ renormalization scale. In our case, we have only one physical bare operator $Q$ and a single renormalization constant $Z_{QQ}(\nu)$. In dimensional regularization there are additional operators $E_i$, as a consequence of the larger Dirac algebra in $D \not=4$ compared to $D = 4$. They can be chosen such that they vanish at tree-level in $D=4$ and are known as evanescent operators. For the four-quark operator $Q$ defined in Eq.\eqref{eq:Q}, we have
\begin{equation}\label{eq:Qren}
	Q^{\overline{\rm MS}}(\nu) = Z_{QQ}(\nu) Q + \sum_{i} Z_{QE_i}(\nu) E_i ,
\end{equation}
here and in the following $\overline{\rm MS}$ is the $\overline{\rm MS}$ $\rm{NDR}$ scheme, where `NDR' stands for naive dimensional regularization and specifies anti-commuting $\gamma^5$. $Z_{QQ}(\nu)$ and $Z_{Q E_i}(\nu)$ are renormalization constants.  The Z factors are defined such that the strong gauge coupling $g_s(\nu)$ and the matrix element $\langle Q^{\overline{\rm MS}} \rangle$ (for renormalized quark fields) have a finite limit $\epsilon \to 0$. The Z factors are singular as $\epsilon \to 0$ and, for $\overline{\rm MS}$, are taken equal to the principal parts of their Laurent expansions (i.e. containing only poles in $\epsilon$). In addition, we define the quark field renormalization $Z_q$ as
\begin{equation}
	\psi = (Z_q)^{1/2}\psi^{\overline{\rm MS}},
\end{equation}
where $\psi$ denotes the bare field and  $\psi^{\overline{\rm MS}}$ is the renormalized field.

At loop level, evanescent operators $E_i$  require renormalization just like the physical operators
\begin{equation}
	E_i^{\overline{\rm MS}}(\nu) = Z_{E_i E_j}(\nu) E_j+ Z_{E_i Q}(\nu) Q .
\end{equation}
While the $Z_{E_i E_j}(\nu)$ can again be chosen to be the
principal parts of their
Laurent series, a finite
$Z_{E_iQ}(\nu)$ is generally required in order to
have $\langle E_i^{\overline{\rm MS}} \rangle = 0$ also at loop level. The
$E_i^{\overline{\rm MS}}(\nu)$ renormalized in such a fashion
give vanishing contributions to physical matrix elements. 

\subsection{Scheme conversion}

For any two renormalization schemes $A$ and $B$, the 
operator $Q$ and the quark field $\psi$ are related by a finite conversion factors ${C}^{B \to A}_{B_K}$ and $C_q^{B\to A}$ respectively, defined as
\begin{eqnarray}
   Q^A &=& {C}^{B \to A}_{B_K} Q^B\,,   \label{eq:COdef}\\
   \psi^A&=&(C_q^{B\to A})^{1/2}\psi^B\,,
\end{eqnarray}
which can be computed from the scalar two-point function $\sigma$ and the scalar four-point function $\lambda$ in the two schemes.

For conversion between an RI-(S)MOM scheme and the  $\overline{\rm MS}$ scheme, the field conversion
factor is
 simply given by $C_q^{s \to \overline{\rm MS}} = \sigma^{(\overline{\rm MS},s)}$,
 where $s$ = $\gamma_\mu$ or $\qslash$ depending on the RI-(S)MOM scheme.
The operator conversion factors from RI-SMOM to $\overline{\rm MS}$   are
\begin{equation}
  {C}^{(l,s)\to \overline{\rm MS}}_{B_K} = \lambda^{(\overline{\rm MS},l)} \left( \sigma^{(\overline{\rm MS},s)} \right)^2 , \qquad
  s,l \in \left\{ \slashed{q},\gamma_\mu \right\}\label{eq:COsmom} \,,
\end{equation}
where all $\lambda$ and $\sigma$ are evaluated for SMOM kinematics.
For the RI-MOM and RI${}^\prime$-MOM conversion to $\overline{\rm MS}$ we have
\begin{equation}
	    { C}^{ \mathrm{RI(')\mbox{-}MOM}\to \overline{\rm MS}}_{B_K} = \lambda^{(\overline{\rm MS},\gamma_\mu)} \left( \sigma^{(\overline{\rm MS}, m)} \right)^2 \label{eq:COmom} \,,
\end{equation}
where $m= \slashed{q}$ or $\gamma_\mu$ for (respectively) $\mom$ and $\mathrm{RI\mbox{-}MOM}$,
 and $\lambda$ and $\sigma$ are evaluated for MOM kinematics. 

\section{Matching calculation}\label{sec:matchingcalc}

In this section, we introduce a method to obtain the RI-(S)MOM projections of  the four-point amplitude. This method enables us to bypass tensor reduction at two-loop order by contracting spinor indices in the beginning. Naively, traces over $\gamma_5$ and more than three gamma matrices lead to ambiguities, which we show how to circumvent, using appropriate projectors.

The section is organised as follows. In Section~\ref{sec:treeamp} we examine the structure of $\Lambda^{{\rm(tree)}}$. In Section~\ref{sec:bil} we investigate the Lorentz structures appearing in the two-loop calculation. As part of our method, we also obtain additional Lorentz structures involving external momenta. Hence, in Section~\ref{sec:ops} we define further tree-level matrix elements corresponding to those structures. We also express the amplitude in terms of tree-level matrix elements with corresponding coefficients. In Section~\ref{sec:proje} we define a set of projectors that do not introduce epsilon tensors and hence do not lead to $\gamma_5$  ambiguities. Along with those, we present an extended basis of evanescent operators. In Section~\ref{sec:AAt}, we proceed to write down the amplitude and explain how to obtain the coefficients in front of the tree-level matrix elements. Finally, in Section~\ref{sec:smomp} we discuss how to extract the projections $P_{(\gamma_\mu)}$ and $P_{(\qslash)}$ of the four-point amplitude from those coefficients.

\subsection{Tree-level amplitude}\label{sec:treeamp}

\begin{figure}[bt]
	\centering
	\begin{tabular}{ m{6cm} m{6cm}}
		\begin{tikzpicture} 
			\tikzfeynmanset{
				my dot/.style={
					/tikzfeynman/dot,
					/tikz/minimum size=7pt,
				},
				every vertex/.style={my dot},
			}
			\begin{feynman}[small]
				\node[my dot] (a);
				\node [right=0.5cm of a, my dot] (b);
				\vertex [right=0.25cm of a] (z);
				\vertex [below=1cm of z] (y){$2q$};
				\node [above=1cm of a] (c); 
				\node [below=1cm of a] (d); 
				\vertex [left=2cm of c] (e){$p_1,j,\beta$};
				\vertex [left=2cm of d] (f){$p_2,i,\alpha$};
				\vertex [right=2.5cm of c] (g){$p_1,l,\delta$};
				\vertex [right=2.5cm of d] (h){$p_2,k,\gamma$};

				\diagram* {
					(e) --[fermion,] (a), (a) --[fermion] (f), (g) --[fermion,](b), (b) --[fermion] (h), (z)--[charged scalar](y)
					
				};
			\end{feynman}
		\end{tikzpicture}&
		\begin{tikzpicture} 
			\tikzfeynmanset{
				my dot/.style={
					/tikzfeynman/dot,
					/tikz/minimum size=7pt,
				},
				every vertex/.style={my dot},
			}
			\begin{feynman}[small]
				\node[my dot] (a);
				\node [right=0.5cm of a, my dot] (b);
				\vertex [right=0.25cm of a] (z);
				\vertex [below=1cm of z] (y){$2q$};
				\node [above=1cm of a] (c); 
				\node [below=1cm of a] (d); 
				\vertex [left=2cm of c] (e){$p_1,j,\beta$};
				\vertex [left=2cm of d] (f){$p_2,k,\gamma$};
				\vertex [right=2.5cm of c] (g){$p_1,l,\delta$};
				\vertex [right=2.5cm of d] (h){$p_2,i,\alpha$};

				\diagram* {
					(e) --[fermion,] (a), (a) --[fermion] (f), (g) --[fermion,](b), (b) --[fermion] (h), (z)--[charged scalar](y)
					
				};
			\end{feynman}
		\end{tikzpicture}\\
		\begin{tikzpicture} 
			\tikzfeynmanset{
				my dot/.style={
					/tikzfeynman/dot,
					/tikz/minimum size=7pt,
				},
				every vertex/.style={my dot},
			}
			\begin{feynman}[small]
				\node[my dot] (a);
				\node [right=0.5cm of a, my dot] (b);
				\vertex [right=0.25cm of a] (z);
				\vertex [below=1cm of z] (y){$2q$};
				\node [above=1cm of a] (c); 
				\node [below=1cm of a] (d); 
				\vertex [left=2cm of c] (e){$p_1,j,\beta$};
				\vertex [left=2cm of d] (f){$p_2,k,\alpha$};
				\vertex [right=2.5cm of c] (g){$p_1,l,\delta$};
				\vertex [right=2.5cm of d] (h){$p_2,i,\gamma$};

				\diagram* {
					(e) --[fermion,] (a), (a) --[fermion] (f), (g) --[fermion,](b), (b) --[fermion] (h), (z)--[charged scalar](y)
					
				};
			\end{feynman}
		\end{tikzpicture}&
		\begin{tikzpicture} 
			\tikzfeynmanset{
				my dot/.style={
					/tikzfeynman/dot,
					/tikz/minimum size=7pt,
				},
				every vertex/.style={my dot},
			}
			\begin{feynman}[small]
				\node[my dot] (a);
				\node [right=0.5cm of a, my dot] (b);
				\vertex [right=0.25cm of a] (z);
				\vertex [below=1cm of z] (y){$2q$};
				\node [above=1cm of a] (c); 
				\node [below=1cm of a] (d); 
				\vertex [left=2cm of c] (e){$p_1,j,\beta$};
				\vertex [left=2cm of d] (f){$p_2,i,\gamma$};
				\vertex [right=2.5cm of c] (g){$p_1,l,\delta$};
				\vertex [right=2.5cm of d] (h){$p_2,k,\alpha$};

				\diagram* {
					(e) --[fermion,] (a), (a) --[fermion] (f), (g) --[fermion,](b), (b) --[fermion] (h), (z)--[charged scalar](y)
					
				};
			\end{feynman}
		\end{tikzpicture}
	\end{tabular}
	\caption{The four configurations of the external indices, corresponding to the following structures described in the text: $(\Gamma \otimes  \Gamma \; 1 \otimes 1)$, $(\Gamma \tilde{\otimes}  \Gamma \; 1 \tilde{\otimes} 1)$, $(\Gamma \otimes  \Gamma \; 1 \tilde{\otimes} 1)$, $(\Gamma \tilde{\otimes}  \Gamma \; 1 \otimes 1)$ (left to right, top to bottom). Each dot corresponds to an insertion of a current $\Gamma$ and arrows indicate fermion and momentum flow. }
	\label{fig:config}	
\end{figure}
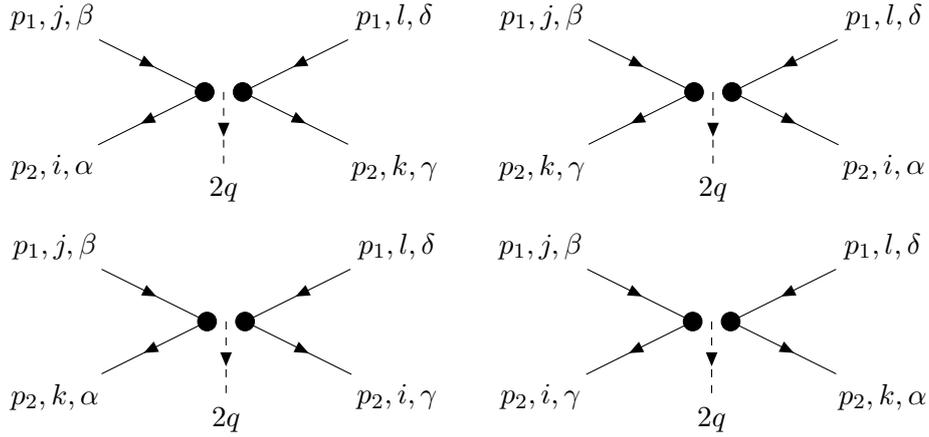

Let us start by considering the tree-level matrix element corresponding to the insertion of the operator $Q$ at (S)MOM kinematics as
\begin{align}\label{eq:LOQ1} 
	\begin{split}
		\langle Q \rangle &\equiv \Lambda^{{\rm(tree)},ijkl}_{\alpha\beta\gamma\delta}(Q) = 2 \left(
		(\gamma^\mu P_L)_{\alpha \beta} (\gamma_\mu P_L)_{\gamma \delta}  \delta^{ij} \delta^{kl}
		- (\gamma^\mu P_L)_{\alpha \delta} (\gamma_\mu P_L)_{\gamma \beta}  \delta^{il} \delta^{kj}
		\right)			 \\
		&\equiv  2 \gamma^\mu P_L \otimes \gamma_\mu P_L \; 1 \otimes 1 \,
		- 2\, \gamma^\mu P_L \tilde{\otimes} \gamma_\mu P_L \; 1 \tilde{\otimes} 1\\
		&\equiv 2\, Q^s \, 1 \otimes 1 - 2\, \tilde{Q}^s \, 1 \tilde{\otimes} 1, 
	\end{split}
\end{align}
where we use the superscript `s' to denote the Dirac structures defined above.
The factor of~2 comes from the fact that we can interchange the two currents. The pictorial representation of the two structures is given in the top row of Fig.(\ref{fig:config}). A further operator~$\tilde{Q}$, corresponding the bottom two diagrams in Fig.(\ref{fig:config}) can be defined as
\begin{equation}
	\tilde{Q}= ( \bar s^i \gamma^\mu P_L  d^l  ) (\bar s^k  \gamma_\mu P_L d^j ),
\end{equation}
with the matrix element given by
\begin{equation}\label{eq:tilde}
	\langle \tilde{Q} \rangle= 2\, Q^s \, 1 \tilde{\otimes} 1 - 2\, \tilde{Q}^s \, 1 \otimes 1.
\end{equation} 
The Greens functions $\langle \tilde Q \rangle$
differ from those of $\langle Q \rangle$ only
by interchange of the two colour structures. 

The expressions for the matrix elements can be split into two parts
\begin{equation}\label{eq:struct}
	\langle Q \rangle = \langle Q \rangle_1 +\langle Q \rangle_2,
\end{equation}
where $\langle Q \rangle_1=-2(\gamma^\mu P_L)_{\alpha \delta} (\gamma_\mu P_L)_{\gamma \beta}  \delta^{il} \delta^{kj}$ and $\langle Q \rangle_2= 2(\gamma^\mu P_L)_{\alpha \beta} (\gamma_\mu P_L)_{\gamma \delta}  \delta^{ij} \delta^{kl}$. Next, we recall that our projectors, defined in Eqs.(\ref{eq:proj}, \ref{eq:proj2}) are of the form
\begin{equation}
	P^{ijkl}_{{(X)},\alpha\beta\gamma\delta}\propto \Gamma_{\beta \alpha} \tilde{\Gamma}_{\delta \gamma}\delta_{ij}\delta_{kl},
\end{equation}
for $X = \qslash$ or $\gamma_\mu$. Projecting the two structures in Eq.\eqref{eq:struct} results in two types of spinor index contractions
\begin{align}
	&P_{(X)} (\langle Q \rangle_1)\propto\Tr(\Gamma\gamma^\mu P_L \tilde{\Gamma} \gamma_\mu P_L),\label{eq:tr2}\\
	&P_{(X)} (\langle Q \rangle_2)\propto\Tr(\Gamma\gamma^\mu P_L)\Tr(\tilde{\Gamma}\gamma_\mu P_L).\label{eq:tr1}
\end{align}
The same traces can be obtained for $\langle \tilde{Q} \rangle$, but with different colour contractions. We will be denoting the structures that result in double traces, like in Eq.\eqref{eq:tr1} as ``crossed", corresponding to diagrams on the LHS of Figure \ref{fig:config}. The structures that lead to single traced contractions as in Eq.\eqref{eq:tr2} and the related diagrams on the RHS of Figure~\ref{fig:config} we will call ``direct". 

\subsection{Bilinears and invariants}\label{sec:bil}

A complete set of bilinears can be chosen to be: 
\begin{align*}
	&\pslash_i P_L, \qquad \gamma^\mu P_L, \qquad \pslash_1 \pslash_2 \gamma^\mu P_L,
\end{align*}
as  in four
dimensions any structure involving a bilinear of length three or more  can be reduced to these three structures.  Consequently, in $D$ dimensions, any Dirac structure
can be written as a linear combination of evanescent structures $E_i^s$ and the following four structures
\begin{equation*}
Q^s = \gamma^\mu P_L \otimes \gamma_\mu P_L,
\end{equation*}
\begin{equation*}
M_{11}^s = \pslash_1 P_L \otimes \pslash_1 P_L,
\quad M_{12}^s =  \frac{1}{2} \left( \pslash_1 P_L \otimes \pslash_2 P_L + \pslash_2 P_L \otimes \pslash_1 P_L \right),
\quad M_{22}^s = \pslash_2 P_L \otimes \pslash_2 P_L.
\end{equation*}

\subsection{Matrix elements entering the total amplitude}\label{sec:ops}

Let us define the tree-level matrix elements, corresponding to the structures above, at generalised (S)MOM kinematics as
\begin{eqnarray}
\langle Q_1 \rangle &=& 2\, Q^s \, 1 \otimes 1 - 2\, \tilde{Q}^s \, 1 \tilde{\otimes} 1,   \label{eq:Q1}   
\\
\langle Q_2 \rangle &=& 
2\, M_{11}^s \, 1 \otimes 1 - 2\, \tilde{M}_{11}^s \, 1 \tilde{\otimes} 1 ,
\\ 
\langle Q_3 \rangle &=&
2\, M_{12}^s \, 1 \otimes 1 - 2\, \tilde{M}_{12}^s \, 1 \tilde{\otimes} 1 ,
\\
\langle Q_4 \rangle &=&
2\, M_{22}^s \, 1 \otimes 1 - 2\, \tilde{M}_{22}^s \, 1 \tilde{\otimes} 1 ,    \label{eq:Q4} 
\end{eqnarray}
We can also define the four further matrix elements $\langle\tilde{Q}_i\rangle$, $i=1, \dots, 4$, with identical Lorentz structures but $\otimes\leftrightarrow \tilde{\otimes}$ for the colour contractions.

The total amplitude up to two loops can then be written in the form
\begin{align}\label{eq:lambda}
\begin{split}
\Lambda &= \sum_{i=1}^4 ( A_i \langle Q_i \rangle + \tilde{A}_i \langle \tilde{Q}_i \rangle )\\
&\quad+\mbox{linear combinations of evanescent Lorentz structures},
\end{split}
\end{align}
where $A_i$ and $\tilde{A}_i$ denote the coefficients in front of the tree-level matrix elements, obtained after reducing the structures appearing in the diagrams that make up the amplitude. The full set of diagrams contains both direct and crossed diagrams, such that the full $\Lambda$ satisfies $\Lambda^{ijkl}_{\alpha\beta\gamma\delta} = - \Lambda^{ilkj}_{\alpha\delta\gamma\beta}$, as required by Fermi statistics, and accounted for by the form of $\langle Q_i \rangle$ and $\langle \tilde{Q}_i \rangle$.

\subsection{Projectors and evanescent structures}\label{sec:proje}

After including all counterterm diagrams the coefficients $A_i$ and $\tilde{A}_i$ are all finite, such that the projectors $P_{(\qslash)}$ and $P_{(\gamma_\mu)}$ can be
directly applied to $\Lambda^{\overline{\rm{MS}}}$.
However, we will not compute all the counterterms (renormalization constants) required to 
obtain finite coefficients for all the evanescent operators and therefore need a method of removing them in the presence
of UV poles. In addition, we would like to use trace techniques to evaluate individual diagrams, which may be divergent. We are able to achieve both aims by choosing a set of projectors which are unambiguous in $D$ dimensions and a set of
evanescent operators which is projected to zero by all projectors.

We choose as projectors
\begin{eqnarray}\label{eq:4proj}
\Pi_\mu (\Gamma_1 P_L \otimes \Gamma_2 P_L) &=& {\rm tr}\, \gamma^\mu \Gamma_1 \gamma_\mu \Gamma_2, \\
\Pi_{11}(\Gamma_1 P_L \otimes \Gamma_2 P_L) &=& {\rm tr}\, \pslash_1 \Gamma_1 \pslash_1 \Gamma_2, \\
\Pi_{12}(\Gamma_1 P_L \otimes \Gamma_2 P_L) &=& {\rm tr}\, \pslash_1 \Gamma_1 \pslash_2 \Gamma_2, \\
\Pi_{22}(\Gamma_1 P_L \otimes \Gamma_2 P_L) &=& {\rm tr}\,  \pslash_2 \Gamma_1 \pslash_2 \Gamma_2, \label{eq:4proj2}
\end{eqnarray}
with no trace over colour.
We have defined them only for the direct diagrams, specified earlier, because, as we will show in the next section, this is sufficient to reconstruct
the entire result. To evaluate them, any chiral projector or $\gamma_5$
in any Dirac line should first be moved to the right end of that line. The traces are unambiguous in $D$ dimensions
because no Levi-Civita symbols are generated by them, nor by the tensor loop integrals we encounter.

We then define evanescent structures $E_j^s$ such that $\Pi_i(E_j^s) = 0$ for all projectors $\Pi_i$ as
\begin{align}
\begin{split}
E_1^s &= \gamma^\mu \gamma^\nu \gamma^\rho P_L \otimes \gamma_\mu \gamma_\nu \gamma_\rho P_L
- ((D-10)D+8) Q^s, \label{eq:E1}
\end{split}\\
\begin{split}
E_3^s &= \gamma^\mu \gamma^\nu \gamma^\rho \gamma^\sigma \gamma^\tau P_L \otimes
\gamma_\mu \gamma_\nu \gamma_\rho \gamma_\sigma \gamma_\tau P_L\\
&\quad-(D-2)(D((D-26)D+152)-128) Q^s, \label{eq:E3}
\end{split}\\
\begin{split}
F_{ij}^s &= \frac{1}{2} \left( \pslash_i \gamma^\mu \gamma^\nu P_L \otimes
\pslash_j \gamma_\mu \gamma_\nu P_L \; + (i \leftrightarrow j) \right)
+ (D-2)(D-4)  M_{ij}^s\\
&\quad-4 p_i \cdot p_j Q^s,
\end{split}\\
\begin{split}
H_{ij}^s &= \frac{1}{2} \left( \pslash_i \gamma^\mu \gamma^\nu \gamma^\rho \gamma^\sigma P_L \otimes
\gamma_\mu \gamma_\nu \gamma_\rho \gamma_\sigma \pslash_j P_L
+ (i \leftrightarrow j) \right)\\
&\quad-(D(D-14)+32)(D-2)(D-4) M_{ij}^s + 8 (D-8) (D-2) p_i \cdot p_j \, Q^s,
\end{split}\\
\begin{split}
G_1^s &= \frac{1}{2} \left( \pslash_1 \pslash_2 \gamma^\mu P_L \otimes \gamma_\mu P_L
+ \gamma^\mu P_L \otimes \pslash_1 \pslash_2 \gamma_\mu P_L \right) -  p_1 \cdot p_2 \, Q^s,
\end{split}\\
\begin{split}
G_2^s &= \pslash_1 \pslash_2 \gamma^\mu \gamma^\nu \gamma^\rho P_L \otimes
\pslash_1 \pslash_2 \gamma_\mu \gamma_\nu \gamma_\rho  P_L \\
&\quad - (D-4) (D (D-14) + 32) \left( p_2^2 M_{11}^s + p_1^2 M_{22}^s - 2 p_1 \cdot p_2 M_{12}^s \right) \\
&\quad+ (D (D-10) + 8) p_1^2 p_2^2 Q^s,
\end{split}\\
\begin{split}
G_3^s &=\frac{1}{2} \left( \pslash_1 \pslash_2 \gamma^\mu \gamma^\nu \gamma^\rho P_L
\otimes \gamma_\mu \gamma_\nu \gamma_\rho P_L
+ \gamma^\mu \gamma^\nu \gamma^\rho P_L
\otimes \pslash_1 \pslash_2 \gamma_\mu \gamma_\nu \gamma_\rho P_L \right)\\
&\quad+ (D (D-10) + 8) p_1 \cdot p_2 Q^s ,
\end{split}\\
\begin{split}
G_4^s &=  \pslash_1 \pslash_2 \gamma^\mu P_L \otimes \pslash_1 \pslash_2 \gamma_\mu P_L  \\
&\quad+ (D-4) \left( p_2^2 M_{11}^s + p_1^2 M_{22}^s - 2 p_1 \cdot p_2 M_{12}^s\right) - p_1^2 p_2^2 Q^s .\qquad\qquad\qquad
\end{split}
\end{align}
All 12 evanescent structures (24 when including colour) are symmetric under the exchange of both bilinears and therefore (upon adding the piece required by Fermi symmetry) are the matrix elements of suitably chosen operators. Only the evanescent structures $E_1^s$ and $E_3^s$ are relevant for the renormalisation at two-loop order. The remaining evanescent structures are needed to derive our projections in the following.

\subsection{Obtaining coefficients $A_i$ and \texorpdfstring{$\tilde{A}_i$}{aO}}\label{sec:AAt}

As before, let us split $\langle Q_i \rangle = \langle Q_i \rangle_1 + \langle Q_i \rangle_2$, where $\langle Q_i \rangle_1$ denotes the
second term on the r.h.s.\ of each of Eqs.(\ref{eq:Q1}--\ref{eq:Q4}), which is due to the direct Feynman diagram. Moreover, we apply the same procedure to $\langle \tilde{Q}_i \rangle$.
The entire amplitude $\Lambda$ then splits in a similar manner into a direct and a crossed contribution. The direct
contribution is due to all the direct diagrams, and has the form
\begin{align}\label{eq:lambda1}
\begin{split}
\Lambda_1  &= \sum_{i=1}^4 ( A_i \langle Q_i \rangle_1 + \tilde{A}_i \langle \tilde{Q}_i \rangle_1)\\
&\quad+\mbox{linear combinations of evanescent Lorentz structures}.
\end{split}
\end{align}
$\Lambda_1$ enjoys the property that our projectors $\Pi_i$ are defined on it, on a diagram-by-diagram basis. We can therefore compute
\begin{align}
\begin{split}
\Pi_i(\Lambda_1) &= \sum_{j=1}^4 \left( A_j \Pi_i(\langle Q_j \rangle_1) + \tilde{A}_j \Pi_i( \langle \tilde{Q}_j \rangle_1) \right)\\
&= \sum_{j=1}^4 B_{ij} \left( A_j 1 \tilde{\otimes} 1 + \tilde{A}_j 1 \otimes 1 \right)\\
&= \sum_{j=1}^4 \left( C_i 1 \tilde{\otimes} 1 +  \tilde{C}_i 1 \otimes 1 \right),
\end{split}
\end{align}
wherein contributions proportional to the tree-level matrix elements of the evanescent operators have disappeared, and
the matrix $B$ is readily found by applying the projectors $\Pi_i$ to the basis Dirac structures $Q^s$, $M_{11}^s$, $M_{12}^s$ and
$M_{22}^s$.
After summing over diagrams and counterterm diagrams, we should find explicitly that $C_i$ and $\tilde{C}_i$ are finite,
and can compute $A_i$ and $\tilde{A}_i$ via the inverse of $B$. $B$ is nonsingular except for $q^2=0$; if we want
a result directly at $q^2=0$ we need to redo the procedure with a subset of basis structures and a $2 \times 2$
$B$-matrix which then should be nonsingular.

Computation of the $B_{ij}=\Pi_i Q_j^s$ for our choice of projectors and operators is given in Table~\ref{tab:Bij}. One can make $B$ dimensionless by rescaling the momentum-dependent basis structures and projectors
by some scalar product(s) of momenta.

\begin{table}
	\centering
	\renewcommand{\arraystretch}{1.2}
	\begin{tabular}{|c | c c c c|}
		\hline
		& $\Pi_\mu$ & $\Pi_{11}$ & $\Pi_{12}$ & $\Pi_{22}$\\
		\hline
		$Q^s$ & $-2(D-2)D$ & $-2(D-2)p_1^2$ & $-2(D-2) (p_1\cdot p_2)$ &  $-2(D-2)p_2^2$\\
		$M_{11}^s$ & $-2(D-2)p_1^2$ & $2p_1^4$ & $2  p_1^2 (p_1\cdot p_2) $ &  $4 (p_1\cdot p_2)^2-2p_1^2p_2^2$\\
		$M_{12}^s$ & $-2(D-2)(p_1\cdot p_2)$ & $2p_1^2 (p_1\cdot p_2)$ & $2p_1^2 p_2^2$ &  $2p_2^2(p_1\cdot p_2)$\\
		$M_{22}^s$ & $-2(D-2)p_2^2$ & $4 (p_1\cdot p_2)^2-2p_1^2p_2^2$ & $2p_2^2 (p_1\cdot p_2)$ &  $2p_2^2$\\
		\hline
	\end{tabular}
\caption{$B_{ij}=\Pi_i Q_j^s$ for projectors, defined in Eqs.(\ref{eq:proj}-\ref{eq:proj2}), and the structures, given in Section~\ref{sec:bil}.}
\label{tab:Bij}
\end{table}

\subsection{Obtaining the (S)MOM projections}\label{sec:smomp}
Once the $A_i$ and $\tilde{A}_i$ are found, calculating the projections $P_{(X)} ( \Lambda ) $, where $X=\qslash$ or $\gamma_\mu$,
(or any other projections) amounts to simply computing
\begin{equation}  \label{eq:finalproj}
P_{(X)} ( \Lambda ) = \sum_{i=1}^4 \left( \sum A_i P_{(X)} (\langle Q_i \rangle)
+ \sum \tilde{A}_i P_{(X)} (\langle \tilde{Q}_i \rangle) \right) ,
\end{equation} 
which is a $D=4$ exercise. Here one needs to include both direct and crossed part and colour. The projections of the tree-level basis structures are given by
\begin{align}
\begin{split}
P_{(\gamma_{\mu})}(\langle Q_i \rangle )&=\left\{\frac{1}{4}-\frac{3\epsilon}{16}+\frac{\epsilon^2}{32},-\frac{\mu^2}{16}\left(1-\frac{\epsilon}{4}\right),\frac{\mu^2}{32}(\omega-2)\left(1-\frac{\epsilon}{4}\right),-\frac{\mu^2}{16}\left(1-\frac{\epsilon}{4}\right)\right\},
\end{split}\\
\begin{split}
P_{(\qslash)}(\langle Q_i \rangle )&=\left\{\frac{1}{4}-\frac{\epsilon}{16},-\frac{\mu^2}{32}(\omega+1),\frac{\mu^2}{64}(3\omega-2),-\frac{\mu^2}{32}(\omega+1)\right\},
\end{split}\\
\begin{split}
P_{(\gamma_{\mu})}(\langle \tilde{Q}_i \rangle )&=\left\{\frac{1}{4}-\frac{5\epsilon}{16}+\frac{3\epsilon^2}{32},-\frac{\mu^2}{16}\left(1-\frac{3\epsilon}{4}\right),\frac{\mu^2}{32}(\omega-2)\left(1-\frac{3\epsilon}{4}\right),\right.\\
&\quad-\left.\frac{\mu^2}{16}\left(1-\frac{3\epsilon}{4}\right)\right\},
\end{split}\\
\begin{split}
P_{(\qslash)}(\langle \tilde{Q}_i \rangle )&=\left\{\frac{1}{4}-\frac{3\epsilon}{16},\frac{\mu^2}{32}(\omega-3),\frac{\mu^2}{64}(\omega-6),\frac{\mu^2}{32}(\omega-3)\right\}.
\end{split}
\end{align}\label{eq:Cs}
where $p_1^2 = p_2^2 = - \mu^2, q^2 = - \omega \mu^2$. For RI-MOM $\omega=0$ and for RI-SMOM $\omega=1$. These expressions are exact, i.e. they include all orders of $\epsilon$.

\section{Loop calculation}\label{sec:loopcalc}

In this section, we present the calculation of the one- and two-loop contributions to the scalar four-point amplitude with (S)MOM kinematics, renormalized in the $\overline{\rm MS}$ scheme. The general procedure for solving loop integrals involves first reducing them into a set of master integrals, which can then be evaluated. This is achieved using the integration by parts (IBP) method \cite{Chetyrkin:1981qh}. In the end, we find that all diagrams reduce to approximately 30 master integrals. These integrals can then be evaluated using a combination of numerical and, where available, analytical results. To obtain the highest precision, it is beneficial to minimize the basis to a set of linearly independent integrals, as numerically computed integrals have an inherent uncertainty. Once the bare amplitude is computed its poles have to cancel against the counterterms. Our computation of counterterms involves one-loop calculations as well as extractions of two-loop renormalization constants from the anomalous dimensions. The result depends on the basis of the evanescent operators and can be converted into a different basis, if needed.  Together with the matching calculation outlined in Section~\ref{sec:matchingcalc} this yields  the projected four-point amplitude, which is finite in $D=4$ dimensions. 

This section is organised as follows. In Section~\ref{sec:NLOd} we briefly discuss the one-loop amplitude.  In Section~\ref{sec:AmpNNLO} we review the diagrams that comprise the bare two-loop amputated Green's functions.  As the two-loop calculation is much more involved, we give further details on how we define the topologies in Section~\ref{sec:top} and evaluate the master integrals in Section~\ref{sec:mast}. In Section~\ref{sec:bgbasis} we describe the change from our basis of evanescent operators to the basis used in the literature, followed by the definition of $\overline{\rm MS}$  renormalized four-point amplitude in Section~\ref{sec:renamp}.

\subsection{Amplitude at NLO}\label{sec:NLOd}

The one-loop amplitude can be obtained from the sum of the following 6 diagrams, including the permutations of external legs (corresponding to Fig.(\ref{fig:config})) and exchange of the two currents:
\begin{figure}[H]
	\centering
	\scalebox{0.85}{\begin{tikzpicture} 
			\tikzfeynmanset{
				my dot/.style={
					/tikzfeynman/dot,
					/tikz/minimum size=7pt,
				},
				every vertex/.style={my dot},
			}
			\begin{feynman}[small]
				\node[my dot] (a);
				\node [right=0.5cm of a, my dot] (b);
				\vertex [right=0.25cm of a] (z);
				
				\node [above=1cm of a] (c); 
				\node [below=1cm of a] (d); 
				\vertex [left=2cm of c] (e);
				\vertex [left=2cm of d] (f);
				\vertex [right=2.5cm of c] (g);
				\vertex [right=2.5cm of d] (h);
				
				\node [left=1cm of a] (k);
				\vertex [above=0.5cm of k] (l);
				\vertex [below=0.5cm of k] (r);
				\vertex [right=2.5cm of l] (m);
				\vertex [below=1.0cm of m] (s);
				\node [left=0.4cm of k] (n);
				\vertex [above=0.7cm of n] (o);
				\vertex [below=0.7cm of n] (p);
				\node [right=1.4cm of b] (t);
				\vertex [above=0.7cm of t] (u);
				\vertex [below=0.7cm of t] (w);
				\node [below=1cm of z] (x) {$a_1$};

				\diagram* {
					(e) --[fermion,] (a), (a) --[fermion] (f), (g) --[fermion,](b), (b) --[fermion] (h),(o)--[gluon](p)
					
				};
			\end{feynman}
	\end{tikzpicture}}\quad
	\scalebox{0.85}{\begin{tikzpicture} 
			\tikzfeynmanset{
				my dot/.style={
					/tikzfeynman/dot,
					/tikz/minimum size=7pt,
				},
				every vertex/.style={my dot},
			}
			\begin{feynman}[small]
				\node[my dot] (a);
				\node [right=0.5cm of a, my dot] (b);
				\vertex [right=0.25cm of a] (z);
				
				\node [above=1cm of a] (c); 
				\node [below=1cm of a] (d); 
				\vertex [left=2cm of c] (e);
				\vertex [left=2cm of d] (f);
				\vertex [right=2.5cm of c] (g);
				\vertex [right=2.5cm of d] (h);
				
				\node [left=1cm of a] (k);
				\vertex [above=0.5cm of k] (l);
				\vertex [below=0.5cm of k] (r);
				\vertex [right=2.5cm of l] (m);
				\vertex [below=1.0cm of m] (s);
				\node [left=0.4cm of k] (n);
				\vertex [above=0.7cm of n] (o);
				\vertex [below=0.7cm of n] (p);
				\node [right=1.4cm of b] (t);
				\vertex [above=0.7cm of t] (u);
				\vertex [below=0.7cm of t] (w);
				\node [below=1cm of z] (x) {$a_2$};

				\diagram* {
					(e) --[fermion,] (a), (a) --[fermion] (f), (g) --[fermion,](b), (b) --[fermion] (h),(w)--[gluon](u)
					
				};
			\end{feynman}
	\end{tikzpicture}}\quad
	\scalebox{0.85}{\begin{tikzpicture} 
			\tikzfeynmanset{
				my dot/.style={
					/tikzfeynman/dot,
					/tikz/minimum size=7pt,
				},
				every vertex/.style={my dot},
			}
			\begin{feynman}[small]
				\node[my dot] (a);
				\node [right=0.5cm of a, my dot] (b);
				\vertex [right=0.25cm of a] (z);
				
				\node [above=1cm of a] (c); 
				\node [below=1cm of a] (d); 
				\vertex [left=2cm of c] (e);
				\vertex [left=2cm of d] (f);
				\vertex [right=2.5cm of c] (g);
				\vertex [right=2.5cm of d] (h);
				
				\node [left=1cm of a] (k);
				\vertex [above=0.5cm of k] (l);
				\vertex [right=2.5cm of l] (m);
				\node [left=0.4cm of k] (n);
				\vertex [above=0.7cm of n] (o);
				\vertex [below=0.7cm of n] (p);
				\node [below=0.9cm of z] (x) {$b_1$};
				
				\diagram* {
					(e) --[fermion,] (a), (a) --[fermion] (f), (g) --[fermion,](b), (b) --[fermion] (h),(u)--[gluon](o)
					
				};
			\end{feynman}
	\end{tikzpicture}}\\
	\scalebox{0.85}{\begin{tikzpicture} 
			\tikzfeynmanset{
				my dot/.style={
					/tikzfeynman/dot,
					/tikz/minimum size=7pt,
				},
				every vertex/.style={my dot},
			}
			\begin{feynman}[small]
				\node[my dot] (a);
				\node [right=0.5cm of a, my dot] (b);
				\vertex [right=0.25cm of a] (z);
				
				\node [above=1cm of a] (c); 
				\node [below=1cm of a] (d); 
				\vertex [left=2cm of c] (e);
				\vertex [left=2cm of d] (f);
				\vertex [right=2.5cm of c] (g);
				\vertex [right=2.5cm of d] (h);
				
				\node [left=1cm of a] (k);
				\vertex [above=0.5cm of k] (l);
				\vertex [right=2.5cm of l] (m);
				\node [left=0.4cm of k] (n);
				\vertex [above=0.7cm of n] (o);
				\vertex [below=0.7cm of n] (p);
				\node [below=0.9cm of z] (x) {$b_2$};
				
				\diagram* {
					(e) --[fermion,] (a), (a) --[fermion] (f), (g) --[fermion,](b), (b) --[fermion] (h),(p)--[gluon](w)
					
				};
			\end{feynman}
	\end{tikzpicture}}\quad
	\scalebox{0.85}{\begin{tikzpicture} 
			\tikzfeynmanset{
				my dot/.style={
					/tikzfeynman/dot,
					/tikz/minimum size=7pt,
				},
				every vertex/.style={my dot},
			}
			\begin{feynman}[small]
				\node[my dot] (a);
				\node [right=0.5cm of a, my dot] (b);
				\vertex [right=0.25cm of a] (z);
				
				\node [above=1cm of a] (c); 
				\node [below=1cm of a] (d); 
				\vertex [left=2cm of c] (e);
				\vertex [left=2cm of d] (f);
				\vertex [right=2.5cm of c] (g);
				\vertex [right=2.5cm of d] (h);
				
				\node [left=1cm of a] (k);
				\vertex [above=0.5cm of k] (l);
				\vertex [below=0.5cm of k] (r);
				\vertex [right=2.5cm of l] (m);
				\vertex [below=1.0cm of m] (s);
				\node [left=0.4cm of k] (n);
				\vertex [above=0.7cm of n] (o);
				\vertex [below=0.7cm of n] (p);
				\node [right=1.4cm of b] (t);
				\vertex [above=0.7cm of t] (u);
				\vertex [below=0.7cm of t] (w);
				\node [below=1cm of z] (x) {$c_1$};

				\diagram* {
					(e) --[fermion,] (a), (a) --[fermion] (f), (g) --[fermion,](b), (b) --[fermion] (h),(o)--[half right,looseness=0.55, gluon](w)
					
				};
			\end{feynman}
	\end{tikzpicture}}\quad
	\scalebox{0.85}{\begin{tikzpicture} 
			\tikzfeynmanset{
				my dot/.style={
					/tikzfeynman/dot,
					/tikz/minimum size=7pt,
				},
				every vertex/.style={my dot},
			}
			\begin{feynman}[small]
				\node[my dot] (a);
				\node [right=0.5cm of a, my dot] (b);
				\vertex [right=0.25cm of a] (z);
				
				\node [above=1cm of a] (c); 
				\node [below=1cm of a] (d); 
				\vertex [left=2cm of c] (e);
				\vertex [left=2cm of d] (f);
				\vertex [right=2.5cm of c] (g);
				\vertex [right=2.5cm of d] (h);
				
				\node [left=1cm of a] (k);
				\vertex [above=0.5cm of k] (l);
				\vertex [below=0.5cm of k] (r);
				\vertex [right=2.5cm of l] (m);
				\vertex [below=1.0cm of m] (s);
				\node [left=0.4cm of k] (n);
				\vertex [above=0.7cm of n] (o);
				\vertex [below=0.7cm of n] (p);
				\node [right=1.4cm of b] (t);
				\vertex [above=0.7cm of t] (u);
				\vertex [below=0.7cm of t] (w);
				\node [below=1cm of z] (x) {$c_2$};

				\diagram* {
					(e) --[fermion,] (a), (a) --[fermion] (f), (g) --[fermion,](b), (b) --[fermion] (h),(p)--[half right,looseness=0.55,gluon](u)
					
				};
			\end{feynman}
	\end{tikzpicture}}
\end{figure}
\noindent The computation of this amplitude for SMOM kinematics can be found in \cite{BK}, where it was done with open indices, by employing the Passarino-Veltman technique and performing the projections in four dimensions. We will reproduce these and the NLO results for MOM kinematics \cite{Ciuchini:1997bw, Buras:2000if} using  our technique as part of the validation.

In order to obtain finite $C_i$ and $\tilde{C}_i$, defined in Eq.(\ref{eq:Cs}), we have to renormalize the amplitude $\Lambda$. At fixed momentum space subtraction scale, the $\overline{\rm MS}$ renormalized amplitude is given by 
\begin{equation}\label{rA}
	\Lambda^{\overline{\rm{MS}}}(\nu)=Z_q^2(\nu)\left(Z_{QQ}(\nu) \langle Q\rangle+Z_{QE_n}(\nu) \langle E_{n}\rangle\right),
\end{equation}
where $\nu$ is the $\overline{\rm MS}$ renormalization scale and $\langle Q\rangle$ and $\langle E_{n}\rangle$ are the bare matrix elements, which can be expanded in $\alpha_s^{\rm bare}$ as
\begin{equation}
	\langle Q\rangle=\langle Q\rangle^{\rm tree}+\frac{\alpha_s^{\rm bare}}{4\pi}\langle Q\rangle^{\rm 1-loop}+\left(\frac{\alpha_s^{\rm bare}}{4\pi}\right)^2\langle Q\rangle^{\rm 2-loop},
\end{equation}
and similarly for $\langle E_{n}\rangle$. The one-loop renormalized amplitude is given by
\begin{align}
	\begin{split}\label{eq:L1}
		\Lambda^{{\overline{\rm{MS}}},(1)}&=\langle Q\rangle^{\rm 1-loop}+\left(2Z^{(1)}_q+Z_{QQ}^{(1)}\right)\langle Q\rangle^{\rm tree}+Z_{QE_F}^{(1)}\langle E_F\rangle^{\rm tree}.
	\end{split}
\end{align}
where $\langle Q\rangle^{\rm 1-loop}$ is the one-loop matrix element, obtained from the  sum of the six diagrams. The wave function renormalization constant $Z_q(\nu)$ is given in Eq.\eqref{eq:MSZq}. For our choice of evanescent operators in Eqs.(\eqref{eq:E1},\eqref{eq:E3}),  the remaining one-loop constants can be found in Appendix \ref{app:olc}. The rest of the evanescent operators do not enter this amplitude as by definition their tree-level matrix elements project to zero.

\subsection{Amplitude at NNLO}\label{sec:AmpNNLO}

\begin{figure}
	\centering
	\begin{tabular}{ m{3cm} m{3cm}  m{3cm}  m{3cm} }
		\scalebox{0.7}{\begin{tikzpicture}
				\tikzfeynmanset{
					my dot/.style={
						/tikzfeynman/dot,
						/tikz/minimum size=7pt,
					},
					every vertex/.style={my dot},
				}
				\begin{feynman}[small]
					\node[my dot] (a);
					\node [right=0.5cm of a, my dot] (b);
					\vertex [right=0.25cm of a] (z);
					
					\node [above=1cm of a] (c); 
					\node [below=1cm of a] (d); 
					\vertex [left=2cm of c] (e);
					\vertex [left=2cm of d] (f);
					\vertex [right=2.5cm of c] (g);
					\vertex [right=2.5cm of d] (h);
					
					\node [left=1cm of a] (k);
					\vertex [above=0.5cm of k] (l);
					\vertex [below=0.5cm of k] (r);
					\vertex [right=2.5cm of l] (m);
					\node [left=0.4cm of k] (n);
					\vertex [above=0.7cm of n] (o);
					\vertex [below=0.7cm of n] (p);
					\node [below=1cm of z] (x) {A1};

					\diagram* {
						(e) --[fermion,] (a), (a) --[fermion] (f), (g) --[fermion,](b), (b) --[fermion] (h),(l)--[gluon](r),(o)--[gluon](p)
						
					};
				\end{feynman}
		\end{tikzpicture}} &
		\scalebox{0.7}{\begin{tikzpicture} 
				\tikzfeynmanset{
					my dot/.style={
						/tikzfeynman/dot,
						/tikz/minimum size=7pt,
					},
					every vertex/.style={my dot},
				}
				\begin{feynman}[small]
					\node[my dot] (a);
					\node [right=0.5cm of a, my dot] (b);
					\vertex [right=0.25cm of a] (z);
					
					\node [above=1cm of a] (c); 
					\node [below=1cm of a] (d); 
					\vertex [left=2cm of c] (e);
					\vertex [left=2cm of d] (f);
					\vertex [right=2.5cm of c] (g);
					\vertex [right=2.5cm of d] (h);
					
					\node [left=1cm of a] (k);
					\vertex [above=0.5cm of k] (l);
					\vertex [below=0.5cm of k] (r);
					\vertex [right=2.5cm of l] (m);
					\vertex [below=1.0cm of m] (s);
					\node [left=0.4cm of k] (n);
					\vertex [above=0.7cm of n] (o);
					\vertex [below=0.7cm of n] (p);
					\node [right=1.4cm of b] (t);
					\vertex [above=0.7cm of t] (u);
					\vertex [below=0.7cm of t] (w);
					\node [below=1cm of z] (x) {A2};

					\diagram* {
						(e) --[fermion,] (a), (a) --[fermion] (f), (g) --[fermion,](b), (b) --[fermion] (h),(o)--[gluon](p),(w)--[gluon](u)
						
					};
				\end{feynman}
		\end{tikzpicture}}&\scalebox{0.7}{\begin{tikzpicture} 
				\tikzfeynmanset{
					my dot/.style={
						/tikzfeynman/dot,
						/tikz/minimum size=7pt,
					},
					every vertex/.style={my dot},
				}
				\begin{feynman}[small]
					\node[my dot] (a);
					\node [right=0.5cm of a, my dot] (b);
					\vertex [right=0.25cm of a] (z);
					
					\node [above=1cm of a] (c); 
					\node [below=1cm of a] (d); 
					\vertex [left=2cm of c] (e);
					\vertex [left=2cm of d] (f);
					\vertex [right=2.5cm of c] (g);
					\vertex [right=2.5cm of d] (h);
					
					\node [left=1cm of a] (k);
					\vertex [above=0.5cm of k] (l);
					\vertex [below=0.5cm of k] (r);
					\vertex [right=2.5cm of l] (m);
					\vertex [below=1.0cm of m] (s);
					\node [left=0.4cm of k] (n);
					\vertex [above=0.7cm of n] (o);
					\vertex [below=0.7cm of n] (p);
					\node [right=1.4cm of b] (t);
					\vertex [above=0.7cm of t] (u);
					\vertex [below=0.7cm of t] (w);
					\node [below=1cm of z] (x) {A3};

					\diagram* {
						(e) --[fermion,] (a), (a) --[fermion] (f), (g) --[fermion,](b), (b) --[fermion] (h),(l)--[gluon](r),(p)--[half right,looseness=0.55, gluon](u)
						
					};
				\end{feynman}
		\end{tikzpicture}}&
		\scalebox{0.7}{\begin{tikzpicture} 
				\tikzfeynmanset{
					my dot/.style={
						/tikzfeynman/dot,
						/tikz/minimum size=7pt,
					},
					every vertex/.style={my dot},
				}
				\begin{feynman}[small]
					\node[my dot] (a);
					\node [right=0.5cm of a, my dot] (b);
					\vertex [right=0.25cm of a] (z);
					
					\node [above=1cm of a] (c); 
					\node [below=1cm of a] (d); 
					\vertex [left=2cm of c] (e);
					\vertex [left=2cm of d] (f);
					\vertex [right=2.5cm of c] (g);
					\vertex [right=2.5cm of d] (h);
					
					\node [left=1cm of a] (k);
					\vertex [above=0.5cm of k] (l);
					\vertex [below=0.5cm of k] (r);
					\vertex [right=2.5cm of l] (m);
					\vertex [below=1.0cm of m] (s);
					\node [left=0.4cm of k] (n);
					\vertex [above=0.7cm of n] (o);
					\vertex [below=0.7cm of n] (p);
					\node [right=1.4cm of b] (t);
					\vertex [above=0.7cm of t] (u);
					\vertex [below=0.7cm of t] (w);
					\node [below=1cm of z] (x) {A4};

					\diagram* {
						(e) --[fermion,] (a), (a) --[fermion] (f), (g) --[fermion,](b), (b) --[fermion] (h),(l)--[half right,looseness=0.55,gluon](s),(o)--[half right,looseness=0.7, gluon](w)
						
					};
				\end{feynman}
		\end{tikzpicture}}\\
		\scalebox{0.7}{\begin{tikzpicture} 
				\tikzfeynmanset{
					my dot/.style={
						/tikzfeynman/dot,
						/tikz/minimum size=7pt,
					},
					every vertex/.style={my dot},
				}
				\begin{feynman}[small]
					\node[my dot] (a);
					\node [right=0.5cm of a, my dot] (b);
					\vertex [right=0.25cm of a] (z);
					
					\node [above=1cm of a] (c); 
					\node [below=1cm of a] (d); 
					\vertex [left=2cm of c] (e);
					\vertex [left=2cm of d] (f);
					\vertex [right=2.5cm of c] (g);
					\vertex [right=2.5cm of d] (h);
					
					\node [left=1cm of a] (k);
					\vertex [above=0.5cm of k] (l);
					\vertex [below=0.5cm of k] (r);
					\vertex [right=2.5cm of l] (m);
					\vertex [below=1.0cm of m] (s);
					\node [left=0.4cm of k] (n);
					\vertex [above=0.7cm of n] (o);
					\vertex [below=0.7cm of n] (p);
					\node [right=1.4cm of b] (t);
					\vertex [above=0.7cm of t] (u);
					\vertex [below=0.7cm of t] (w);
					\node [below=1cm of z] (x) {A5};

					\diagram* {
						(e) --[fermion,] (a), (a) --[fermion] (f), (g) --[fermion,](b), (b) --[fermion] (h),(p)--[half right,looseness=0.55,gluon](u),(o)--[half right,looseness=0.55, gluon](w)
						
					};
				\end{feynman}
		\end{tikzpicture}}&
		\scalebox{0.7}{\begin{tikzpicture} 
				\tikzfeynmanset{
					my dot/.style={
						/tikzfeynman/dot,
						/tikz/minimum size=7pt,
					},
					every vertex/.style={my dot},
				}
				\begin{feynman}[small]
					\node[my dot] (a);
					\node [right=0.5cm of a, my dot] (b);
					\vertex [right=0.25cm of a] (z);
					
					\node [above=1cm of a] (c); 
					\node [below=1cm of a] (d); 
					\vertex [left=2cm of c] (e);
					\vertex [left=2cm of d] (f);
					\vertex [right=2.5cm of c] (g);
					\vertex [right=2.5cm of d] (h);
					
					\node [left=1cm of a] (k);
					\vertex [above=0.5cm of k] (l);
					\vertex [right=2.5cm of l] (m);
					\node [left=0.4cm of k] (n);
					\vertex [above=0.7cm of n] (o);
					\vertex [below=0.7cm of n] (p);
					\node [below=1cm of z] (x) {A6};
					
					\diagram* {
						(e) --[fermion,] (a), (a) --[fermion] (f), (g) --[fermion,](b), (b) --[fermion] (h),(s)--[half right,looseness=0.55,gluon](l),(o)--[gluon](p)
						
					};
				\end{feynman}
		\end{tikzpicture}}&
		\scalebox{0.7}{\begin{tikzpicture} 
				\tikzfeynmanset{
					my dot/.style={
						/tikzfeynman/dot,
						/tikz/minimum size=7pt,
					},
					every vertex/.style={my dot},
				}
				\begin{feynman}[small]
					\node[my dot] (a);
					\node [right=0.5cm of a, my dot] (b);
					\vertex [right=0.25cm of a] (z);
					
					\node [above=1cm of a] (c); 
					\node [below=1cm of a] (d); 
					\vertex [left=2cm of c] (e);
					\vertex [left=2cm of d] (f);
					\vertex [right=2.5cm of c] (g);
					\vertex [right=2.5cm of d] (h);
					
					\node [left=1cm of a] (k);
					\vertex [above=0.5cm of k] (l);
					\vertex [right=2.5cm of l] (m);
					\node [left=0.4cm of k] (n);
					\vertex [above=0.7cm of n] (o);
					\vertex [below=0.7cm of n] (p);
					\node [below=1cm of z] (x) {B1};
					
					\diagram* {
						(e) --[fermion,] (a), (a) --[fermion] (f), (g) --[fermion,](b), (b) --[fermion] (h),(m)--[gluon](l),(o)--[gluon](p)
						
					};
				\end{feynman}
		\end{tikzpicture}}&
		\scalebox{0.7}{\begin{tikzpicture} 
				\tikzfeynmanset{
					my dot/.style={
						/tikzfeynman/dot,
						/tikz/minimum size=7pt,
					},
					every vertex/.style={my dot},
				}
				\begin{feynman}[small]
					\node[my dot] (a);
					\node [right=0.5cm of a, my dot] (b);
					\vertex [right=0.25cm of a] (z);
					
					\node [above=1cm of a] (c); 
					\node [below=1cm of a] (d); 
					\vertex [left=2cm of c] (e);
					\vertex [left=2cm of d] (f);
					\vertex [right=2.5cm of c] (g);
					\vertex [right=2.5cm of d] (h);
					
					\node [left=1cm of a] (k);
					\vertex [above=0.5cm of k] (l);
					\vertex [right=2.5cm of l] (m);
					\node [left=0.4cm of k] (n);
					\vertex [above=0.7cm of n] (o);
					\vertex [below=0.7cm of n] (p);
					\node [below=1cm of z] (x) {B2};
					
					\diagram* {
						(e) --[fermion,] (a), (a) --[fermion] (f), (g) --[fermion,](b), (b) --[fermion] (h),(o)--[half right,looseness=0.55,gluon](w),(m)--[gluon](l)
						
					};
				\end{feynman}
		\end{tikzpicture}}\\
		\scalebox{0.7}{\begin{tikzpicture} 
				\tikzfeynmanset{
					my dot/.style={
						/tikzfeynman/dot,
						/tikz/minimum size=7pt,
					},
					every vertex/.style={my dot},
				}
				\begin{feynman}[small]
					\node[my dot] (a);
					\node [right=0.5cm of a, my dot] (b);
					\vertex [right=0.25cm of a] (z);
					
					\node [above=1cm of a] (c); 
					\node [below=1cm of a] (d); 
					\vertex [left=2cm of c] (e);
					\vertex [left=2cm of d] (f);
					\vertex [right=2.5cm of c] (g);
					\vertex [right=2.5cm of d] (h);
					
					\node [left=0.80cm of a] (k);
					\vertex [above=0.4cm of k] (l);
					\vertex [right=2.1cm of l] (m);
					\node [left=0.6cm of k] (n);
					\vertex [above=0.7cm of n] (o);
					\vertex [below=0.7cm of n] (p);
					\node [below=1cm of z] (x) {C1};
					
					\diagram* {
						(e) --[fermion,] (a), (a) --[fermion] (f), (g) --[fermion,](b), (b) --[fermion] (h),(u)--[gluon](o),(m)--[gluon](l)
						
					};
				\end{feynman}
		\end{tikzpicture}}&
		\scalebox{0.7}{\begin{tikzpicture} 
				\tikzfeynmanset{
					my dot/.style={
						/tikzfeynman/dot,
						/tikz/minimum size=7pt,
					},
					every vertex/.style={my dot},
				}
				\begin{feynman}[small]
					\node[my dot] (a);
					\node [right=0.5cm of a, my dot] (b);
					\vertex [right=0.25cm of a] (z);
					
					\node [above=1cm of a] (c); 
					\node [below=1cm of a] (d); 
					\vertex [left=2cm of c] (e);
					\vertex [left=2cm of d] (f);
					\vertex [right=2.5cm of c] (g);
					\vertex [right=2.5cm of d] (h);
					
					\node [left=1cm of a] (k);
					\vertex [above=0.5cm of k] (l);
					\vertex [right=2.5cm of l] (m);
					\node [left=0.4cm of k] (n);
					\vertex [above=0.7cm of n] (o);
					\vertex [below=0.7cm of n] (p);
					\node [below=1cm of z] (x) {C2};
					
					\diagram* {
						(e) --[fermion,] (a), (a) --[fermion] (f), (g) --[fermion,](b), (b) --[fermion] (h),(u)--[gluon](o),(p)--[gluon](w)
						
					};
				\end{feynman}
		\end{tikzpicture}}&
		\scalebox{0.7}{\begin{tikzpicture} 
				\tikzfeynmanset{
					my dot/.style={
						/tikzfeynman/dot,
						/tikz/minimum size=7pt,
					},
					every vertex/.style={my dot},
				}
				\begin{feynman}[small]
					\node[my dot] (a);
					\node [right=0.5cm of a, my dot] (b);
					\vertex [right=0.25cm of a] (z);
					
					\node [above=1cm of a] (c); 
					\node [below=1cm of a] (d); 
					\vertex [left=2cm of c] (e);
					\vertex [left=2cm of d] (f);
					\vertex [right=2.5cm of c] (g);
					\vertex [right=2.5cm of d] (h);
					
					\node [left=1cm of a] (k);
					\vertex [above=0.5cm of k] (l);
					\vertex [right=2.5cm of l] (m);
					\node [left=0.4cm of k] (n);
					\vertex [above=0.7cm of n] (o);
					\vertex [below=0.7cm of n] (p);
					\node [below=1cm of z] (x) {C3};
					
					\diagram* {
						(e) --[fermion,] (a), (a) --[fermion] (f), (g) --[fermion,](b), (b) --[fermion] (h),(l)--[gluon](r),(u)--[gluon](o)
						
					};
				\end{feynman}
		\end{tikzpicture}}&
		\scalebox{0.7}{\begin{tikzpicture} 
				\tikzfeynmanset{
					my dot/.style={
						/tikzfeynman/dot,
						/tikz/minimum size=7pt,
					},
					every vertex/.style={my dot},
				}
				\begin{feynman}[small]
					\node[my dot] (a);
					\node [right=0.5cm of a, my dot] (b);
					\vertex [right=0.25cm of a] (z);
					
					\node [above=1cm of a] (c); 
					\node [below=1cm of a] (d); 
					\vertex [left=2cm of c] (e);
					\vertex [left=2cm of d] (f);
					\vertex [right=2.5cm of c] (g);
					\vertex [right=2.5cm of d] (h);
					
					\node [left=1cm of a] (k);
					\vertex [above=0.5cm of k] (l);
					\vertex [right=2.5cm of l] (m);
					\node [left=0.4cm of k] (n);
					\vertex [above=0.7cm of n] (o);
					\vertex [below=0.7cm of n] (p);
					\node [below=1cm of z] (x) {C4};
					
					\diagram* {
						(e) --[fermion,] (a), (a) --[fermion] (f), (g) --[fermion,](b), (b) --[fermion] (h),(l)--[half right,looseness=0.55,gluon](s),(u)--[gluon](o)
						
					};
				\end{feynman}
		\end{tikzpicture}}\\
		\scalebox{0.7}{\begin{tikzpicture} 
				\tikzfeynmanset{
					my dot/.style={
						/tikzfeynman/dot,
						/tikz/minimum size=7pt,
					},
					every vertex/.style={my dot},
				}
				\begin{feynman}[small]
					\node[my dot] (a);
					\node [right=0.5cm of a, my dot] (b);
					\vertex [right=0.25cm of a] (z);
					
					\node [above=1cm of a] (c); 
					\node [below=1cm of a] (d); 
					\vertex [left=2cm of c] (e);
					\vertex [left=2cm of d] (f);
					\vertex [right=2.5cm of c] (g);
					\vertex [right=2.5cm of d] (h);
					
					\node [left=0.80cm of a] (k);
					\vertex [above=0.4cm of k] (l);
					\vertex [right=2.1cm of l] (m);
					\node [left=0.6cm of k] (n);
					\vertex [above=0.7cm of n] (o);
					\vertex [below=0.7cm of n] (p);
					\node [below=1cm of z] (x) {D1};
					
					\diagram* {
						(e) --[fermion,] (a), (a) --[fermion] (f), (g) --[fermion,](b), (b) --[fermion] (h),(u)--[gluon](l),(m)--[gluon](o)
						
					};
				\end{feynman}
		\end{tikzpicture}}&
		\scalebox{0.7}{\begin{tikzpicture}
				\tikzfeynmanset{
					my dot/.style={
						/tikzfeynman/dot,
						/tikz/minimum size=7pt,
					},
					every vertex/.style={my dot},
				}
				\begin{feynman}[small]
					\node[my dot] (a);
					\node [right=0.5cm of a, my dot] (b);
					\vertex [right=0.25cm of a] (z);
					
					\node [above=1cm of a] (c); 
					\node [below=1cm of a] (d); 
					\vertex [left=2cm of c] (e);
					\vertex [left=2cm of d] (f);
					\vertex [right=2.5cm of c] (g);
					\vertex [right=2.5cm of d] (h);
					
					\node [left=1cm of a] (k);
					\vertex [above=0.5cm of k] (l);
					\vertex [below=0.5cm of k] (r);
					\vertex [right=2.5cm of l] (m);
					\node [left=0.4cm of k] (n);
					\vertex [above=0.7cm of n] (o);
					\vertex [below=0.7cm of n] (p);
					\node [below=1cm of z] (x) {D2};

					\diagram* {
						(e) --[fermion,] (a), (a) --[fermion] (f), (g) --[fermion,](b), (b) --[fermion] (h),(l)--[gluon](p),(o)--[gluon](r)
						
					};
				\end{feynman}
		\end{tikzpicture}}&
		\scalebox{0.7}{\begin{tikzpicture} 
				\tikzfeynmanset{
					my dot/.style={
						/tikzfeynman/dot,
						/tikz/minimum size=7pt,
					},
					every vertex/.style={my dot},
				}
				\begin{feynman}[small]
					\node[my dot] (a);
					\node [right=0.5cm of a, my dot] (b);
					\vertex [right=0.25cm of a] (z);
					
					\node [above=1cm of a] (c); 
					\node [below=1cm of a] (d); 
					\vertex [left=2cm of c] (e);
					\vertex [left=2cm of d] (f);
					\vertex [right=2.5cm of c] (g);
					\vertex [right=2.5cm of d] (h);
					
					\node [left=1cm of a] (k);
					\vertex [above=0.5cm of k] (l);
					\vertex [below=0.5cm of k] (r);
					\vertex [right=2.5cm of l] (m);
					\vertex [below=1.0cm of m] (s);
					\node [left=0.4cm of k] (n);
					\vertex [above=0.7cm of n] (o);
					\vertex [below=0.7cm of n] (p);
					\node [right=1.4cm of b] (t);
					\vertex [above=0.7cm of t] (u);
					\vertex [below=0.7cm of t] (w);
					\node [below=1cm of z] (x) {D3};

					\diagram* {
						(e) --[fermion,] (a), (a) --[fermion] (f), (g) --[fermion,](b), (b) --[fermion] (h),(w)--[half right,looseness=0.55,gluon](l),(o)--[half right,looseness=0.55, gluon](s)
						
					};
				\end{feynman}
		\end{tikzpicture}}&
		\scalebox{0.7}{\begin{tikzpicture}
				\tikzfeynmanset{
					my dot/.style={
						/tikzfeynman/dot,
						/tikz/minimum size=7pt,
					},
					every vertex/.style={my dot},
				}
				\begin{feynman}[small]
					\node[my dot] (a);
					\node [right=0.5cm of a, my dot] (b);
					\vertex [right=0.25cm of a] (z);
					
					\node [above=1cm of a] (c); 
					\node [below=1cm of a] (d); 
					\vertex [left=2cm of c] (e);
					\vertex [left=2cm of d] (f);
					\vertex [right=2.5cm of c] (g);
					\vertex [right=2.5cm of d] (h);
					
					\node [left=1cm of a] (k);
					\vertex [above=0.5cm of k] (l);
					\vertex [below=0.5cm of k] (r);
					\vertex [right=2.5cm of l] (m);
					\vertex [right=1.25cm of l] (v);
					\node [left=0.4cm of k] (n);
					\vertex [above=0.7cm of n] (o);
					\vertex [below=0.7cm of n] (p);
					\node [below=1cm of z] (x) {D4};

					\diagram* {
						(e) --[fermion,] (a), (a) --[fermion] (f), (g) --[fermion,](b), (b) --[fermion] (h),(v)--[half right,looseness=0.3,gluon](p),(m)--[gluon](v)--[gluon](l)
						
					};
				\end{feynman}
		\end{tikzpicture}}\\
		\scalebox{0.7}{\begin{tikzpicture} 
				\tikzfeynmanset{
					my dot/.style={
						/tikzfeynman/dot,
						/tikz/minimum size=7pt,
					},
					every vertex/.style={my dot},
				}
				\begin{feynman}[small]
					\node[my dot] (a);
					\node [right=0.5cm of a, my dot] (b);
					\vertex [right=0.25cm of a] (z);
					
					\node [above=1cm of a] (c); 
					\node [below=1cm of a] (d); 
					\vertex [left=2cm of c] (e);
					\vertex [left=2cm of d] (f);
					\vertex [right=2.5cm of c] (g);
					\vertex [right=2.5cm of d] (h);
					
					\node [left=1cm of a] (k);
					\vertex [above=0.5cm of k] (l);
					\vertex [right=2.5cm of l] (m);
					\node [left=0.4cm of k] (n);
					\vertex [above=0.7cm of n] (o);
					\vertex [below=0.7cm of n] (p);
					\node [left=0.5cm of a] (q);
					\vertex [above=0.25cm of q] (s);
					\node [below=1cm of z] (x) {T1};
					
					\diagram* {
						(e) --[fermion,] (a), (a) --[fermion] (f), (g) --[fermion,](b), (b) --[fermion] (h),(m)--[gluon](l),(o)--[half right,looseness=1.0,gluon](s)
						
					};
				\end{feynman}
		\end{tikzpicture}}&
		\scalebox{0.7}{\begin{tikzpicture} 
				\tikzfeynmanset{
					my dot/.style={
						/tikzfeynman/dot,
						/tikz/minimum size=7pt,
					},
					every vertex/.style={my dot},
				}
				\begin{feynman}[small]
					\node[my dot] (a);
					\node [right=0.5cm of a, my dot] (b);
					\vertex [right=0.25cm of a] (z);
					
					\node [above=1cm of a] (c); 
					\node [below=1cm of a] (d); 
					\vertex [left=2cm of c] (e);
					\vertex [left=2cm of d] (f);
					\vertex [right=2.5cm of c] (g);
					\vertex [right=2.5cm of d] (h);
					
					\node [left=1cm of a] (k);
					\vertex [above=0.5cm of k] (l);
					\vertex [right=2.5cm of l] (m);
					\node [left=0.4cm of k] (n);
					\vertex [above=0.7cm of n] (o);
					\vertex [below=0.7cm of n] (p);
					\node [left=0.5cm of a] (q);
					\vertex [above=0.25cm of q] (s);
					\node [below=1cm of z] (x) {T2};
					
					\diagram* {
						(e) --[fermion,] (a), (a) --[fermion] (f), (g) --[fermion,](b), (b) --[fermion] (h),(l)--[gluon](r),(s)--[half right,looseness=1.0,gluon](o)
						
					};
				\end{feynman}
		\end{tikzpicture}}&
		\scalebox{0.7}{\begin{tikzpicture} 
				\tikzfeynmanset{
					my dot/.style={
						/tikzfeynman/dot,
						/tikz/minimum size=7pt,
					},
					every vertex/.style={my dot},
				}
				\begin{feynman}[small]
					\node[my dot] (a);
					\node [right=0.5cm of a, my dot] (b);
					\vertex [right=0.25cm of a] (z);
					
					\node [above=1cm of a] (c); 
					\node [below=1cm of a] (d); 
					\vertex [left=2cm of c] (e);
					\vertex [left=2cm of d] (f);
					\vertex [right=2.5cm of c] (g);
					\vertex [right=2.5cm of d] (h);
					
					\node [left=1cm of a] (k);
					\vertex [above=0.5cm of k] (l);
					\vertex [right=2.5cm of l] (m);
					\vertex [below=1.0cm of m] (r);
					\node [left=0.4cm of k] (n);
					\vertex [above=0.7cm of n] (o);
					\vertex [below=0.7cm of n] (p);
					\node [left=0.5cm of a] (q);
					\vertex [above=0.25cm of q] (s);
					\node [below=1cm of z] (x) {T3};
					
					\diagram* {
						(e) --[fermion,] (a), (a) --[fermion] (f), (g) --[fermion,](b), (b) --[fermion] (h),(l)--[half right,looseness=0.55,gluon](r),(s)--[half right,looseness=1.0,gluon](o)
						
					};
				\end{feynman}
		\end{tikzpicture}}&
		\scalebox{0.7}{\begin{tikzpicture} 
				\tikzfeynmanset{
					my dot/.style={
						/tikzfeynman/dot,
						/tikz/minimum size=7pt,
					},
					every vertex/.style={my dot},
				}
				\begin{feynman}[small]
					\node[my dot] (a);
					\node [right=0.5cm of a, my dot] (b);
					\vertex [right=0.25cm of a] (z);
					
					\node [above=1cm of a] (c); 
					\node [below=1cm of a] (d); 
					\vertex [left=2cm of c] (e);
					\vertex [left=2cm of d] (f);
					\vertex [right=2.5cm of c] (g);
					\vertex [right=2.5cm of d] (h);
					
					\node [left=1cm of a] (k);
					\vertex [above=0.5cm of k] (l);
					\vertex [right=2.5cm of l] (m);
					\vertex [left=0.4cm of k] (n);
					\vertex [above=0.7cm of n] (o);
					\vertex [below=0.7cm of n] (p);
					\node [left=0.5cm of a] (q);
					\vertex [above=0.25cm of q] (s);
					\node [below=1cm of z] (x) {T4};
					
					\diagram* {
						(e) --[fermion,] (a), (a) --[fermion] (f), (g) --[fermion,](b), (b) --[fermion] (h),(o)--[gluon](n)--[gluon](p),(s)--[gluon](n)
						
					};
				\end{feynman}
		\end{tikzpicture}}\\
		\scalebox{0.7}{\begin{tikzpicture} 
				\tikzfeynmanset{
					my dot/.style={
						/tikzfeynman/dot,
						/tikz/minimum size=7pt,
					},
					every vertex/.style={my dot},
				}
				\begin{feynman}[small]
					\node[my dot] (a);
					\node [right=0.5cm of a, my dot] (b);
					\vertex [right=0.25cm of a] (z);
					
					\node [above=1cm of a] (c); 
					\node [below=1cm of a] (d); 
					\vertex [left=2cm of c] (e);
					\vertex [left=2cm of d] (f);
					\vertex [right=2.5cm of c] (g);
					\vertex [right=2.5cm of d] (h);
					
					\node [left=1cm of a] (k);
					\vertex [above=0.5cm of k] (l);
					\vertex [right=2.5cm of l] (m);
					\vertex [below=1.0cm of m] (r);
					\node [left=0.4cm of k] (n);
					\vertex [above=0.7cm of n] (o);
					\vertex [below=0.7cm of n] (p);
					\node [left=0.5cm of a] (q);
					\vertex [above=0.25cm of q] (s);
					\vertex [right=1.7cm of l] (v);
					\node [below=1cm of z] (x) {T5};
					
					\diagram* {
						(e) --[fermion,] (a), (a) --[fermion] (f), (g) --[fermion,](b), (b) --[fermion] (h),(r)--[half right,looseness=0.7,gluon](v)--[half right,looseness=0.7,gluon](l),(v)--[gluon](s)
						
					};
				\end{feynman}
		\end{tikzpicture}}&
		\scalebox{0.7}{\begin{tikzpicture} 
				\tikzfeynmanset{
					my dot/.style={
						/tikzfeynman/dot,
						/tikz/minimum size=7pt,
					},
					every vertex/.style={my dot},
				}
				\begin{feynman}[small]
					\node[my dot] (a);
					\node [right=0.5cm of a, my dot] (b);
					\vertex [right=0.25cm of a] (z);
					
					\node [above=1cm of a] (c); 
					\node [below=1cm of a] (d); 
					\vertex [left=2cm of c] (e);
					\vertex [left=2cm of d] (f);
					\vertex [right=2.5cm of c] (g);
					\vertex [right=2.5cm of d] (h);
					
					\node [left=1cm of a] (k);
					\vertex [above=0.5cm of k] (l);
					\vertex [right=2.5cm of l] (m);
					\vertex [below=1.0cm of m] (r);
					\node [left=0.4cm of k] (n);
					\vertex [above=0.7cm of n] (o);
					\vertex [below=0.7cm of n] (p);
					\node [left=0.5cm of a] (q);
					\vertex [above=0.25cm of q] (s);
					\vertex [right=1.7cm of l] (v);
					\vertex [left=1.6cm of u] (cc);
					\node [below=1cm of z] (x) {T6};
					
					\diagram* {
						(e) --[fermion,] (a), (a) --[fermion] (f), (g) --[fermion,](b), (b) --[fermion] (h),(u)--[gluon](cc)--[gluon](s),(cc)--[gluon](o)
						
					};
				\end{feynman}
		\end{tikzpicture}}&
		\scalebox{0.7}{\begin{tikzpicture}
				\tikzfeynmanset{
					my dot/.style={
						/tikzfeynman/dot,
						/tikz/minimum size=7pt,
					},
					every vertex/.style={my dot},
				}
				\begin{feynman}[small]
					\node[my dot] (a);
					\node [right=0.5cm of a, my dot] (b);
					\vertex [right=0.25cm of a] (z);
					
					\node [above=1cm of a] (c); 
					\node [below=1cm of a] (d); 
					\vertex [left=2cm of c] (e);
					\vertex [left=2cm of d] (f);
					\vertex [right=2.5cm of c] (g);
					\vertex [right=2.5cm of d] (h);
					
					\node [left=1cm of a] (k);
					\node [above=0.5cm of k] (l);
					\vertex [below=0.5cm of k] (r);
					\vertex [right=2.5cm of l] (m);
					\node [left=0.4cm of k] (n);
					\vertex [above=0.7cm of n] (o);
					\vertex [below=0.7cm of n] (p);
					\node [below=1cm of z] (x) {OL1};
					\node [left=0.5cm of a] (q);
					\node [blob,above=0.25cm of q] (s);

					\diagram* {
						(e) --[fermion,] (a), (a) --[fermion] (f), (g) --[fermion,](b), (b) --[fermion] (h),(o)--[gluon](p)
						
					};
				\end{feynman}
		\end{tikzpicture}}&
		\scalebox{0.7}{\begin{tikzpicture}
				\tikzfeynmanset{
					my dot/.style={
						/tikzfeynman/dot,
						/tikz/minimum size=7pt,
					},
					every vertex/.style={my dot},
				}
				\begin{feynman}[small]
					\node[my dot] (a);
					\node [right=0.5cm of a, my dot] (b);
					\vertex [right=0.25cm of a] (z);
					
					\node [above=1cm of a] (c); 
					\node [below=1cm of a] (d); 
					\vertex [left=2cm of c] (e);
					\vertex [left=2cm of d] (f);
					\vertex [right=2.5cm of c] (g);
					\vertex [right=2.5cm of d] (h);
					
					\node [left=1cm of a] (k);
					\node [above=0.5cm of k] (l);
					\vertex [below=0.5cm of k] (r);
					\vertex [right=2.5cm of l] (m);
					\node [left=0.4cm of k] (n);
					\vertex [above=0.7cm of n] (o);
					\vertex [below=0.7cm of n] (p);
					\node [below=1cm of z] (x) {OL2};
					\node [left=0.5cm of a] (q);
					\node [blob,above=0.25cm of q] (s);

					\diagram* {
						(e) --[fermion,] (a), (a) --[fermion] (f), (g) --[fermion,](b), (b) --[fermion] (h),(o)--[half right,looseness=0.55,gluon](w)
						
					};
				\end{feynman}
		\end{tikzpicture}}\\
		\scalebox{0.7}{\begin{tikzpicture}
				\tikzfeynmanset{
					my dot/.style={
						/tikzfeynman/dot,
						/tikz/minimum size=7pt,
					},
					every vertex/.style={my dot},
				}
				\begin{feynman}[small]
					\node[my dot] (a);
					\node [right=0.5cm of a, my dot] (b);
					\vertex [right=0.25cm of a] (z);
					
					\node [above=1cm of a] (c); 
					\node [below=1cm of a] (d); 
					\vertex [left=2cm of c] (e);
					\vertex [left=2cm of d] (f);
					\vertex [right=2.5cm of c] (g);
					\vertex [right=2.5cm of d] (h);
					
					\node [left=1cm of a] (k);
					\node [above=0.5cm of k] (l);
					\vertex [below=0.5cm of k] (r);
					\vertex [right=2.5cm of l] (m);
					\node [left=0.4cm of k] (n);
					\vertex [above=0.7cm of n] (o);
					\vertex [below=0.7cm of n] (p);
					\node [below=1cm of z] (x) {OL3};
					\node [left=0.5cm of a] (q);
					\node [blob,above=0.25cm of q] (s);

					\diagram* {
						(e) --[fermion,] (a), (a) --[fermion] (f), (g) --[fermion,](b), (b) --[fermion] (h),(u)--[gluon](o)
						
					};
				\end{feynman}
		\end{tikzpicture}}&
		\scalebox{0.7}{\begin{tikzpicture}
				\tikzfeynmanset{
					my dot/.style={
						/tikzfeynman/dot,
						/tikz/minimum size=7pt,
					},
					every vertex/.style={my dot},
				}
				\begin{feynman}[small]
					\node[my dot] (a);
					\node [right=0.5cm of a, my dot] (b);
					\vertex [right=0.25cm of a] (z);
					
					\node [above=1cm of a] (c); 
					\node [below=1cm of a] (d); 
					\vertex [left=2cm of c] (e);
					\vertex [left=2cm of d] (f);
					\vertex [right=2.5cm of c] (g);
					\vertex [right=2.5cm of d] (h);
					
					\node [left=1cm of a] (k);
					\node [above=0.5cm of k] (l);
					\vertex [below=0.5cm of k] (r);
					\vertex [right=2.5cm of l] (m);
					\node [blob,left=0.4cm of k] (n);
					\vertex [above=0.7cm of n] (o);
					\vertex [below=0.7cm of n] (p);
					\node [below=1cm of z] (x) {OL4};

					\diagram* {
						(e) --[fermion,] (a), (a) --[fermion] (f), (g) --[fermion,](b), (b) --[fermion] (h),(o)--[gluon](n)--[gluon](p)
						
					};
				\end{feynman}
		\end{tikzpicture}}&
		\scalebox{0.7}{\begin{tikzpicture}
				\tikzfeynmanset{
					my dot/.style={
						/tikzfeynman/dot,
						/tikz/minimum size=7pt,
					},
					every vertex/.style={my dot},
				}
				\begin{feynman}[small]
					\node[my dot] (a);
					\node [right=0.5cm of a, my dot] (b);
					\vertex [right=0.25cm of a] (z);
					
					\node [above=1cm of a] (c); 
					\node [below=1cm of a] (d); 
					\vertex [left=2cm of c] (e);
					\vertex [left=2cm of d] (f);
					\vertex [right=2.5cm of c] (g);
					\vertex [right=2.5cm of d] (h);
					
					\node [left=1cm of a] (k);
					\node [above=0.5cm of k] (l);
					\vertex [below=0.5cm of k] (r);
					\vertex [right=2.5cm of l] (m);
					\node [left=0.4cm of k] (n);
					\vertex [above=0.7cm of n] (o);
					\vertex [below=0.7cm of n] (p);
					\node [below=1cm of z] (x) {OL5};
					\node [right=0.25cm of a] (q);
					\node [blob,above=0.7cm of q] (s);

					\diagram* {
						(e) --[fermion,] (a), (a) --[fermion] (f), (g) --[fermion,](b), (b) --[fermion] (h),(u)--[gluon](s)--[gluon](o)
						
					};
				\end{feynman}
		\end{tikzpicture}}&
		\scalebox{0.7}{\begin{tikzpicture}
				\tikzfeynmanset{
					my dot/.style={
						/tikzfeynman/dot,
						/tikz/minimum size=7pt,
					},
					every vertex/.style={my dot},
				}
				\begin{feynman}[small]
					\node[my dot] (a);
					\node [right=0.5cm of a, my dot] (b);
					\vertex [right=0.25cm of a] (z);
					
					\node [above=1cm of a] (c); 
					\node [below=1cm of a] (d); 
					\vertex [left=2cm of c] (e);
					\vertex [left=2cm of d] (f);
					\vertex [right=2.5cm of c] (g);
					\vertex [right=2.5cm of d] (h);
					
					\node [left=1cm of a] (k);
					\node [above=0.5cm of k] (l);
					\vertex [below=0.5cm of k] (r);
					\vertex [right=2.5cm of l] (m);
					\node [left=0.4cm of k] (n);
					\vertex [above=0.7cm of n] (o);
					\vertex [below=0.7cm of n] (p);
					\node [below=1cm of z] (x) {OL6};
					\node [left=0cm of a] (q);
					\node [blob,below=0.6cm of q] (s);

					\diagram* {
						(e) --[fermion,] (a), (a) --[fermion] (f), (g) --[fermion,](b), (b) --[fermion] (h),(o)--[gluon](s)--[gluon](w)
						
					};
				\end{feynman}
		\end{tikzpicture}}\\
		
	\end{tabular}
	\caption{28 classes of diagrams corresponding to the two-loop radiative corrections to the $\Lambda^{ijkl}_{\alpha\beta\gamma\delta}$. The hatched blobs correspond to the sum of one-loop insertions into the propagators.  Kinematics are defined in Figure \ref{Fig:SMOM}.}
	\label{fig:scalar_diag}
\end{figure}
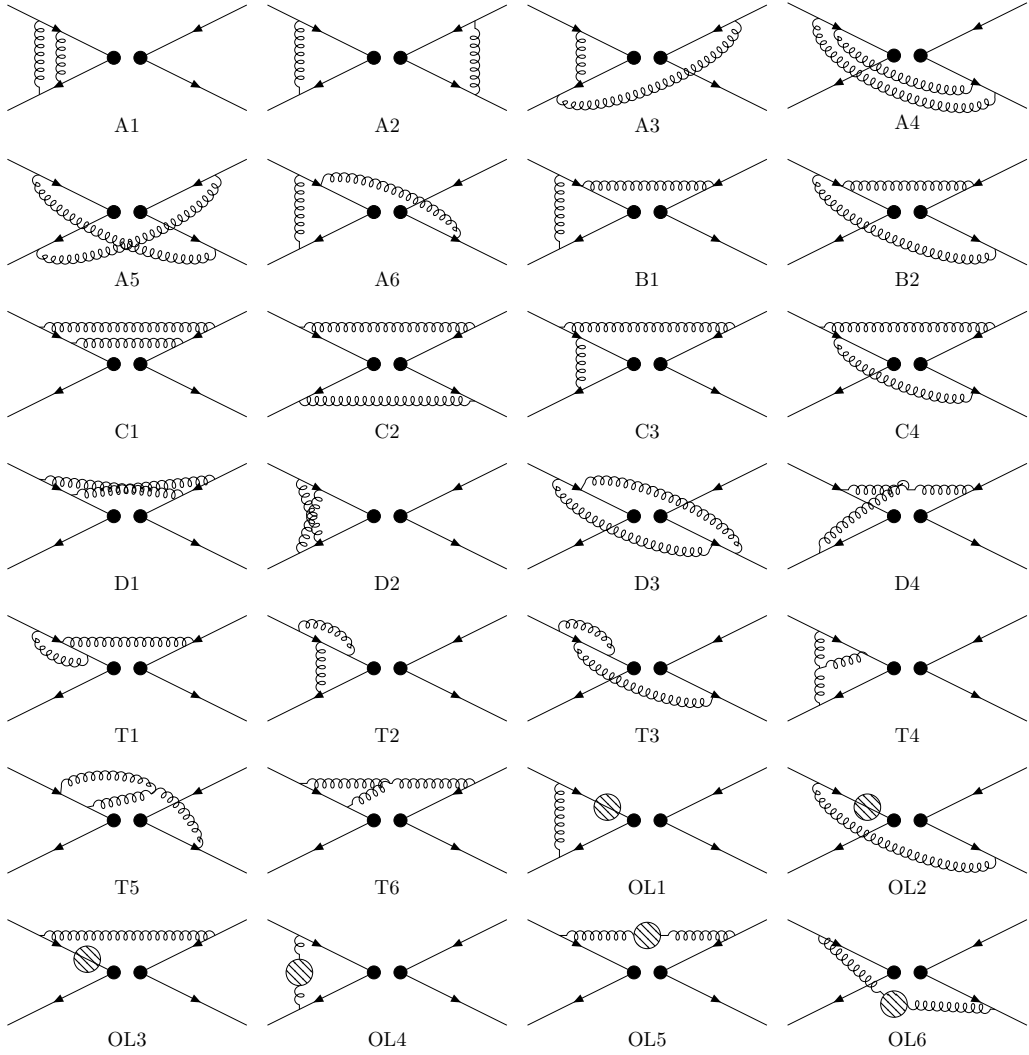

The direct part of the two-loop amplitude is comprised of 103 diagrams.  36 of these are recursively one-loop, i.e. they involve insertions of self-energies into the propagators of one-loop diagrams. The remaining 67 diagrams are the true two-loop diagrams. In Figure~\ref{fig:scalar_diag} we give the pictorial representation of the unique 28 diagrams. The rest of the diagrams can be obtained by exchanging the external legs and the currents. 

\subsection{Defining two-loop topologies}\label{sec:top}

As we perform projections first it is sufficient to consider scalar integrals. Thus, in this section we focus on the scalar part of our four-point function $\Lambda^{ijkl}_{\alpha\beta\gamma\delta}$. At tree level, this amplitude can be pictured as shown in Figure \ref{Fig:SMOM}: we have two particles with momentum $p_1$ going into the vertex and two particles with momentum $p_2$, as well as $q = p_2-p_1$, that ensures momentum conservation in the case of $p_1\neq p_2$, going out of the vertex.  

\begin{figure}
	\begin{tikzpicture} 
	\begin{feynman}[small]
	\vertex (a) ;
	\vertex [left=0.8cm of a] (e) {$p_1$}; 
	\vertex [right=3.2 of a] (c) ; 
	\vertex [below=1.6cm of a] (b);
	\vertex [left=0.8cm of b] (f) {$-(q+p_1)$};
	\vertex [right=3.2 of b] (d);
	\vertex [right=0.8cm of d] (h) {$p_1$};	
	\vertex [right=0.8cm of c] (g) {-$p_2$};
	\vertex [right=1.6cm of a] (i);
	\vertex [right=1.6cm of b] (j);
	\node [below=1cm of j] (l) {Top A};
	
	\diagram* {
		(a) -- (e), (a) --[edge label=1](i)-- [edge label=2](c), (a) -- [edge label=4](b), (b) -- (f),
		(d) --[edge label=7](j)--[edge label=6] (b), (c) -- [edge label=5](d), (c) -- (g),
		(d) -- (h),(i)-- [edge label=3](j)
	};
	\end{feynman}
	\end{tikzpicture}\quad
\begin{tikzpicture} 
	\begin{feynman}[small]
	\vertex (a) ;
	\vertex [left=0.8cm of a] (e) {$p_1$}; 
	\vertex [right=3.2 of a] (c) ; 
	\vertex [below=1.6cm of a] (b);
	\vertex [left=0.8cm of b] (f) {$-(q+p_1)$};
	\vertex [right=3.2 of b] (d);
	\vertex [right=0.8cm of d] (h) {$-p_2$};	
	\vertex [right=0.8cm of c] (g) {$p_1$};
	\vertex [right=1.6cm of a] (i);
	\vertex [right=1.6cm of b] (j);
	\node [below=1cm of j] (l) {Top B};
	
	\diagram* {
		(a) --  (e), (a) --[edge label=1](i)-- [edge label=2](c), (a) -- [edge label=4](b), (b) -- (f),
		(d) --[edge label=7](j)--[edge label=6] (b), (c) -- [edge label=5](d), (c) -- (g),
		(d) -- (h),(i)--[edge label=3](j)
	};
	\end{feynman}
	\end{tikzpicture}
\begin{tikzpicture} 
	\begin{feynman}[small]
		\vertex (a) ;
		\vertex [left=0.8cm of a] (e) {$-p_2$}; 
		\vertex [right=3.2 of a] (c) ; 
		\vertex [below=1.6cm of a] (b);
		\vertex [left=0.8cm of b] (f) {$-(q+p_1)$};
		\vertex [right=3.2 of b] (d);
		\vertex [right=0.8cm of d] (h) {$2q$};	
		\vertex [right=0.8cm of c] (g) {$2p_2$};
		\vertex [right=1.6cm of a] (i);
		\vertex [right=1.6cm of b] (j);
		\node [below=1.0cm of j] (l) {Top C2};
			
		\diagram* {
			(a) -- (e), (a) --[edge label=1](i)--[edge label=2] (c), (a) -- [edge label=4](b), (b) -- (f),
			(d) --[edge label=7](j)-- [edge label=6](b), (c) --[edge label=5] (d), (c) -- (g),
			(d) -- (h),(i)--[edge label=3](j)
		};
	\end{feynman}
	\end{tikzpicture}
\begin{tikzpicture}
	\begin{feynman}[small]
		\vertex (a) ;
		\vertex [left=0.8cm of a] (e) {$-2p_1$}; 
		\vertex [right=3.2 of a] (c) ; 
		\vertex [below=1.6cm of a] (b);
		\vertex [left=0.8cm of b] (f) {$q$};
		\vertex [right=3.2 of b] (d);
		\vertex [right=0.8cm of d] (h) {$p_2$};	
		\vertex [right=0.8cm of c] (g) {$p_1$};
		\vertex [right=1.6cm of a] (i);
		\vertex [right=1.6cm of b] (j);
		\node [below=1cm of j] (l) {Top C3};
		
		\diagram* {
			(a) -- (e), (a) --[edge label=1](i)--[edge label=2] (c), (a) -- [edge label=4](b), (b) -- (f),				(d) --[edge label=7](j)-- [edge label=6](b), (c) --[edge label=5] (d), (c) -- (g),
			(d) -- (h),(i)--[edge label=3](j)
		};
	\end{feynman}
	\end{tikzpicture}
\begin{tikzpicture} 
	\begin{feynman}[small]
		\vertex (d) ;
		\vertex [right=0.5cm of d] (h) {$p_1$};
		\node [left=3.2cm of d] (j);
		\node [above=0.47cm of j] (j1);
		\node [below=0.47cm of j] (j2);
		\vertex [left=1.75cm of d] (c) ; 
		\vertex [below=0.8cm of j] (b);
		\vertex [above=0.8cm of j] (a);
		\vertex [left=0.8cm of a] (e){$p_1$};
		\vertex [left=0.8cm of b] (f){$-(q+p_1)$};	
		\vertex [right=0.4cm of c] (g){$-p_2$};
		\vertex [right=1.6cm of j1] (k);
		\vertex [right=1.6cm of j2] (l);
		\node [below=1.35cm of l] (m) {Top D};
		
		\diagram* {
			(a) -- (e), (a) --[edge label=1](k), (c)--[edge label=3](k), (b) --[edge label=4] (a), (k) -- [edge label=2](d),
			(b) -- (f), (d) -- [edge label=7](l), (l) --[edge label=6](b),(l)--[edge label=5] (c), (g) -- (c),
			(d) -- (h),
		};
	\end{feynman}
	\end{tikzpicture}
	\caption{Topologies A, B, C2, C3 and D.  All external momenta are defined as incoming. Numbers correspond to propagator labels, as defined in Section~\ref{sec:top}.}
	\label{fig:ABDtopologies}
\end{figure}
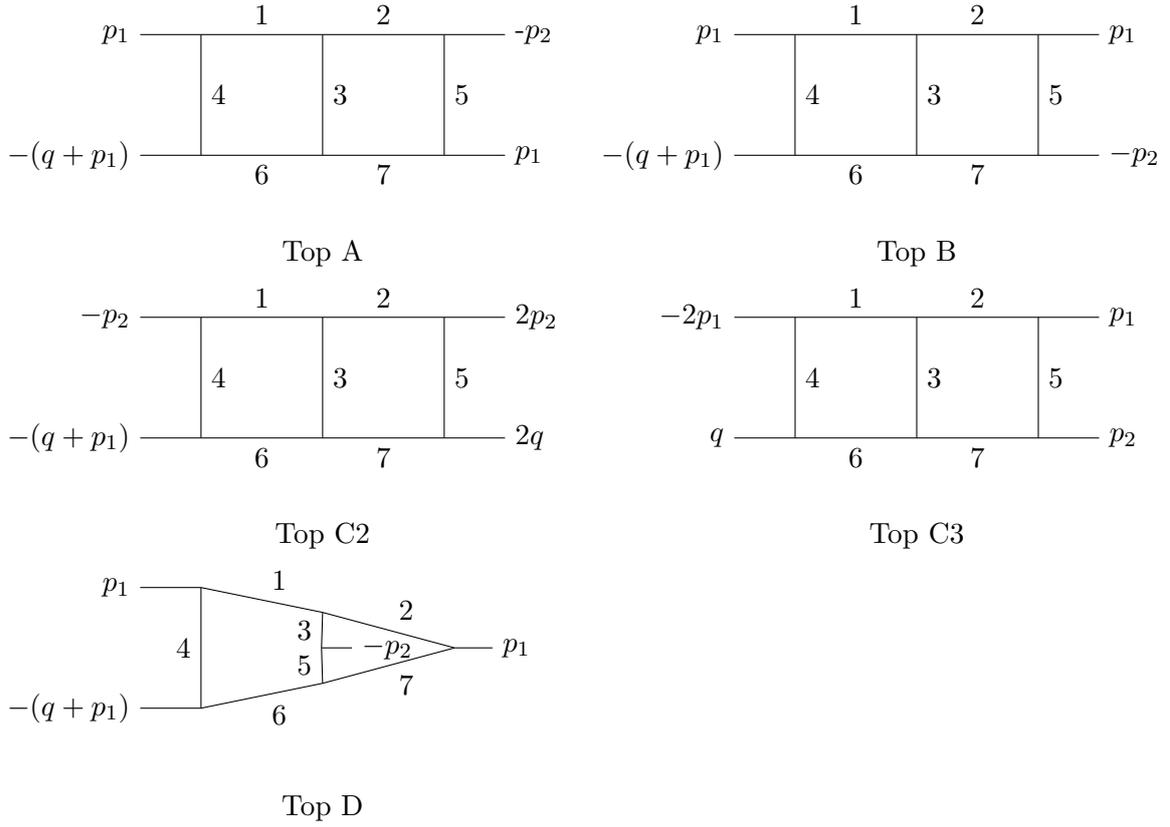

In order to compute the NNLO corrections for this diagram, we have to consider all of the possible two-loop radiative corrections, given in Figure \ref{fig:scalar_diag}. The diagrams can be divided into six groups: A, B, C, D, T and OL. Diagrams A, B and D correspond to the topologies with the same name. The diagrams C have linearly dependent propagators and can be written in terms of topologies A, B, C2 and C3. T stands for the remaining triangle integrals and OL for the integrals with one-loop insertions, both of which can be expressed in terms of topology A, B, C and D integrals. All of the relevant topologies are shown in Figure \ref{fig:ABDtopologies} and the propagators are given in Table \ref{array1}. Further details on the mapping of the integrals onto the topologies  and the reduction of integrals with linearly dependent propagators can be found in \cite{Kvedaraite:2021ymv}.

\begin{table}[b]
	\renewcommand{\arraystretch}{1.0}
	\centering
	\begin{tabular}{|m{0.4 cm}| m{2cm}| m{2cm}| m{2cm} | m{2.5cm} | m{2cm}| }
		\hline
		&Top A& Top B&Top C2&Top C3&Top D \\
		\hline
		1.&$k_1$& $k_1$&$k_1$& $k_1$& $k_1$\\
		2.&$k_2$& $k_2$&$k_2$& $k_2$& $k_2$\\
		3.&$k_1-k_2$& $k_1-k_2$&$k_1-k_2$& $k_1-k_2$& $k_1-k_2$\\
		4.&$k_1+p_1$& $k_1-p_1$&$k_1+p_2$& $k_1+2p_1$& $k_1-p_1$\\
		5.&$k_2+p_2$& $k_2+p_1$&$k_2+2p_2$& $k_2+p_1$& $k_1-k_2-p_2 $\\
		6.&$k_1-q$& $k_1+q$&$k_1+2p_1$& $k_1+p_1+p_2$& $k_1+q$\\
		7.&$k_2-q$& $k_2+q$&$k_2+2p_1$& $k_2+p_1+p_2$& $k_2+p_1$\\
		\hline
	\end{tabular}
	\caption{List of propagators for topologies in Figure \ref{fig:ABDtopologies}. The numbers correspond to the numbering of propagators in the corresponding topology.}
	\label{array1}
\end{table}

\subsection{Master integrals and evaluation}\label{sec:mast}

\begin{figure}[t]
	\centering
	\includegraphics[trim=3.3cm 12.6cm 2.8cm 9.5cm, clip=true, width=15.6cm]{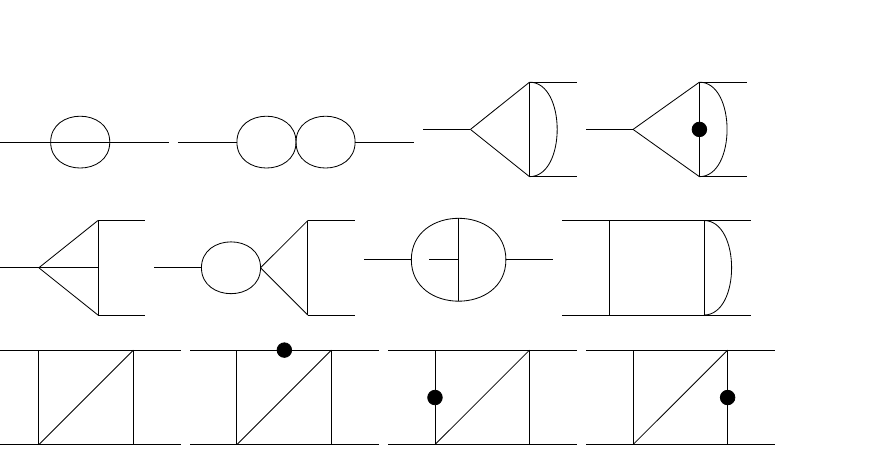}
	\caption{Two-loop master integrals. Black dots represent squared propagators.}
	\label{fig:mast}
\end{figure}

After expressing all the integrals in terms of the five topologies we can proceed to calculate the IBP identities, using \texttt{Reduze} 2 \cite{Reduze2,Laporta:2001dd}. This allows us to express the results in terms of a minimal set of master integrals. We find that our set contains 15 unique two-loop diagrams, shown in Figure~\ref{fig:mast}.

The computation on the Lattice is done at a fixed renormalization scale, hence it is sufficient to obtain the matching coefficient numerically at the corresponding scale.  All of the necessary bubble diagrams are available up to any order in $\epsilon$. The two-loop triangle diagrams have been calculated analytically up to finite order in the literature, whereas the results for the one-loop triangle are available to $\mathcal{O}(\epsilon)$ \cite{Usyukina:1994iw,Gehrmann:1999as}. In addition to these we also need the one-loop triangle up to $\mathcal{O}(\epsilon^2)$ for the one-loop matrix elements. There are no analytic results for the box diagrams with four off-shell legs available.  We calculate the missing pieces and the box diagrams using sector decomposition method. We use \texttt{PySecDec} \cite{Borowka:2017idc} to facilitate the evaluation of two-loop off-shell box diagrams as well as obtain the missing $\mathcal{O}(\epsilon)$ and $\mathcal{O}(\epsilon^2)$ of triangle diagrams. A more detailed discussion of the master integrals for the SMOM case can be found in \cite{Kvedaraite:2021ymv}.

\subsection{Operators in Brod--Gorbahn basis}\label{sec:bgbasis}

Our choice of evanescent operators differs from the ones used by J. Brod and M. Gorbahn (BG)~\cite{Brod:2010mj} for the Wilson coefficients. Here and in the next section we will briefly refer to our scheme as  $\text{GJK}$. The relevant operators are given by
\begin{align}\label{eq:evabasisBG}
	\begin{split}
		E_1^{(2),\rm BG} &= ( \bar s^i\gamma^{\mu_1\mu_2\mu_3\mu_4\mu_5} P_L  d^j  ) (\bar s^k 
		\gamma_{\mu_1\mu_2\mu_3\mu_4\mu_5} P_L  d^l  )
		- \left(256 - 224 \epsilon - \frac{108\, 816}{325} \epsilon^2\right) Q   ,
		\\
		E_2^{(2),\rm BG} &= ( \bar s^i\gamma^{\mu_1\mu_2\mu_3\mu_4\mu_5} P_L  d^l  ) (\bar s^k 
		\gamma_{\mu_1\mu_2\mu_3\mu_4\mu_5} P_L  d^j  )
		- \left(256 - 224 \epsilon - \frac{108\, 816}{325} \epsilon^2\right) ( Q+E_F ),
		\nonumber
	\end{split}
\end{align}
where $\gamma^{\mu_1\mu_2\ldots}=\gamma^{\mu_1}  \gamma^{\mu_2}\dots$.  Comparing with Eq.(\ref{eq:E3}), we can see that the difference is in the $\epsilon^2$ parts of $E_3$. Our choice of evanescent operators can be translated to the ones in which BG have
obtained the NNLO Wilson coefficients and anomalous dimension matrices as follows:
\begin{align}
	E_3 &=E_1^{(2), \rm BG} + \kappa \, \epsilon^2 Q ,
	\\
	E_4 &= E_2^{(2), \rm BG} + \kappa \, \epsilon^2 (Q + E_F) ,
\end{align}
where $\kappa = -62\,016/325$. As a result, the renormalized Green's functions of $Q$ to NNLO differs between the two
schemes as
\begin{align}\label{eq:QBG}
		\langle Q^{\overline{\rm MS},\text{GJK}} \rangle = \langle Q^{\overline{\rm MS}, \rm BG} \rangle
		+ \kappa Z^{(2, 2)}_{QE_3}  \langle Q \rangle^{\rm tree}+ \kappa Z^{(2, 2)}_{QE_4}   ( \langle Q \rangle^{\rm tree} +  \langle E_F \rangle^{\rm tree}),
\end{align}
with the $Z$ factors given in Appendix \ref{app:tlc}.

\subsection{Renormalized amplitude}\label{sec:renamp}

The renormalized two-loop amplitude can be written as
\begin{align}
\begin{split}
\Lambda^{{\overline{\rm MS}} ,(2)}&=\langle Q\rangle^{\rm 2-loop}+2Z^{(1)}_q \Lambda^{{\overline{\rm MS}} ,(1)}+\left( Z_g^{(1)}+Z_\xi^{(1)}\xi\frac{\partial}{\partial \xi}+Z_{QQ}^{(1)}\right)\langle Q\rangle^{\rm 1-loop}\\
&+Z_{QE_n}^{(1)}\langle E_n\rangle^{\rm 1-loop}+\left(-3(Z^{(1)}_q)^2+2Z^{(2)}_q+Z_{QQ}^{(2)}\right)\langle Q\rangle^{\rm tree}+Z_{QE_n}^{(2)}\langle E_n\rangle^{\rm tree}.
\end{split}
\end{align}
where  $n=\{F,1,2\}$. $\langle Q\rangle^{\rm 2-loop}$ is the two-loop matrix element, obtained from the sum of all diagrams, discussed in Section \ref{sec:AmpNNLO}. $\Lambda^{{\overline{\rm MS}} ,(1)}$ is the renormalized one-loop amplitude, defined in Eq.\eqref{eq:L1}. $Z_{QQ}^{(1)}$ and $Z_{QE_n}^{(1)}$ are the one-loop and  $Z_{QQ}^{(2)}$ and $Z_{QE_n}^{(2)}$  are the two-loop counterterms, given in Appendices \ref{app:olc} and \ref{app:tlc} respectively, for our scheme. The one- and two-loop wave function renormalization constants $Z_q^{(1)}$ and $Z_q^{(2)}$  are listed in Eq.\eqref{eq:MSZq}. The gauge and gauge parameter $Z$-factors $Z_g^{(1)}$ and $Z_\xi^{(1)}$ are provided in Eqs.(\ref{eq:MSZg}, \ref{eq:MSZxi}) respectively. Including the scheme change outlined in the previous section along with $Z$ factors in Eqs.(\ref{eq:BG1}, \ref{eq:BG2}), the amplitude in the BG scheme is given by
\begin{equation}
	\Lambda^{{\overline{\rm MS}} ,\text{BG},(2)}=\Lambda^{{\overline{\rm MS}},\text{GJK},(2)}- \left(\left(\kappa Z^{(2, 2)}_{QE_3}+\kappa Z^{(2, 2)}_{QE_4}\right) \langle Q \rangle^{\rm tree}+ \kappa Z^{(2, 2)}_{QE_4} \langle E_F \rangle^{\rm tree}\right).
\end{equation}
All of our $\text{RI-(S)MOM}$ to $\overline{\text{MS}}$ conversion results in the following sections are presented for the BG $\overline{\text{MS}}$  scheme, to which we simply refer to as $\overline{\text{MS}}$  scheme.

\section{Results}
\label{sec:results}
We present our results for the two-loop conversion factors from RI-(S)MOM to the $\overline{\text{MS}}$ scheme in Section~\ref{sec:results1} along with an analytic expression for the coefficients at general MOM and  $\overline{\text{MS}}$ renormalization scales. In Section~\ref{sec:results2}  we present the formulas for obtaining $C_{B_K}^{S \to \rm{RGI}}$ and study its residual scale dependence in the 3 and 4 flavour theories.
  In Section~\ref{sec:BK}  we use our results for the conversion factors to obtain $B_K^{\overline{\text{MS}}}$ and $\hat{B}_K$ and investigate the uncertainty due to unknown N${}^3$LO corrections.
In Section \ref{sec:results4}  we discuss conversion between numbers of flavours for  $\hat{B}_K$. In Section \ref{sec:results5} we proceed to obtain our main result for $\hat{B}_K$ by updating the current FLAG average. Finally, we  update the value of $\epsilon_K$ in Section~\ref{sec:results6} and the bag coefficient for D~mixing in Section~\ref{sec:results7}.

\subsection{$C_{B_K}^{\text{(S)MOM}\rightarrow\overline{\text{MS}}}$ up to NNLO}\label{sec:results1}

We present the conversion factors $C_{B_K}^{\text{(S)MOM}\rightarrow\overline{\text{MS}}}$ for matching between RI-(S)MOM schemes to $\overline{\text{MS}}$ scheme up to two-loop order computed at Landau gauge and $N_c=3$.  The $\overline{\text{MS}}$ scheme is defined with anti-communting $\gamma_5$ and evanescent operators in line with \cite{Brod:2010mj}. In deriving the conversion factors via Eqs.(\ref{eq:COsmom}, \ref{eq:COmom}), we take  NLO coefficients of  $C_q^{\qslash\to\overline{\text{MS}}}$ and $C_q^{\gamma_{\mu}\to\overline{\text{MS}}}$ from \cite{BK} and the NNLO coefficients from \cite{Gracey:2003yr} (setting $C_A=N_c$, $T_F=1/2$), \cite{Gorbahn:2010bf} (setting $w=1$, $r=1$) and \cite{Chetyrkin:1999pq}. We compute the projected amplitudes $P_{(\qslash)}(\Lambda)$ and $P_{(\gamma_\mu)}(\Lambda)$, as outlined in Section~\ref{sec:matchingcalc}, using the renormalized amplitudes defined in Section~\ref{sec:loopcalc} together with the countertems, given in Appendices~\ref{app:olc} and \ref{app:tlc}. Expanding the conversion factors in $\alpha_s$  while keeping the RI-(S)MOM and $\overline{\text{MS}}$ renormalization scales general gives
\begin{equation}
	\label{eq:CBKexp}
	\begin{split}
		C_{B_K}^{S\rightarrow\overline{\text{MS}}}(\mu,\nu)&=1+\frac{\alpha_s(\nu)}{4\pi} \left[C_{B_K,\text{NLO}}^{S}-4 L(\mu,\nu) \right]+ \frac{\alpha_s^2(\nu)}{16\pi^2}
		\left[ C_{B_K,\text{NNLO}}^{S} \right.\\
		&+
	\left.	C_{B_K,\text{NLO}}^{S} L(\mu,\nu) (18-\tfrac{3}{9}f) +
		L(\mu,\nu) \left\{ (7-\tfrac{4}{9}f) - L(\mu,\nu) (72-\tfrac{12}{9}f) \right\} \right],\quad\qquad
	\end{split}
\end{equation}
where $S$ is the momentum subtraction scheme,
$L(\mu,\nu)=\log(\nu/\mu)$, $\mu = \sqrt{-p^2}$ is the momentum subtraction scale, $\nu$ is the $\overline{\text{MS}}$ scale, $
f$ is the number of flavours, and $C_{B_K,\text{NLO}}^{S}$ and $C_{B_K,\text{NNLO}}^{S}$ are the NLO and NNLO coefficients of  $C_{B_K}^{S\rightarrow\overline{\text{MS}}}$, values of which are presented in Table \ref{tab:twoloop}. The coefficients of the $\alpha_s$ expansion are independent of the momentum subtraction scale  as is the fully analytic NLO result. However, we evaluate the NNLO coefficient in part numerically, and this depends on the ratio $\mu/\nu$ (but not
on the scales individually). In Table \ref{tab:twoloop} we present numerical results for $\mu=\nu$,
keeping the number of flavours $f$ general, assuming all quarks are massless. In deriving the NLO results we have also checked that our method correctly reproduces the corresponding values in \cite{BK}. The main uncertainties in these results arise from the numerical evaluation of the integrals. We have checked that the coefficients of all of the poles in the NNLO calculation are consistent with zero within the uncertainties as well as with analytic expressions in \cite{Buras:1989xd}, hence they have been dropped.

\begin{table}[t]
	\centering
	\renewcommand{\arraystretch}{1.3}
	\begin{tabular}{|l|l|c|}
	\hline
	 $S$&\phantom{11111}$C_{B_K,\text{NLO}}^{S}$& $C_{B_K,\text{NNLO}}^{S}$ \\
	\hline
	SMOM$(\gamma_\mu,\qslash)$ & $8\log 2 -8=-2.45482...$&$\phantom{-}3.88 f +\phantom{1}21.05\pm0.08$\\
	SMOM$(\gamma_\mu,\gamma_\mu)$ &$8\log 2 -16/3=0.211844...$&$-0.42 f+\phantom{1}86.41\pm0.08$\\
	SMOM$(\qslash,\qslash)$ &$8\log 2 -6=-0.454823...$&$\phantom{-}0.90 f+\phantom{1}52.78\pm0.09$\\
	SMOM$(\qslash,\gamma_\mu)$ &$8\log 2 -10/3=2.21184...$&$-3.39 f+123.47\pm0.09$\\
	RI$^\prime$-MOM & $8\log 2-14/3=0.878511...$&$\phantom{-}0.17 f+\phantom{1}61.71\pm0.07$\\
	RI-MOM & $ 8\log 2-14/3=0.878511...$&$-4.49 f+\phantom{1}94.04\pm0.07 $\\
	\hline
	\end{tabular}	

	\caption{$C_{B_K,\text{NLO}}^{S}$ and $C_{B_K,\text{NNLO}}^{S}$ as defined in Eq.(\ref{eq:CBKexp}). The values presented are in Landau  gauge with $N_c=3$  and the NNLO result computed at $\nu=\mu=\sqrt{-p^2}$ for two RI-MOM schemes ($S=\{\text{RI-MOM},\text{RI}^\prime\text{-MOM}$\})  and four RI-SMOM schemes ($S=\text{SMOM}(X,Y)$ with $X=\gamma_{\mu},\qslash$ and $Y=\gamma_{\mu},\qslash$). The uncertainty in these results arises due to partly numerical evaluation of the two-loop integrals.} 
	\label{tab:twoloop}
\end{table}

In Table~\ref{tab:twoloopfull} we present values for the LO+NLO and LO+NLO+NNLO conversion factors (LO=1), as well as the difference between the NNLO and NLO corrections at  $\nu=\mu$,  for $f=3$.
We use the world average of $\alpha_s(M_Z)=0.1180\pm0.0009$ \cite{ParticleDataGroup:2024cfk}, which we evolve
down to the scale $\nu =\mu= 3$ GeV
using the 4-loop QCD $\beta$ function and threshold corrections available in \texttt{RunDec} \cite{Chetyrkin:2000yt}. We find that the perturbative series exhibits excellent convergence as the NNLO corrections give relative contributions below 4\% for all schemes. For the $(\gamma_{\mu},\gamma_{\mu})$, $(\qslash,\qslash)$ and RI-MOM schemes, the NNLO contributions are larger than the NLO ones. However, the relative NLO corrections to the series are smaller compared to the other two schemes, while the NNLO contributions are of comparable size.  Hence, we do not consider the perturbative behaviour in these cases to be abnormal. The dominant uncertainties here come from the error on $\alpha_s(\mu)$ at both NLO and NNLO.

\begin{table}[b]
	\centering
	\renewcommand{\arraystretch}{1.3}
	\begin{tabular}{|l| lll |}
		\hline
		$S$&NLO&NNLO&$|\text{diff.}|$\\
		\hline
		SMOM$(\gamma_\mu,\qslash)$ &0.9523(8)&0.9646(4)&0.0124(9)\\
		SMOM$(\gamma_\mu,\gamma_\mu)$
		&1.00412(7)&1.036(1)&0.032(1)\\
		SMOM$(\qslash,\qslash)$ &0.9912(1)&1.0121(6)&0.0210(6)\\
		SMOM$(\qslash,\gamma_\mu)$ &1.0430(7)&1.086(2)&0.043(2)\\
		RI$^\prime$-MOM&1.0171(3)&1.041(1)&0.024(1)\\
		RI-MOM&1.0171(3) &1.048(1) &0.030(1) \\
		\hline
	\end{tabular}
	\caption{Conversion factors $C_{B_K}^{S}$ evaluated with $\alpha_s(\mu)$ including NLO (i.e. 1+NLO) and NNLO (i.e. 1+NLO+NNLO) corrections, as well as the difference $|\text{diff.}|$ between the two (i.e. NNLO) in Landau gauge for  RI-MOM schemes ($S=\{\text{RI-MOM},\text{RI}^\prime\text{-MOM}$\}) and four RI-SMOM schemes ($S=\text{SMOM}(X,Y)$ with $X=\gamma_{\mu},\qslash$ and $Y=\gamma_{\mu},\qslash$). The results are computed at $\nu=\mu=\sqrt{-p^2} = 3$~GeV with $N_c=3$ and $f=3$. The dominant uncertainty in these results is due to the error on $\alpha_s(3~\text{GeV})$.}
	\label{tab:twoloopfull}
\end{table}

\subsection{Conversion to the RGI bag factor}\label{sec:results2}

\begin{figure}
	\centering
	\begin{tabular}{lll}
		\includegraphics[width=4.5cm]{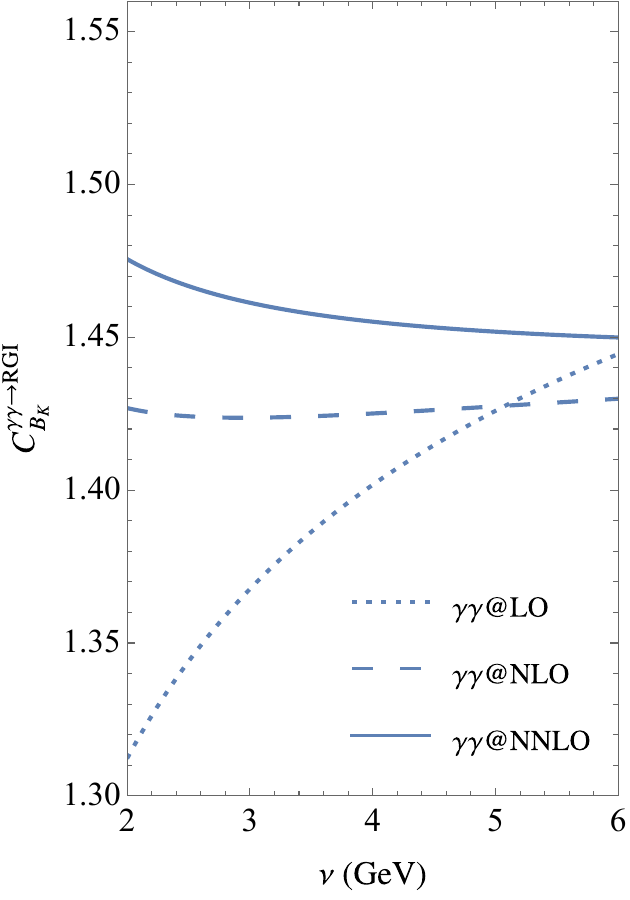}&
		\includegraphics[width=4.5cm]{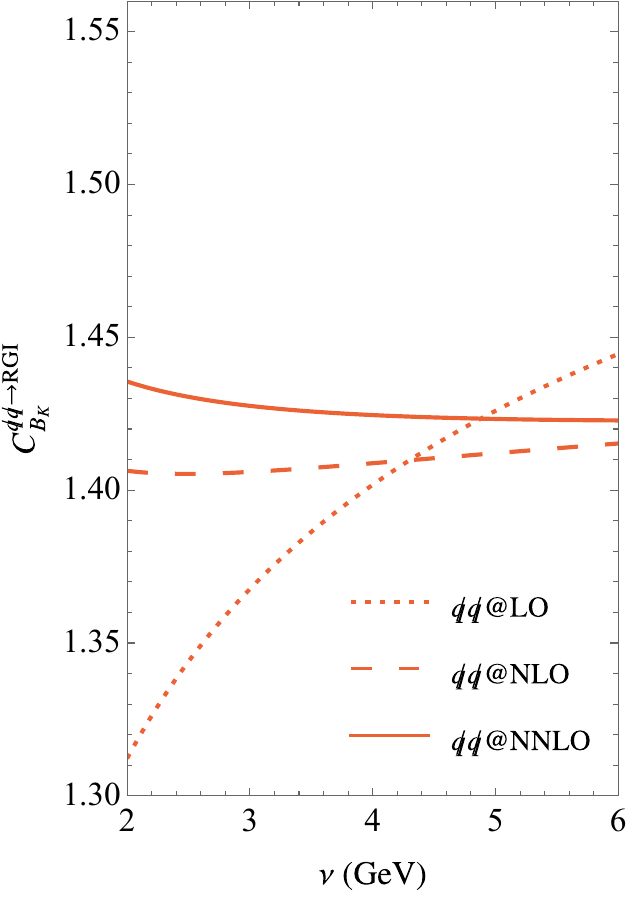}&
		\includegraphics[width=4.5cm]{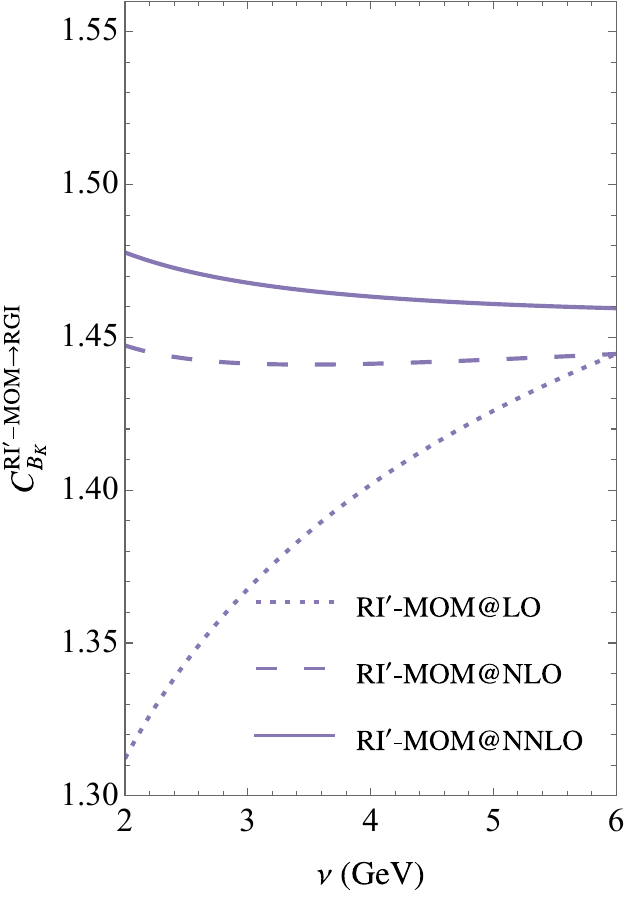}\\
		\includegraphics[width=4.5cm]{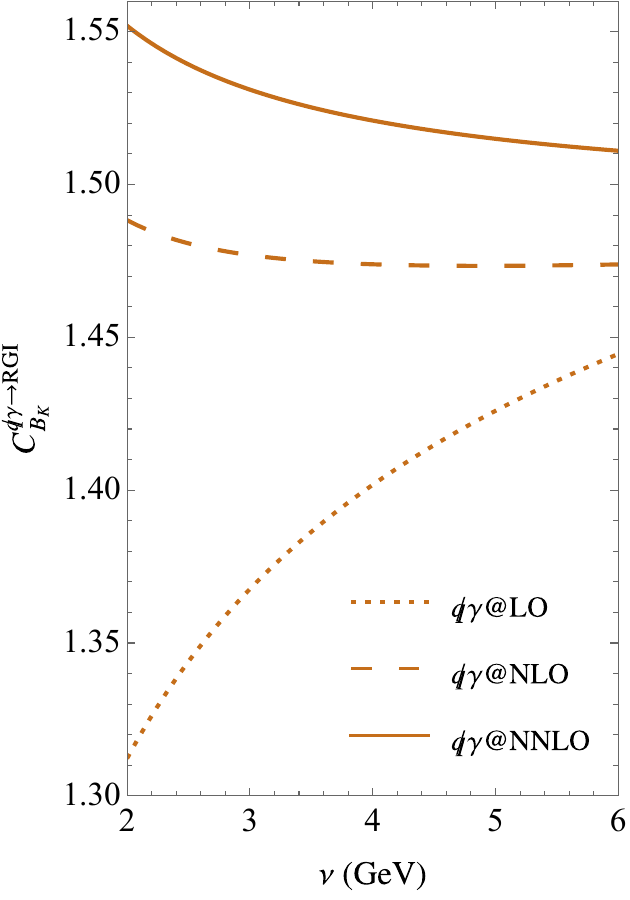}&
		\includegraphics[width=4.5cm]{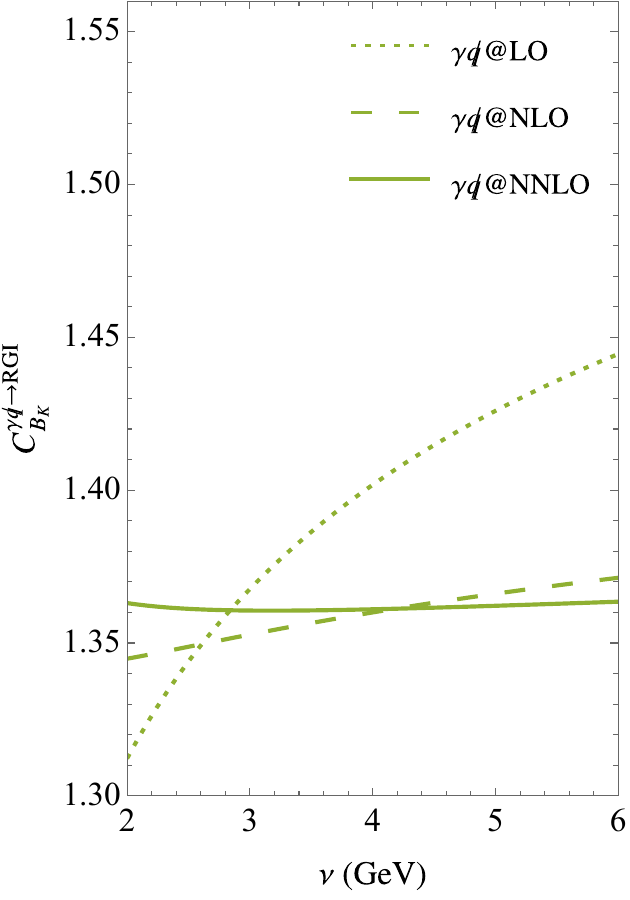}&
		\includegraphics[width=4.5cm]{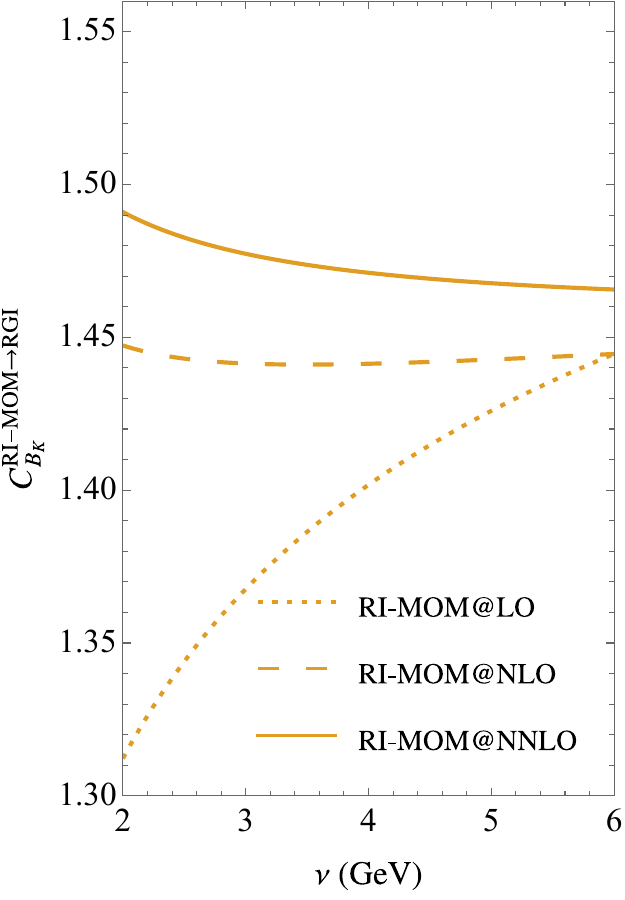}
	\end{tabular}
	\caption{Conversion factors $C_{B_K}^{S \to \rm RGI}$ from   RI-MOM schemes ($S=\{\text{RI-MOM},\text{RI}^\prime\text{-MOM}$\}) and from RI-SMOM schemes ($S=XY$ with $X=\gamma_{\mu},\qslash$ and $Y=\gamma_{\mu},\qslash$ )  schemes to RGI for $\mu=3$~GeV and $f=3$. Here $\gamma_\mu\equiv\gamma$.}
	\label{fig:CBK3GeV}
\end{figure}

\begin{figure}
	\centering
	\begin{tabular}{lll}
		\includegraphics[width=4.5cm]{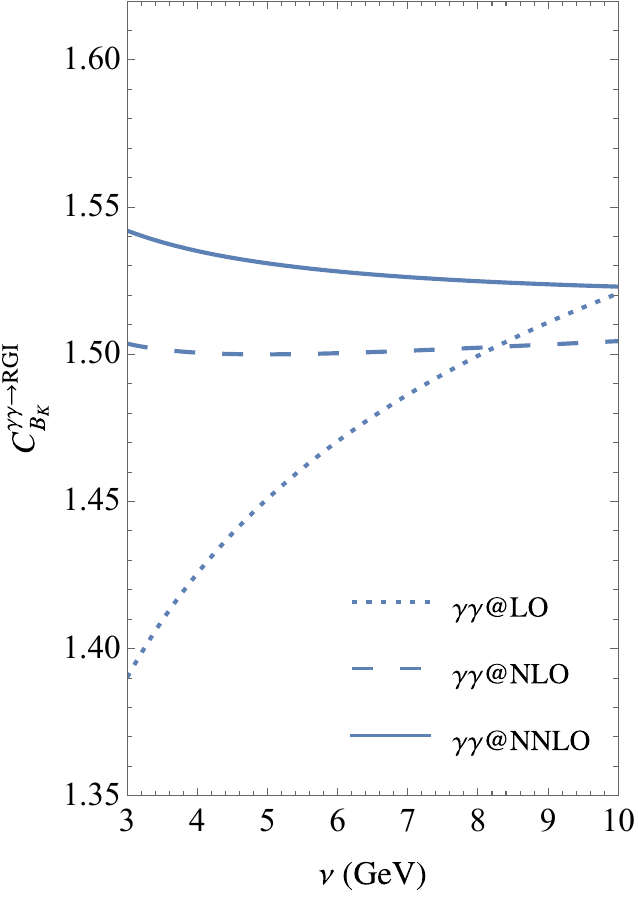}&
		\includegraphics[width=4.5cm]{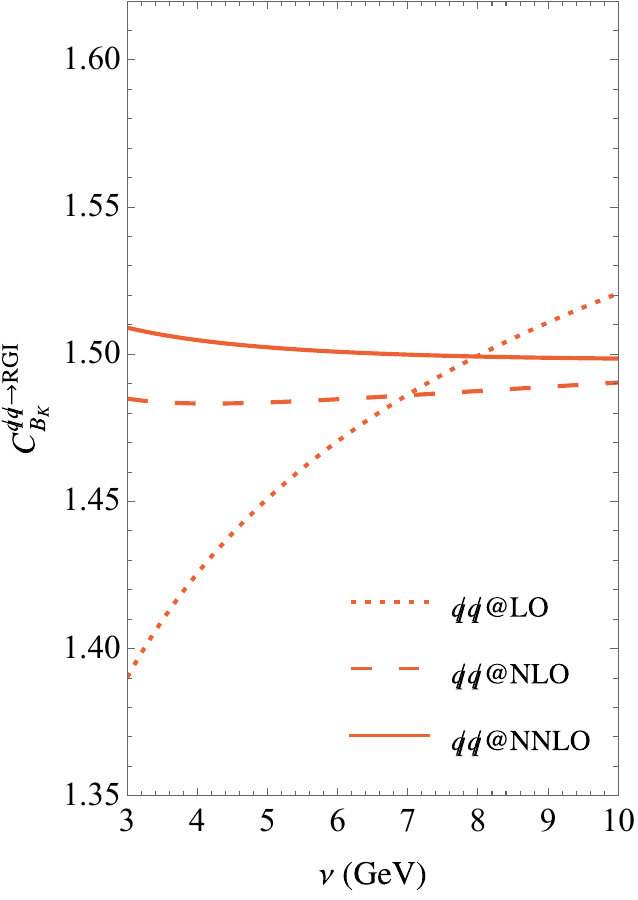}&
		\includegraphics[width=4.5cm]{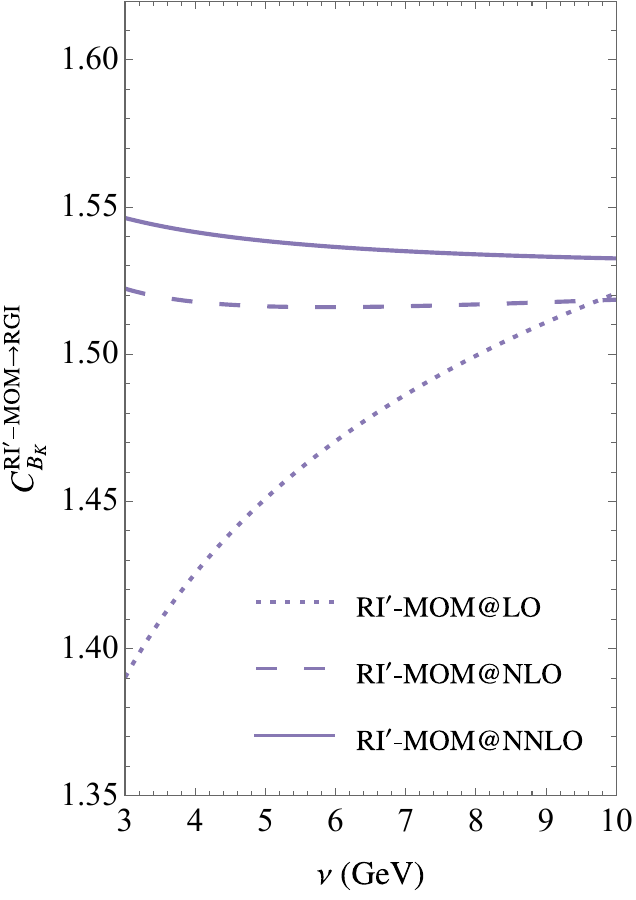}\\
		\includegraphics[width=4.5cm]{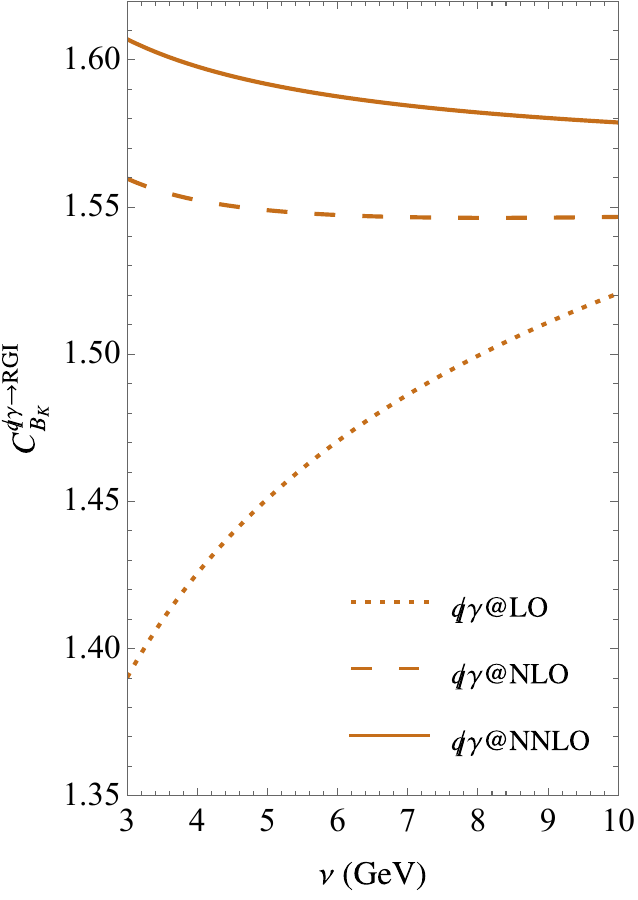}&
		\includegraphics[width=4.5cm]{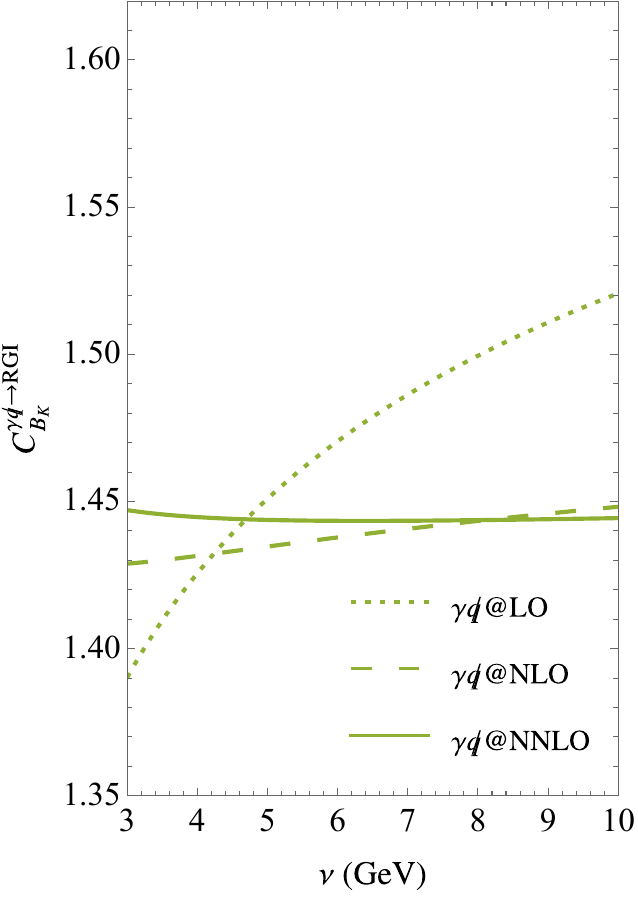}&
		\includegraphics[width=4.5cm]{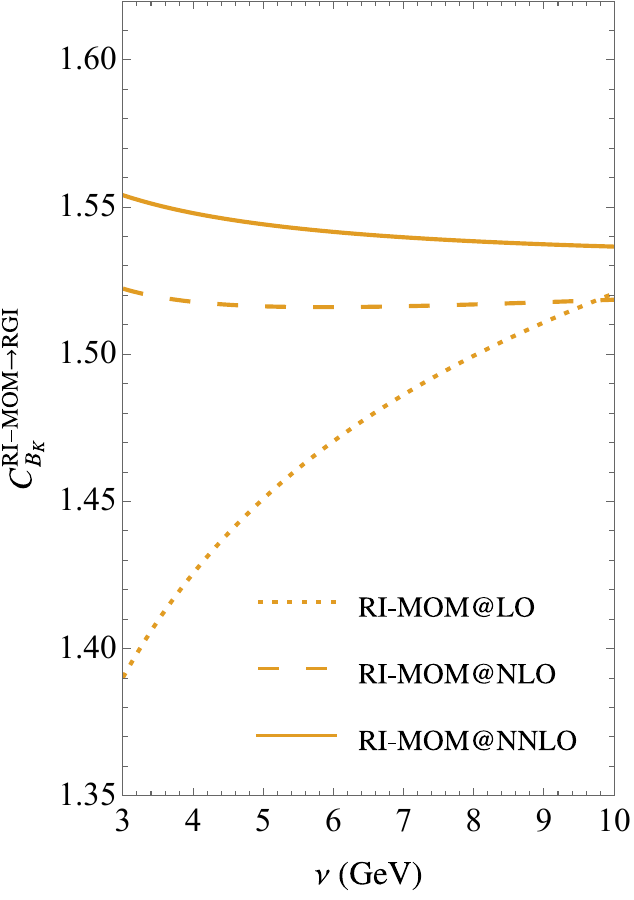}
	\end{tabular}
	\caption{Conversion factors $C_{B_K}^{S \to \rm RGI}$  from  RI-MOM schemes ($S=\{\text{RI-MOM},\text{RI}^\prime\text{-MOM}$\}) and from RI-SMOM schemes ($S=XY$ with $X=\gamma_{\mu},\qslash$ and $Y=\gamma_{\mu},\qslash$)  to RGI  for  $\mu=5$~GeV and $f=4$. Here $\gamma_\mu\equiv\gamma$.}
	\label{fig:CBK5GeV}
\end{figure}

Next, from the conversion factors to the $\overline{\rm MS}$ scheme we construct
the conversion factors $C_{B_K}^{S \to \rm RGI}$ to the scale- and scheme-independent bag factor $\hat B_K$,
 for any number of flavours $f$ as
\begin{equation}
	C_{B_K}^{S \to \rm RGI}(\mu)=U_{(f)}^{(0)}(\nu)\left(1+\frac{\alpha_s^{(f)}(\nu)}{4\pi}J^{(1)}_{(f)}+\left(\frac{\alpha_s^{(f)}(\nu)}{4\pi}\right)^2J^{(2)}_{(f)}\right)C_{B_K}^{S\rightarrow\overline{\text{MS}}}(\mu,\nu),\label{eq:Chat}
\end{equation}
where the relevant parts of  the RG evolution operator \cite{Brod:2010mj} are given by
\begin{align}
	U^{(0)}_{(f)}(\nu)&=(\alpha_s^{(f)}(\nu))^{-\frac{6}{33-2f}},\label{eq:U}\\
	J^{(1)}_{(f)}&=\frac{2f(4f-813)+13\,095}{6(33-2f)^2},\label{eq:J1}\\
	\begin{split}
		J^{(2)}_{(f)}&=\frac{1}{2700(33-2f)^4}(-52\,000 f^5+1\,755\,216 f^4+33\,796\,944 f^3\\
		&-1\,897\,533\,864 f^2+25\,597\,290\,654 f-119\,065\,711\,779\\
		&-21\,600(2f-33)^2(5f+63)\zeta(3)). \label{eq:J2}
	\end{split}
\end{align}
The $C_{B_K}^{S \to \rm RGI}$ are formally independent of the  $\overline{\rm MS}$ scale $\nu$. Their residual scale dependence for the 3 and 4 flavour scenarios for the most commonly used MOM schemes is presented in Figures \ref{fig:CBK3GeV} and \ref{fig:CBK5GeV} for momentum subtraction scales
$\mu=3$~GeV ($f=3$) and $\mu=5$~GeV ($f=4$) respectively.
We can see that the $(\slashed{q},\slashed{q})$ scheme exhibits the best behaviour in terms of perturbative convergence as well as the size of the residual scale dependence. While RI-MOM schemes are the same at NLO, RI'-MOM scheme converges better at NNLO. Comparing Figures  \ref{fig:CBK3GeV} and \ref{fig:CBK5GeV} we can see a further improvement in the convergence of the perturbative series as well as further reduction in the scale dependence for the 4 flavour $\mu=5$~GeV case. In addition,  in Table~\ref{tab:Chat} we present numerical values for $C_{B_K}^{S \to \rm RGI}$ for $\mu=3$~GeV, including the ratios of the uncertainties coming from $\alpha_s$ versus residual scale variation. It is clear that in each case the uncertainty due to scale variation dominates.

\begin{table}
	\centering
	\renewcommand{\arraystretch}{1.3}
	\begin{tabular}{|l| lc|lc |}
		\hline
		$S$&NLO&$ (\delta_{\alpha_s}/\delta_\nu)$&NNLO&$ (\delta_{\alpha_s}/\delta_\nu)$\\
		\hline
		SMOM$(\gamma_\mu,\qslash)$ &1.3528(341)&0.15&1.36061(945)&0.62\\
		SMOM$(\gamma_\mu,\gamma_\mu)$
		&1.4237(157)&0.28&1.4613(146)&0.22\\
		SMOM$(\qslash,\qslash)$ &1.4060(203)&0.23&1.42757(886)&0.49\\
		SMOM$(\qslash,\gamma_\mu)$ &1.4769(120)&0.31&1.5310(275)&0.09\\
		RI$^\prime$-MOM&1.4414(112)&0.39&1.4678(106)&0.33\\
		RI-MOM&1.4414(112) &0.39&1.4773(151) &0.22\\
		\hline
	\end{tabular}
	\caption{The NLO and NNLO conversion factors $C_{B_K}^{S \to \rm RGI}$ in Landau gauge for  RI-MOM schemes ($S=\{\text{RI-MOM},\text{RI}^\prime\text{-MOM}$\}) and four RI-SMOM schemes ($S=\text{SMOM}(X,Y)$ with $X=\gamma_{\mu},\qslash$ and $Y=\gamma_{\mu},\qslash$) computed at $\mu=\sqrt{-p^2} = 3$~GeV with $N_c=3$ and $f=3$, with $\nu$ varied from 2 to 6 GeV (central value of 3 GeV). The uncertainties include the error on $\alpha_s$, and residual $\nu$-depndence, where  $ (\delta_{\alpha_s}/\delta_\nu)$ indicates the ratio of the aforementioned uncertainties at NLO and NNLO for each scheme choice. The dominant errors in these results arise from the variation of the renormalization scale.}
	\label{tab:Chat}
\end{table}

\subsection{$B_K^{\overline{\rm MS}}$, $\hat{B}_K$ and their residual scale dependence}\label{sec:BK}

\begin{table}[b]
	\centering
	\renewcommand{\arraystretch}{1.3}
	\begin{tabular}{|l|lll|l|l|}
		\hline
		Scheme &Lattice&NLO&NNLO&$\mu=\nu$ & $f$\\
		\hline
		SMOM$(\qslash,\qslash)$&0.5342(21)&0.5295(21)&0.5407(21)& 3 GeV & 3\\
		SMOM$(\gamma_\mu,\gamma_\mu)$&0.5164(18)&0.5185(18)&0.5352(20) &3 GeV & 3\\
		$\mbox{RI-MOM}_{16}$&0.517(13)&0.526(13)&0.542(14) & 3 GeV & 3\\
		\hline
		$\mbox{RI-MOM}_{11}$&0.5308(61)&0.5393(62)&0.5536(64) &3.5 GeV &3\\
		\hline
		$\mbox{RI$^\prime$-MOM}$ &0.498(16)&0.507(16)&0.519(17)& 3 GeV & 4\\
		\hline
	\end{tabular}
	\caption{Bag parameter $B_K^{\overline{\rm MS}}$ for the available $f=3$ and $f=4$ lattice inputs.
Lattice results are taken from \cite{Boyle:2024gge, Garron:2016mva, BMW:2011zrh,Carrasco:2015pra}, where RI-MOM$_{16}$ corresponds to RBC/UKQCD~16 and RI-MOM$_{11}$ to BMW~11 results.
The dominant uncertainties here are due to errors on the lattice results. Uncertainties due to higher-order
perturbative corrections are not included in this Table.}
	\label{tab:3gev}
\end{table}
Taking our results for the conversion factors we can perform a matching calculation at 3 and 4 flavours as well as fixed momentum subtraction scale $\mu$ from the  currently available lattice estimates of $B_K$ \cite{Boyle:2024gge, Garron:2016mva, BMW:2011zrh, Carrasco:2015pra}. We quote the lattice inputs and present our computed values of $B_K^{\overline{\rm MS}}(\nu)$ using one-loop and two-loop matching in Table~\ref{tab:3gev}, for $\nu=\mu$. We can see that inclusion of perturbative corrections leave the uncertainties comparable to the initial errors obtained on the lattice, meaning that the results so far are dominated by the lattice errors.

\begin{figure}[t]
	\centering
	\includegraphics[width=4.8cm]{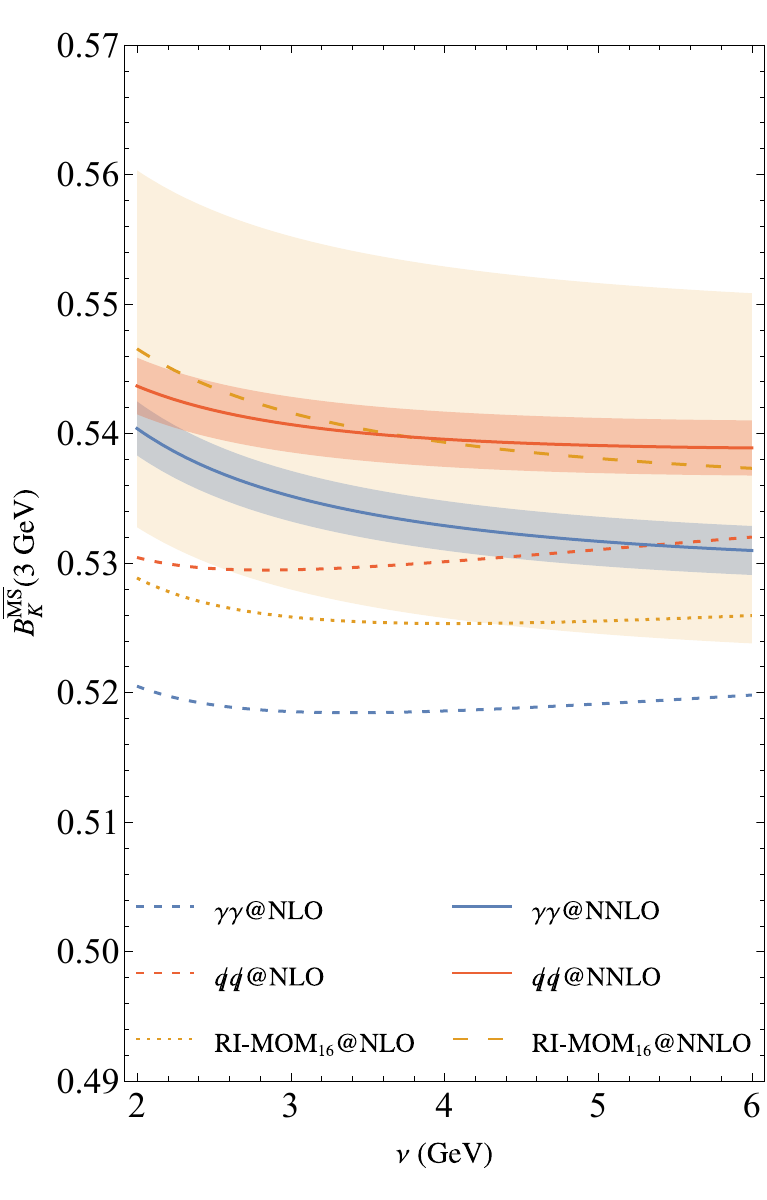}
	\includegraphics[width=4.8cm]{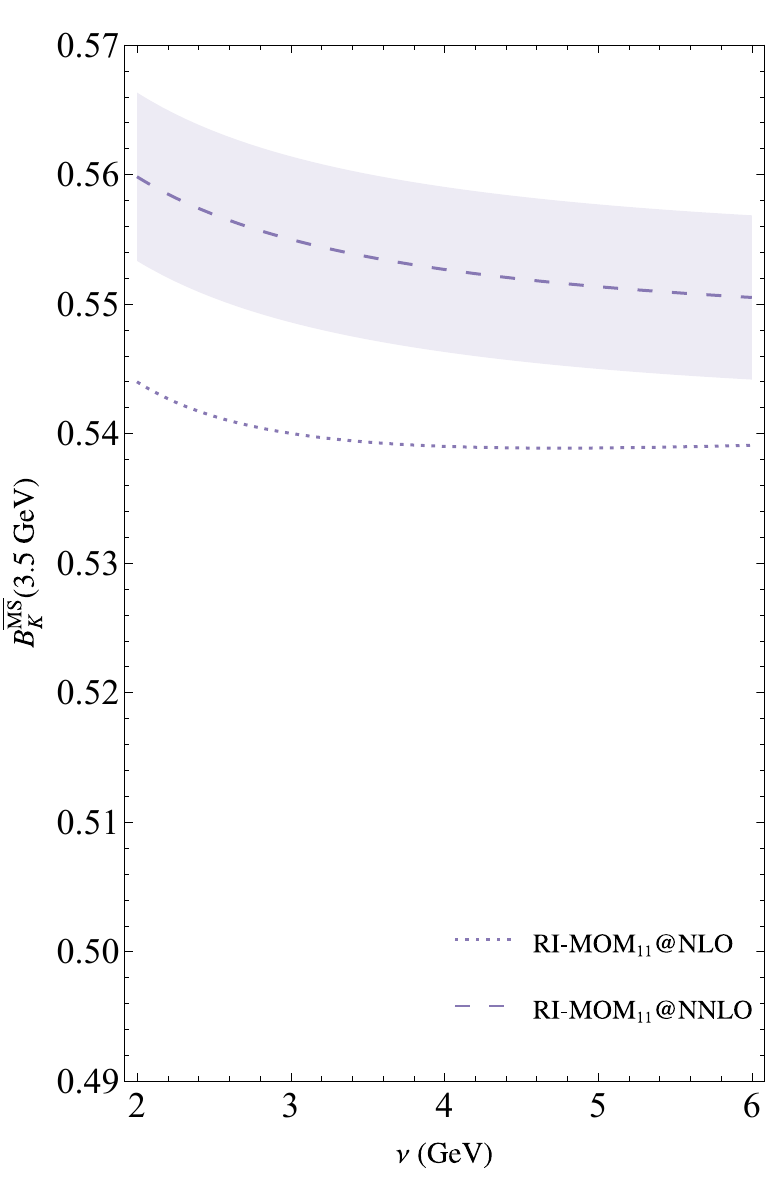}
	\includegraphics[width=4.8cm]{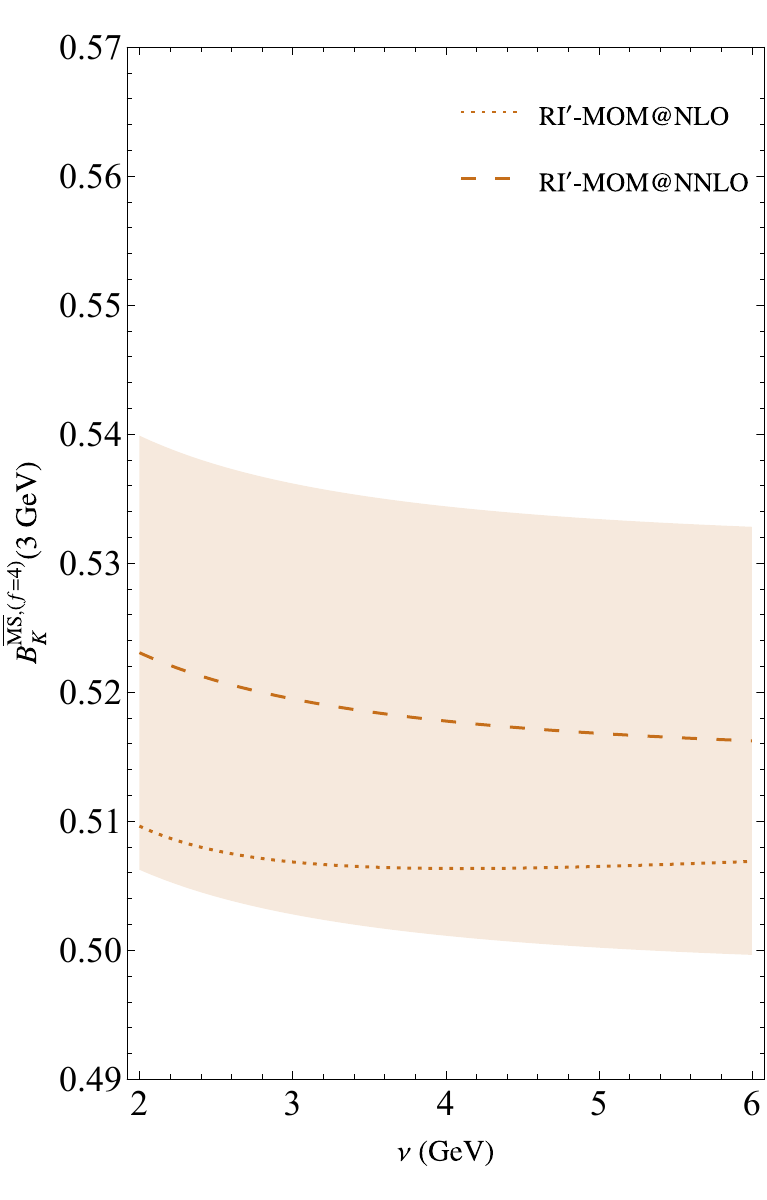}
	\caption{$B_K^{\overline{\rm MS}}$ evaluated at 3~GeV ($f=3$, left), 3.5~GeV ($f=3$, middle) and at 3~GeV ($f=4$, right) as a function of the $\overline{\mbox{MS}}$ renormalization scale $\nu$ in Landau gauge for the two RI-SMOM schemes  $(\gamma_\mu,\gamma_\mu)\equiv\gamma\gamma$ and $(\slashed{q},\slashed{q})\equiv\slashed{q}\slashed{q}$, RI-MOM and RI$^\prime$-MOM. (For explanation, see text). The bands indicate the corresponding maximum and minimum values of $B_K$ given the uncertainties on $\alpha_s$, $C_{B_K,\text{NNLO}}$ and $B_K^{\text{RI-(S)MOM}}=B_K^{\text{lat}}$. Here
	$\gamma\gamma$ and  $\slashed{q}\slashed{q}$ corresponds to $B_K^{\text{lat}}$ from RBC/UKQCD~24~\cite{Boyle:2024gge},
	RI-MOM$_{16}$ - $B_K^{\text{lat}}$ from RBC/UKQCD 16~\cite{Garron:2016mva},
	RI-MOM$_{11}$ - $B_K^{\text{lat}}$ from BMW~11~\cite{BMW:2011zrh},
	RI$^\prime$-MOM - $B_K^{\text{lat}}$ from ETM 15~\cite{Carrasco:2015pra}.}
	\label{fig:BKMS}
\end{figure}

As $B_K^{\overline{\rm MS}}(\nu)$ is not formally scale-invariant, we estimate the error due to uncomputed higher
orders as follows: We use our fixed-order conversion for a range $\nu \in [2, 6]$ GeV and $\overline{\rm MS}$-evolve it back to the scale $\nu=\mu$. The resulting variation gives us an estimate of the uncomputed higher orders. The results are given in Figure~\ref{fig:BKMS}. As we can see the uncertainties associated with the RI-SMOM schemes are significantly lower compared to RI-MOM results, for which the residual scale dependence is fairly negligible compared to lattice errors. For RI-SMOM schemes, on the other hand, the residual scale dependence is significant. One can also observe a nice overlap of the RBC/UKQCD results as opposed to the BMW 11 result, which gives a larger prediction for $B_K$ albeit at a slightly higher scale $\mu$.

\begin{figure}[b]
	\centering
	\includegraphics[width=7.45cm]{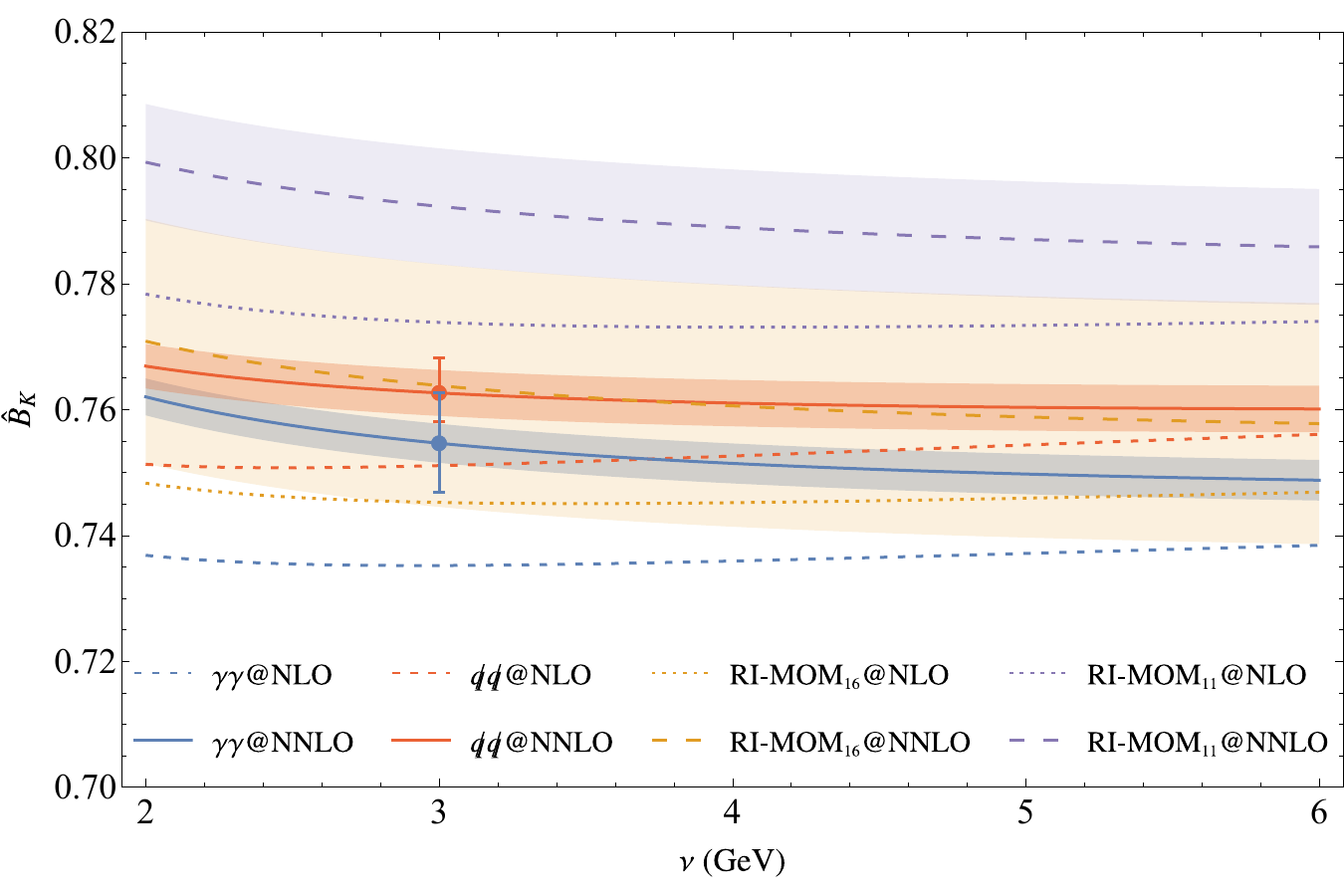}
	\includegraphics[width=7.45cm]{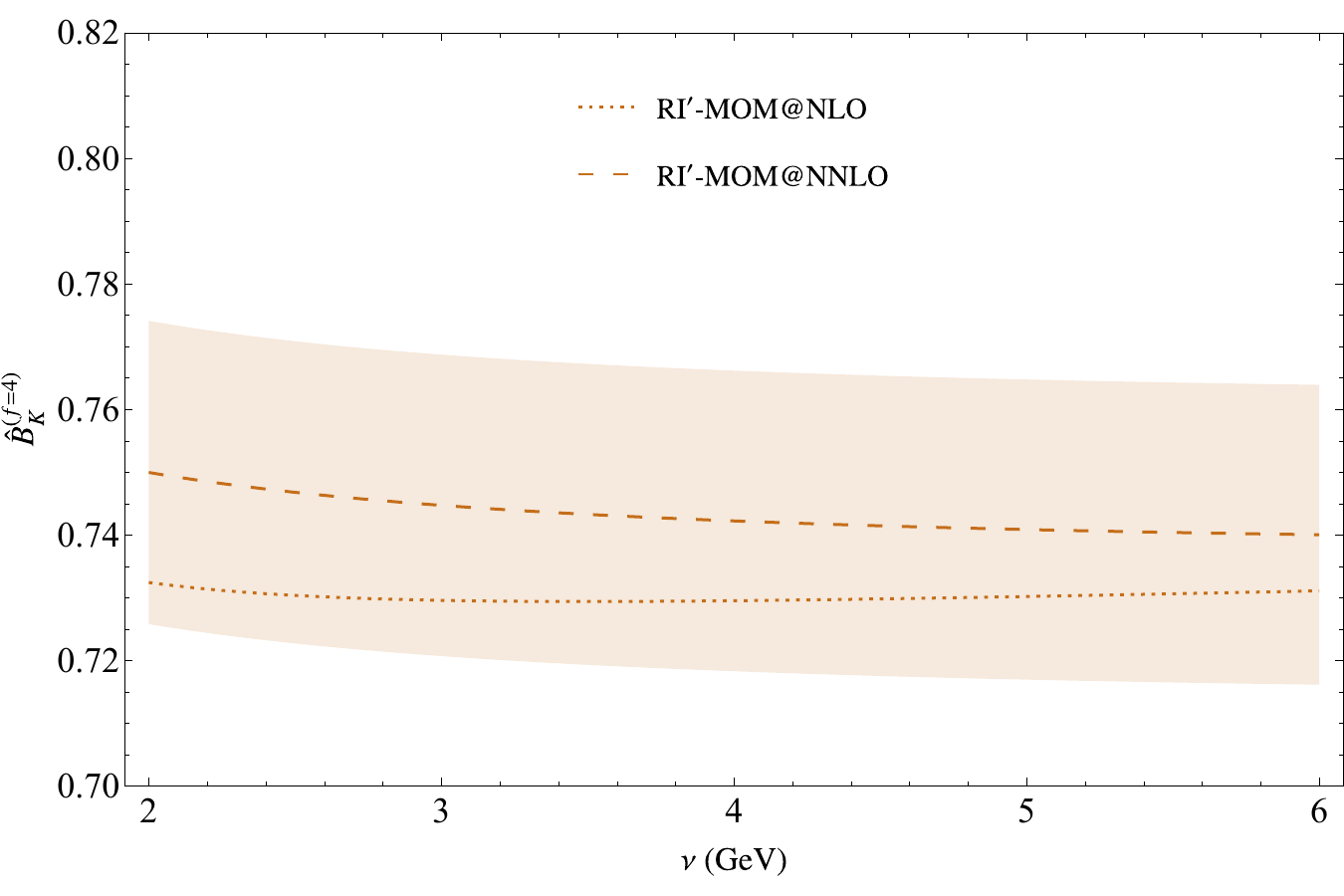}
	\caption{Residual dependence of $\hat{B}_K$ on
		$\overline{\mbox{MS}}$ renormalization scale in Landau gauge for the two RI/SMOM schemes  $(\gamma_\mu,\gamma_\mu)\equiv\gamma\gamma$ and $(\slashed{q},\slashed{q})\equiv\slashed{q}\slashed{q}$ and RI-MOM at 3 flavours (left) and 4 flavours (right). The bands indicate the corresponding maximum and minimum values of $\hat{B}_K$ given the uncertainties on $\alpha_s$, $C_{B_K,\text{NNLO}}$ and $B_K^{\text{RI-(S)MOM}}=B_K^{\text{lat}}$. The error bars for $\gamma\gamma$ and $\slashed{q}\slashed{q}$ also include the uncertainty due to the scale variation. Here
		$\gamma\gamma$ and  $\slashed{q}\slashed{q}$ corresponds to $B_K^{\text{lat}}$ from RBC/UKQCD 24~\cite{Boyle:2024gge},
		RI-MOM$_{16}$ - $B_K^{\text{lat}}$ from RBC/UKQCD 16~\cite{Garron:2016mva},
		RI-MOM$_{11}$ - $B_K^{\text{lat}}$ from BMW 11~\cite{BMW:2011zrh},
		RI$^\prime$-MOM - $B_K^{\text{lat}}$ from ETM 15~\cite{Carrasco:2015pra}.}
	\label{fig:BKh}
\end{figure}

We compute the RG-invariant value of the kaon bag parameter $\hat{B}_K$ using the conversion factors in Eq.(\ref{eq:Chat}) along with the aforementioned Lattice results. The residual scale dependence of $\hat{B}_K$ for the five lattice results is given in Figure~\ref{fig:BKh}. Again we see that the RBC/UKQCD results overlap nicely and the BMW 11 result lies above. Taking into account the scale variation, the two RI-SMOM schemes agree with each other as well. As the ETM result is at 4 flavours it still can't be directly compared to the 3-flavour results, however as for the RI-MOM schemes we observe that residual scale variation introduces only a sub-dominant uncertainty.

\subsection{Matching between number of flavours}
\label{sec:results4}

In this section, we define a matching coefficient $\hat{M}^{(f\rightarrow f-1)}$ for obtaining the $\hat{B}_K^{(f)}$ from $\hat{B}_K^{(f-1)}$,  i.e. 
\begin{equation}
	\hat{B}_K^{(f)}=\hat{M}^{(f\rightarrow f-1)}\hat{B}_K^{(f-1)},
\end{equation}
where $f$ is the number of active flavours. 

\begin{figure}[t]
	\centering
	\includegraphics[width=10cm]{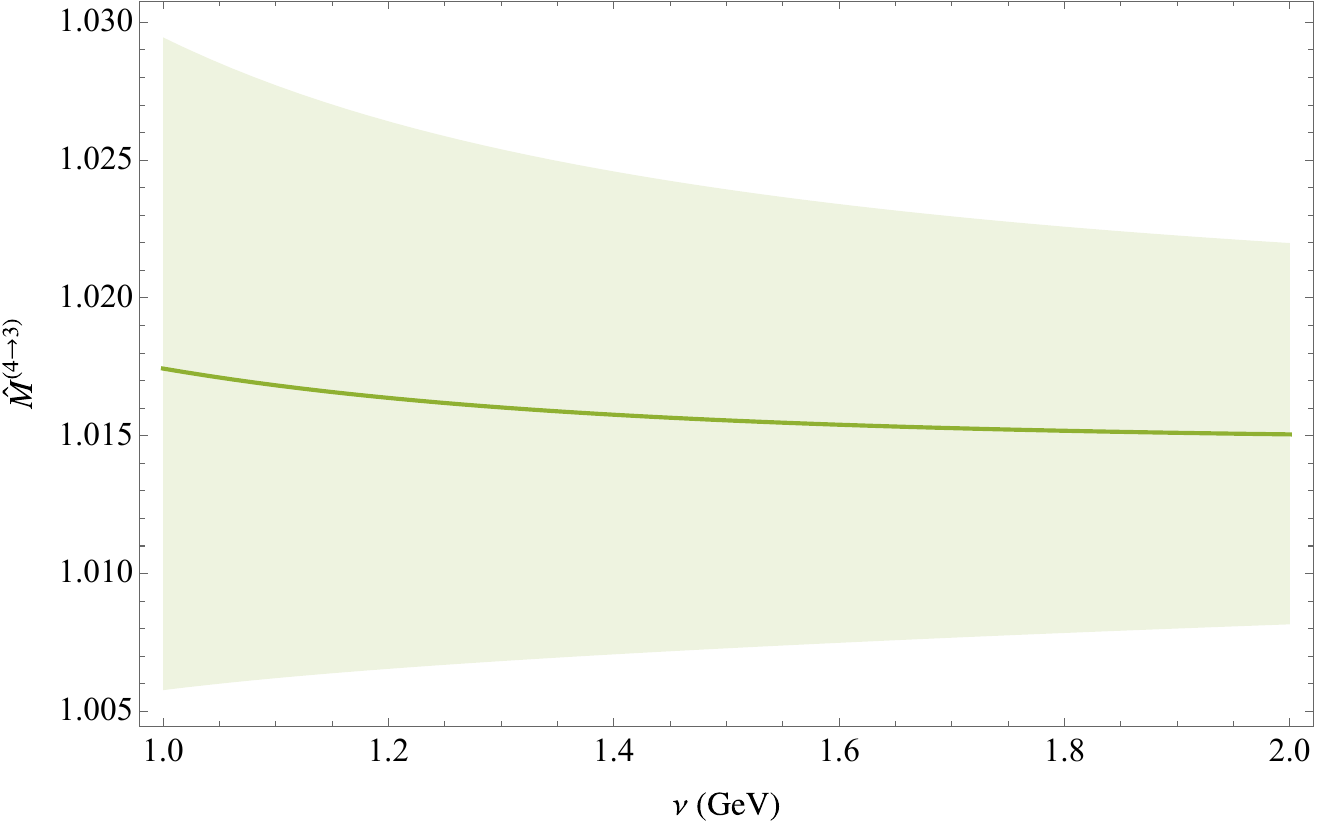}
	\caption{Dependence of $\hat{M}^{(4\rightarrow 3)}$ on the
		$\overline{\mbox{MS}}$ renormalization scale. The bands indicate the corresponding maximum and minimum values of $\hat{M}$ given the uncertainties on $\alpha_s(\nu_f)$ at 3 and 4 flavour. }
	\label{fig:Mh}
\end{figure}

We extract $\hat{M}^{(f\rightarrow f-1)}$ from the results in \cite{Brod:2010mj} and obtain
\begin{align}\label{eq:.Mh}
	\begin{split}
		\hat{M}^{(f\rightarrow f-1)}&=\frac{U^{(0)}_{(f)}(\nu)}{U^{(0)}_{(f-1)}(\nu)}\left(1+\frac{\alpha_s^{(f-1)}(\nu)}{4\pi}(J^{(1)}_{(f)}-J^{(1)}_{(f-1)})\right.\\
		&+\left.\left(\frac{\alpha_s^{(f-1)}(\nu)}{4\pi}\right)^2\Biggl(J^{(2)}_{(f)}-J^{(2)}_{(f-1)}-J^{(1)}_{(f-1)}(J^{(1)}_{(f)}-J^{(1)}_{(f-1)})\right.\\
		&+\left.\left. \frac{2}{3}J^{(1)}_{(f)}\log\frac{\nu^2}{m_f^2(\nu)}-\frac{2}{3}\log^2\frac{\nu^2}{m_f^2(\nu)}-\frac{2}{9}\log\frac{\nu^2}{m_f^2(\nu)}-\frac{59}{54} \right)\right),
	\end{split}
\end{align}
where the relevant parts of the evolution kernel are defined in Eqs.(\ref{eq:U}, \ref{eq:J1} and \ref{eq:J2}). In addition to the running of the matrix element, this expression also takes into account threshold corrections associated with the change in flavours. One can see that $\hat{M}^{(f\rightarrow f-1)}$ is independent of the renormalization scale up to 2-loop order by taking a derivative with respect to $\nu$. Using Eq.(\ref{eq:.Mh}) along with 3-loop $\alpha_s(\nu)$ and $m_c(\nu)$ running we can plot the residual scale dependence of $\hat{M}^{(4\rightarrow 3)}$, as shown in Figure~\ref{fig:Mh}. We observe excellent behaviour as the residual scale dependence is significantly smaller than the error from $\alpha_s(\nu)$.

\subsection{$\hat B_K$ global average }
\label{sec:results5}

\begin{table}[b]
	\centering
	\renewcommand{\arraystretch}{1.3}
	\begin{tabular}{|l|l|lcll|}
		\hline
		$f$&Ref.&Scheme&$\mu$~(GeV)&$B_K^{(\text{Scheme})}(\mu)$&$\hat{B}_{K,\text{NLO}}^{(f)}$\\
		\hline
		4&ETM 15 &RI$^\prime$-MOM&3&0.498(16)&0.717(24)\\
		\hline
		\hline
		3&RBC/&SMOM($\slashed{q},\slashed{q}$)&3&0.5342(21)&0.7436(82)\\
		&UKQCD 24&SMOM($\gamma_{\mu},\gamma_{\mu}$)&3&0.5164(18)&-\\
		3&SWME 15A&$\overline{\text{MS}}$&3&0.519(26)&0.735(36)\\
		3&RBC/&SMOM($\slashed{q},\slashed{q}$)&3&0.5341(18)&0.7499(152)\\
		&UKQCD 14B&SMOM($\gamma_{\mu},\gamma_{\mu}$)&3&0.5166(18)&-\\
		3&Laiho 11&$\overline{\text{MS}}$&2&0.5572(152)&0.7628(208)\\
		3&BMW 11&RI-MOM&3.5&0.5308(61)&0.7727(117)\\
		\hline
	\end{tabular}
	\caption{Summary of lattice results entering the current FLAG average \cite{Carrasco:2015pra, SWME:2015oos,  Blum:2014tka, Laiho:2011np, BMW:2011zrh, Boyle:2024gge}. The $B_K^{(\text{Scheme})}(\mu)$ is taken from the corresponding literature and $\hat{B}_{K,\text{NLO}}^{(f)}$ from FLAG review \cite{FlavourLatticeAveragingGroupFLAG:2024oxs}. }
	\label{tab:BKlit}
\end{table}

We proceed to compute the global averages of $\hat{B}_K$ using our two-loop conversion. In Table~\ref{tab:BKlit} we review the existing results entering into the current 3 and 4 flavour FLAG averages. In Table~\ref{tab:BKres} we present our updated values for the corresponding $\hat{B}_K$ values in Table~\ref{tab:BKlit}. Here we have also converted the 3 flavour results to 4 flavours and 4 flavour ETM 15 result to 3 flavours using Eq.(\ref{eq:.Mh}). 

\begin{table}[t]
	\centering
	\renewcommand{\arraystretch}{1.3}
	\begin{tabular}{|l|l|lll|}
		\hline
		$f$ &Ref.&Scheme&$\hat{B}_K^{(f=3)}$& $\hat{B}_K^{(f=4)}$\\
		\hline
		4&ETM 15 &RI$^\prime$-MOM&0.733(26)&0.745(25)\\
		\hline
		\hline
		3&RBC/&SMOM($\slashed{q},\slashed{q}$)&0.7626(56)&0.7749(93)\\
		&UKQCD 24&SMOM($\gamma_{\mu},\gamma_{\mu}$)&0.7546(80)&0.767(11)\\
		3&SWME 15A&$\overline{\text{MS}}$&0.735(36)$^{*}$&0.747(37)\\
		3&Laiho 11&$\overline{\text{MS}}$&0.7628(208)$^*$&0.7751(224)\\
		3&BMW 11&RI-MOM&0.790(13)&0.804(15)\\
		\hline
	\end{tabular}
	\caption{Summary of our updated results for $\hat{B}_K$. For $B_K^{(\text{Scheme})}(\mu)$ values given in MOM schemes (see Table~\ref{tab:BKlit}) we obtain $\hat{B}_K$ via two-loop matching to $\overline{\text{MS}}$ at $\nu=\mu$ and conversion to RGI value. The RGI errors take into account uncertainties in the lattice value,  $\alpha_s$, $C_{B_K}^S$ and $\overline{\text{MS}}$ residual scale variation. The asterisk ($^{*}$) indicates that
$\hat{B}_K^{(f=3)}$ is obtained from $\overline{\text{MS}}$ value using the conversion factor 1.369 as in \cite{FlavourLatticeAveragingGroupFLAG:2024oxs}, since no intermediate lattice scheme results were provided. The 3~flavour $\hat{B}_K$ results are converted to 4~flavours and vice versa via Eq.(\ref{eq:.Mh}).  }
	\label{tab:BKres}
\end{table}

To obtain the averages, we follow the procedure for averaging correlated data outlined by  Schmelling \cite{Schmelling:1994pz}. As we expect the two results of RBC/UKQCD 24~\cite{Boyle:2024gge} to be correlated we compute the full covariance matrix giving
\begin{equation}\label{eq:BKhRBC}
	f=2+1:\qquad \qquad\hat{B}_K=0.7600(53),\qquad \text{(RBC/UKQCD 24 updated)}
\end{equation}
where we use $\chi^2=0.671$ to estimate the size of the correlation to be $33.9\%$. The error is then simply taken to be the square root of the variance. Unlike FLAG review~\cite{FlavourLatticeAveragingGroupFLAG:2024oxs} we do not include the  RBC/UKQCD 14B~\cite{Blum:2014tka} result in the average as the newer result is a reanalysis of the same lattice measurement as the older one with an improved renormalization. 

Next,  we combine the results from SWME 15A~\cite{SWME:2015oos}, RBC/UKQCD 24~\cite{Boyle:2024gge} Eq.(\ref{eq:BKhRBC}), Laiho 11~\cite{Laiho:2011np} and BMW 11~\cite{BMW:2011zrh}. Here and in the following we neglect correlations between different Lattice results, and get
\begin{equation}\label{eq:BK3flnew}
	f=2+1:\qquad \hat{B}_K=0.7637(62),\qquad \qquad\qquad~~
\end{equation}
with  $\chi^2/\text{dof}=1.739$. According to our procedure, we rescale the error by the square root of $\chi^2/\text{dof}$. This result can be compared to the FLAG average of $\hat{B}_K=0.7533(91)$~\cite{FlavourLatticeAveragingGroupFLAG:2024oxs}, which includes specific correlations between Lattice results. The $f=2+1+1$ average is simply given by the 4 flavour ETM 15~\cite{Carrasco:2015pra} result, i.e.
\begin{equation}
	f=2+1+1:\qquad~ \hat{B}_K=0.745(25), \qquad\qquad\qquad\qquad~
\end{equation}
which can again be compared to the FLAG average of  $\hat{B}_K=0.717(24)$~\cite{FlavourLatticeAveragingGroupFLAG:2024oxs}.

Furthermore, we use our results in Table~\ref{tab:BKres} to obtain the full 3 and 4 flavour averages for $\hat{B}_K$ using combined 3 and 4 flavour results. For 3 flavours we get
\begin{equation}\label{eq:BKh3fl}
\hat{B}_K^{(f=3)}=0.7627(60),\qquad \qquad\qquad~~
\end{equation}
with  $\chi^2/\text{dof}=1.641$. Here, we also rescale the error by square root of $\chi^2/\text{dof}$. For 4 flavours we once more compute the average of RBC/UKQCD 24 values $(\hat{B}_K^{(f=4)})_{\text{RBC/UKQCD}}=0.7715(92)$ with correlation of $68.2\%$, followed by our full 4 flavour result
\begin{equation}\label{eq:BKh4fl}
\hat{B}_K^{(f=4)}=0.7759(84),\qquad \qquad\qquad~~
\end{equation}
 with  $\chi^2/\text{dof}=1.470$. Here, the error is also rescaled by square root of $\chi^2/\text{dof}$. 

\subsection{$\epsilon_{K}$ updated value}
\label{sec:results6}

Using our updated results for $\hat{B}_K$ in Eq.(\ref{eq:BKh3fl}) we can obtain an updated value for $\epsilon_K$. As we observe a small shift in central value and reduction in error, we update the result of~\cite{Brod:2022har} by recomputing the non-perturbative contribution and get
\begin{equation}
	|\epsilon_K|=2.171(65)_\text{pert.}(71)_\text{non-pert.}(153)_\text{param.} \times 10^{-3},
\end{equation}
where the errors refer to perturbative, non-perturbative and parametric. As we can see our result leads to reduction of non-perturbative uncertainty from 3.5\% to roughly 3.28\%. Here we note that the long-distance
contributions to $\epsilon_K$, that now dominate the
non-perturbative uncertainties, can be drastically improved using
future lattice calculations \cite{Bai:2023lkr, Huo:2025bhq}.

\subsection{D meson mixing}
\label{sec:results7}

One can also use our results for conversion factors to obtain the $\overline{\text{MS}}$ and RGI values of the D meson bag parameter computed by ETM 15~\cite{Carrasco:2015pra}. The results are summarized in Table~\ref{tab:Dmix}.

\begin{table}[t]
	\centering
	\renewcommand{\arraystretch}{1.3}
	\begin{tabular}{|l|lllll|}
		\hline
		Scheme&Lattice&NLO&NNLO&$\hat{B}_D^{(f=3)}$&$\hat{B}_D^{(f=4)}$\\
		\hline
		$B_D^{(f=4)}(\mbox{RI$^\prime$-MOM})$&0.744(27)&0.757(27)&0.776(28)&1.095(41)&1.113(41)\\
		\hline
	\end{tabular}
	\caption{Bag parameter for D meson mixing. The Lattice value taken from ETM 15~\cite{Carrasco:2015pra}. The $\overline{\text{MS}}$ and RGI values computed analogously to $B_K$. The errors take into account uncertainties in the Lattice value,  $\alpha_s$ and $C_{B_D}^S=C_{B_K}^S$. The RGI value errors also include $\overline{\text{MS}}$ residual scale variation.} 
	\label{tab:Dmix}
\end{table}

\section{Conclusions}\label{sec:summary3}

Meson–antimeson mixing plays a central role in current particle physics
phenomenology, with the CP-violating parameter $\epsilon_K$
imposing very tight constraints on physics beyond the Standard
Model. The theoretical predictions for these observables rely heavily
on non-perturbative bag parameters, which must be converted from
lattice renormalization schemes to the $\overline{\mathrm{MS}}$ scheme
to ensure compatibility with perturbatively computed short-distance contributions
and allow for phenomenological analyses.

In this work, we derived the scheme conversion factors for several MOM
and SMOM renormalization schemes at NNLO in QCD, verifying and
extending the previously established NLO results. These NNLO
conversion factors are essential for achieving precision in the determination
of the bag parameter $\hat{B}_K$, a critical component of the
$\epsilon_K$ calculation. We confirmed the numerical stability of
our results, with residual scale dependence providing an estimate of
potential higher-order corrections.

We use our results to compute world averages of $\hat{B}_K$ at NNLO 
incorporating diverse lattice inputs obtained in different renormalization schemes and
with either 3 or 4 dynamical quarks.
Notably, unlike previous analyses, by matching as needed at the charm threshold at ${\cal O}(\alpha_s^2)$,
we are able to combine the available 3- and 4-flavour lattice results into a single average. We present this
average for both 3 and 4 active quark flavours, i.e.\ as $\hat{B}_K^{(f=3)}$ and $\hat{B}_K^{(f=4)}$. These
averages can be directly used in
the phenomenology of $\epsilon_K$. Our 3-flavour result can be compared to previous
averages involving only 3-flavour lattice inputs and improves both precision and consistency.

The updated central value of $\hat{B}_K$, with a $1\%$ shift and a
reduced uncertainty of under $0.8\%$, represents a significant
improvement. It translates to an updated prediction for indirect CP violation in the Kaon
system,
$	|\epsilon_K|=2.171(65)_\text{pert.}(71)_\text{non-pert.}(153)_\text{param.} \times 10^{-3}$.

Going forward, the enhanced precision on $\hat B_K$ will play a pivotal role in
future phenomenological applications, further constraining the
Standard Model and probing new physics scenarios.

\section*{Acknowledgments}
The work of MG is partially supported by the UK Science and Technology Facilities Council grant ST/X000699/1. 
The work of SK has been funded by Consejería de Universidad, Investigación e Innovación, Gobierno de España and Unión Europea – NextGenerationEU under grants AST22 6.5 and CNS2022-136024 and by MICIU/AEI/10.13039/501100011033 and FEDER/UE (grant PID2022-139466NB-C21).
The work of SJ is supported in part by the UK Science and Technology Facilities Council grant ST/X000796/1.

\appendix

\section{Fierz-evanescent operators}\label{sec:fbasis}
One can present the results in a slightly different manner, by trading the $\tilde Q_i$ for Fierz-evanescent operators
such that only a minimal number of operators contribute to the renormalized Green's function at $D \to 4$ (and hence
to the SMOM projections).
We find that the following operators are evanescent, including the first one which we already had:
\begin{eqnarray}
	E_{1F} &=& \tilde{Q}_1 - Q_1, \\
	E_{2F} &=& Q_2 + \tilde{Q}_2 - \frac{p_1^2}{2} (Q_1 + E_{1F}), \\
	E_{3F} &=& Q_3 + \tilde{Q}_3 - \frac{p_1 \cdot p_2}{2} (Q_1 + E_{1F}), \\
	E_{4F} &=& Q_4 + \tilde{Q}_4 - \frac{p_2^2}{2} (Q_1  + E_{1F} ).
\end{eqnarray}
One can then rearrange e.g.\ Eq.(\ref{eq:lambda}) in terms of the (direct parts of the) matrix elements of $Q_i$ and
$E_{iF}$, which gives the new coefficients
\begin{eqnarray}
	A'_1 &=& A_1 + \tilde{A}_1 + \frac{p_1^2}{2} \tilde{A}_2 + \frac{p_1 \cdot p_2}{2} \tilde{A}_3 + \frac{p_2^2}{2} \tilde{A}_4, \\
	A'_2 &=& A_2 - \tilde{A}_2, \\
	A'_3 &=& A_3 - \tilde{A}_3, \\
	A'_4 &=& A_4 - \tilde{A}_4 .
\end{eqnarray}
The second sum in Eq.(\ref{eq:finalproj}) then disappears without replacement,
as the evanescent operators have zero tree-level
matrix elements. Both methods give the same result.

\section{$\overline{\rm MS}$ renormalization constants}

$\overline{\rm MS}$ renormalization constants $Z(\nu)$ can written as
\begin{equation}\label{eq:Zexpand}
	Z(\nu) = 1 + Z^{(1)} \frac{\alpha(\nu)}{4\pi}
	+ Z^{(2)} \left( \frac{\alpha(\nu)}{4\pi} \right)^2 +  {\cal O}\left( \frac{\alpha(\nu)}{4\pi} \right)^3,
\end{equation}
where each perturbative order $Z^{(n)}$ is expanded in powers of $\epsilon$
\begin{equation}\label{eq:Zexpand2}
	Z^{(n)} = \sum_{m=0}^n Z^{(n, m)} \frac{1}{\epsilon^m}.
\end{equation}
The $m = 0$ term is non-vanishing only for $Z_{E_i Q}(\nu)$.
The wavefunction renormalization constants, given by
\begin{align}\label{eq:MSZq}
	\begin{split}
		&Z_q^{(1,1)}=-C_F\xi,\\
		&Z_q^{(2,2)}=\frac{C_F}{4 N_c}\xi(-\xi+3 N_c^2+2\xi N_c^2),\\
		&Z_q^{(2,1)}=\frac{C_F}{8N_c}(-3-2\xi^2-22 N_c^2-8\xi N_c^2+\xi^2N_c^2+4N_c f)-\frac{C_F^2}{2}\xi^2,
	\end{split}
\end{align}
where $C_F=(N_c^2-1)/(2 N_c)$ is the  quadratic Casimir invariant for the defining representation of $SU(N_c)$ \cite{FRANCO1998641}. The gauge renormalization constant is defined as 
\begin{equation}\label{eq:MSZg}
	Z^{(1,1)}_g=-\beta_0,
\end{equation}
where $\beta_0=\frac{11}{3} N_c - \frac{2}{3} f$ \cite{PhysRevD.86.096008}, and the gauge parameter $Z$-factor is given by
\begin{equation}\label{eq:MSZxi}
	Z_\xi^{(1,1)}=N_c\left(\frac{5}{3}+\frac{1}{2}(1-\xi)\right)-\frac{2}{3}f.
\end{equation}

\section{Operator anomalous dimensions}

In order to perform the conversion between the two schemes at two-loop order, we will need two-loop $\overline{\mbox{MS}}$ renormalization constants. These can usually be extracted from the computation of the amplitude (or at least its poles). Alternatively, they enter the computation of the two-loop anomalous dimensions. Several such computations have been performed specifically for operator $Q$ and can be found in \cite{Buras:1989xd, Brod:2010mj}. In this section we provide the derivation of the anomalous dimensions in our conventions.

In general, renormalized operators $Q_i^{\overline{\rm MS}}$ can be expressed in terms of bare operators $Q_i$, which have a well-defined meaning during the calculation, as defined in Eq.\eqref{eq:Qrengen}. The renormalization group equations of the operators
\begin{equation}
	\nu \frac{d}{d\nu}Q_i^{\overline{\rm MS}}(\nu)=\gamma_{ij} Q_j^{\overline{\rm MS}}(\nu),
\end{equation}
are determined by the anomalous dimension matrix (ADM) $\gamma_{ij}$, defined via
\begin{equation}
	\gamma_{ij}=\nu \frac{d Z_{ik}(\nu)}{d\nu} Z^{-1}_{kj}(\nu).
\end{equation}
Hence, the one-loop ADM is given by
\begin{equation}\label{eq:ADM1}
	\gamma^{(0)}=-2Z^{(1,1)}.
\end{equation}
The coefficient in front of the $1/\epsilon$ pole has to vanish as the ADM has to be finite since it encodes the change of operators with the renormalization scale. Hence, one can obtain an ADM finiteness (or renormalisability) condition as
\begin{equation}\label{eq:ADMfin}
	-2\beta_0 Z^{(1,1)}-4 Z^{(2,2)}+2Z^{(1,1)}Z^{(1,1)}=0.
\end{equation}
Finally, the two-loop ADM is given by
\begin{equation}\label{eq:ADM2}
	\gamma^{(1)}=(-2\beta_0 Z^{(1,0)} - 4 Z^{(2,1)}+2Z^{(1,0)}Z^{(1,1)}+2Z^{(1,1)}Z^{(1,0)}).
\end{equation}
As long as $\gamma^{(0)}$ arises at one-loop, as is the case in our investigation, it is scheme-independent. The $\gamma^{(1)}$ generally depend on the renormalization scheme as $Z^{(1,0)}$ and $Z^{(2,1)}$ usually depend on the choice of the evanescent operators, conventionally chosen such that their Green's functions vanish in four dimensions. In the following two sections we derive the operator renormalization constants for our choice of scheme, corresponding to the evanescent operators defined in Eqs.(\eqref{eq:E1},\eqref{eq:E3}).

\section{One-loop counterterms}\label{app:olc}

Adding all 6 diagrams, for the one-loop amplitude we obtain
\begin{align}
	\begin{split}
		\langle Q\rangle^{\rm 1-loop}&=\frac{\alpha_s}{4\pi}\frac{1}{\epsilon}
		\left\{\left(2C_F \xi - 3\left(1 - \frac{1}{N_c}\right)\right)\langle Q \rangle
		-3\langle E_F\rangle
		+\frac{1}{2N_c}\langle E_1\rangle
		-\frac{1}{2}\langle E_2\rangle\right\}\\
		&\quad+ {\cal O}(\epsilon^0).
	\end{split}
\end{align}
From the requirement $Z_q^2(\nu) \langle Q^{{\rm \overline{MS}}} \rangle =$ finite, where $Z_q (\nu) = 1 - C_F \xi \alpha_s(\nu) /(4\pi \epsilon)$ is the 1-loop $\overline{\rm MS}$
field renormalization constant in a general covariant gauge \cite{BK},
the $\alpha_s/(4\pi)$ coefficients in the $Z$-factors can then be read-off as 
\begin{align}
	Z_{QQ}^{(1,1)}&=3\left(1-\frac{1}{N_c}\right),\label{eq:ZQQ} &Z_{QE_2}^{(1,1)}&=\frac{1}{2},\phantom{\left(\frac{1}{1}\right)}\\
	Z_{QE_F}^{(1,1)}&=3,\phantom{\left(\frac{1}{1}\right)} &Z_{QE_3}^{(1,1)}&=0,\phantom{\left(\frac{1}{1}\right)}\\
	Z_{QE_1}^{(1,1)}&=-\frac{1}{2N_c },\phantom{\left(\frac{1}{1}\right)} &Z_{QE_4}^{(1,1)}&=0.\phantom{\left(\frac{1}{1}\right)}
\end{align}

In addition, for the two-loop renormalization we also require additional one-loop $Z$ factors corresponding to the evanescent operators $E_F$, $E_1$ and $E_2$. We obtain them via insertions of the evanescent operators into the vertices of the one-loop diagrams. Inserting $E_F$ into the vertex yields
\begin{align}
	\begin{split}
		\langle E_F\rangle^{1-loop}&=\frac{\alpha_s}{4\pi}\frac{1}{\epsilon}\left\{
		\left(2C_F \xi + 3 \left(1+\frac{1}{N_c}\right)\right)\langle E_F \rangle\right.\\
		&\quad-\left.\left(\frac{1}{4}+\frac{1}{2\,N_c}\right)\langle E_1\rangle
		+\left(\frac{1 - C_F}{2}+ \frac{1}{4\,N_c}\right)\langle E_2\rangle\right\} + {\cal O}(\epsilon^0),
	\end{split}
\end{align}
giving the constants
\noindent
\begin{align}
	Z_{E_F Q}^{(1,1)}&=0,\phantom{\left(\frac{1}{1}\right)} &Z_{E_FE_2}^{(1,1)}&=\frac{C_F-1}{2}-\frac{1}{4\,N_c},\\
	Z_{E_FE_F}^{(1,1)}&=-3\left(1+\frac{1}{N_c}\right), &Z_{E_FE_3}^{(1,1)}&=0,\phantom{\left(\frac{1}{1}\right)}\\
	Z_{E_FE_1}^{(1,1)}&=\frac{1}{4}+\frac{1}{2\,N_c}, &Z_{E_FE_4}^{(1,1)}&=0.\phantom{\left(\frac{1}{1}\right)}
\end{align}
Similarly, the remaining $Z$ factors have been obtained from the corresponding insertions into the one-loop amplitudes
\begin{align}
	\begin{split}
		\langle E_1\rangle^{1-loop}=&\frac{\alpha_s}{4\pi}\frac{1}{\epsilon}\left\{
		\left(2\,C_F \xi-\frac{13}{N_c}\right)\langle E_1 \rangle
		+13\langle E_2\rangle
		+\frac{1}{2N_c}\langle E_3\rangle
		-\frac{1}{2}\langle E_4\rangle \right\} \\
		+& 24\left(2\,C_F + 1 -\frac{1}{N_c}\right)\langle Q \rangle
		+ \mbox{evanescent},
	\end{split}\\
	\begin{split}
		\langle E_2\rangle^{1-loop}=&\frac{\alpha_s}{4\pi}\frac{1}{\epsilon}\left\{
		5 \langle E_1 \rangle
		+\left(2\,C_F \xi + 16\,C_F-\frac{5}{N_c}\right)\langle E_2\rangle
		-\frac{1}{4} \langle E_3\rangle \right.
		\\
		+&\left.\frac{1}{4}\left(\frac{1}{N_c}-2C_F\right)\langle E_4\rangle \right\} + 48 \left(1 - \frac{1}{N_c} \right) \langle Q \rangle
		+ \mbox{evanescent} ,
	\end{split}
\end{align}
where ``evanescent'' denotes terms that vanish as $D \to 4$. The renormalization constants can then be read-off as

\noindent
\begin{align}
	Z_{E_1E_F}^{(1,1)}&=0,\phantom{\left(\frac{1}{1}\right)} &Z_{E_2E_F}^{(1,1)}&=0, \phantom{\left(\frac{1}{1}\right)}\\
	Z_{E_1E_1}^{(1,1)}&=\frac{13}{N_c},\phantom{\left(\frac{1}{1}\right)} &Z_{E_2E_1}^{(1,1)}&=-5,\phantom{\left(\frac{1}{1}\right)}\\
	Z_{E_1E_2}^{(1,1)}&=-13,\phantom{\left(\frac{1}{1}\right)} &Z_{E_2E_2}^{(1,1)}&=\frac{5}{N_c} - 16\, C_F,\\
	Z_{E_1E_3}^{(1,1)}&=-\frac{1}{2N_c},\phantom{\left(\frac{1}{1}\right)} &Z_{E_2E_3}^{(1,1)}&=\frac{1}{4},\phantom{\left(\frac{1}{1}\right)}\\
	Z_{E_1E_4}^{(1,1)}&=\frac{1}{2},\phantom{\left(\frac{1}{1}\right)} &Z_{E_2E_4}^{(1,1)}&=\frac{1}{4}\left(2\, C_F - \frac{1}{N_c}\right).
\end{align}

The properly renormalized evanescent operators also require a subtraction
of the finite constants multiplying $\langle Q \rangle$, giving
\noindent
\begin{align}
	Z_{E_1Q}^{(1,0)}&=24\left(2\,C_F + 1 -\frac{1}{N_c}\right), &Z_{E_2Q}^{(1,0)}&=48 \left(1 - \frac{1}{N_c} \right).
\end{align}

\section{Two-loop counterterms}\label{app:tlc}

Two-loop Z factors can be extracted from the 2-loop ADM. The Z factor can be expanded as
\begin{equation}
	Z^{(2)}=\left(\frac{1}{\epsilon}Z^{(2,1)}+\frac{1}{\epsilon^2}Z^{(2,2)}\right),
\end{equation}
Recalling the ADM finiteness limit in Eq.\eqref{eq:ADMfin}, given by
\begin{equation}
	4 Z^{(2,2)}+2 \beta_0 Z^{(1,1)} - 2 Z^{(1,1)}Z^{(1,1)}=0,
\end{equation}
where $\beta_0=(11 N_c - 2 f)/3$, we get
\begin{align}
	\begin{split}
		&Z^{(2,2)}_{QQ}=\frac{1}{2}(Z^{(1,1)}_{QQ})^2 -\frac{1}{2}Z^{(1,1)}_{QQ}\beta_0=-\frac{(N_c-1)(N_c(-2f+11N_c-9)+9)}{2 N_c^2},
	\end{split}\\
	\begin{split}
		&Z^{(2,2)}_{QE_F}=\frac{1}{2}(Z^{(1,1)}_{QQ}Z^{(1,1)}_{QE_F}+Z^{(1,1)}_{E_FE_F}Z^{(1,1)}_{QE_F})-\frac{1}{2}Z^{(1,1)}_{QE_F}\beta_0=f-\frac{11N_c}{2}-\frac{9}{N_c}.
	\end{split}
\end{align}
Using Eq.(5.2) of \cite{Buras:1989xd}, which translated to our conventions is written as
\begin{equation}
	\gamma^{(1)}_{QQ}=\frac{(N_c-1)}{2N_c}\left(21-\frac{57}{N_c}+\frac{19}{3} N_c -\frac{4}{3} f\right),
\end{equation}
and Eq.\eqref{eq:ADM2}, given by
\begin{align}
	&\gamma^{(1)}_{QQ}=-4 Z^{(2,1)}_{QQ}+2Z^{(1,1)}_{QE_1}Z^{(1,0)}_{E_1Q},
\end{align}
we get
\begin{align}
	&Z^{(2,1)}_{QQ}=\frac{N_c(-288 C_F + N_c(19N_c-4f+44)+4f-378)+315}{24 N_c^2}.
\end{align}
Technically, $Z^{(2,1)}_{QE_F}$ also enters the two-loop amplitude, however, we find that it drops out of the computation of $A'$, hence it is not essential.

Finally, the change of basis between our and BG evanescent operator basis can be obtained in terms of $1/\epsilon^2$ parts of renormalization constants. To convert to the BG scheme, we require two two-loop $1/\epsilon^2$ pole
coefficients in the evanescent sector. They can be inferred from the Eq.\eqref{eq:ADMfin} as
\begin{align}
	Z^{(2,2)}_{QE_3} &= Z^{(1,1)}_{QE_1} Z^{(1,1)}_{E_1E_3}
	+ Z^{(1,1)}_{QE_2} Z^{(1,1)}_{E_2E_3} = \frac{1}{4 N_c^2} + \frac{1}{8}, \label{eq:BG1}\\
	Z^{(2,2)}_{QE_4} &= Z^{(1,1)}_{QE_1} Z^{(1,1)}_{E_1E_4}
	+ Z^{(1,1)}_{QE_2} Z^{(1,1)}_{E_2E_4} = \frac{C_F}{4} -\frac{3}{8N_c}. \label{eq:BG2}
\end{align}

\bibliography{References}

\end{document}